\PassOptionsToPackage{unicode}{hyperref}
\PassOptionsToPackage{hyphens}{url}
\documentclass[11pt]{article}
\usepackage{amsmath,amssymb}
\usepackage{lmodern}
\usepackage{iftex}
\ifPDFTeX
  \usepackage[T1]{fontenc}
  \usepackage[utf8]{inputenc}
  \usepackage{textcomp} 
\else 
  \usepackage{unicode-math}
  \defaultfontfeatures{Scale=MatchLowercase}
  \defaultfontfeatures[\rmfamily]{Ligatures=TeX,Scale=1}
\fi
\IfFileExists{upquote.sty}{\usepackage{upquote}}{}
\IfFileExists{microtype.sty}{
  \usepackage[]{microtype}
  \UseMicrotypeSet[protrusion]{basicmath} 
}{}
\makeatletter
\@ifundefined{KOMAClassName}{
  \IfFileExists{parskip.sty}{%
    \usepackage{parskip}
  }{
    \setlength{\parindent}{0pt}
    \setlength{\parskip}{6pt plus 2pt minus 1pt}}
}{
  \KOMAoptions{parskip=half}}
\makeatother
\usepackage{xcolor}
\usepackage{color}
\usepackage{fancyvrb}

\DefineVerbatimEnvironment{Highlighting}{Verbatim}{commandchars=\\\{\}}
\usepackage{framed}
\definecolor{shadecolor}{RGB}{248,248,248}
\newenvironment{Shaded}{\begin{snugshade}}{\end{snugshade}}

\newcommand{\AttributeTok}[1]{\textcolor[rgb]{0.77,0.63,0.00}{{\small #1}}}

\newcommand{\CommentTok}[1]{\textcolor[rgb]{0.56,0.35,0.01}{\textit{{\small #1}}}}

\newcommand{\ConstantTok}[1]{\textcolor[rgb]{0.00,0.00,0.00}{{\small #1}}}
\newcommand{\ControlFlowTok}[1]{\textcolor[rgb]{0.13,0.29,0.53}{\textbf{#1}}}

\newcommand{\DecValTok}[1]{\textcolor[rgb]{0.00,0.00,0.81}{{\small #1}}}

\newcommand{\FloatTok}[1]{\textcolor[rgb]{0.00,0.00,0.81}{{\small #1}}}
\newcommand{\FunctionTok}[1]{\textcolor[rgb]{0.00,0.00,0.00}{{\small #1}}}

\newcommand{\NormalTok}[1]{{\small #1}}

\newcommand{\OtherTok}[1]{\textcolor[rgb]{0.56,0.35,0.01}{{\small #1}}}

\newcommand{\SpecialCharTok}[1]{\textcolor[rgb]{0.00,0.00,0.00}{{\small #1}}}

\newcommand{\StringTok}[1]{\textcolor[rgb]{0.31,0.60,0.02}{{\small #1}}}

\usepackage{longtable,booktabs,array}
\usepackage{calc} 
\usepackage{etoolbox}
\makeatletter
\patchcmd\longtable{\par}{\if@noskipsec\mbox{}\fi\par}{}{}
\makeatother
\IfFileExists{footnotehyper.sty}{\usepackage{footnotehyper}}{\usepackage{footnote}}
\makesavenoteenv{longtable}
\usepackage{graphicx}
\makeatletter
\def\maxwidth{\ifdim\Gin@nat@width>\linewidth\linewidth\else\Gin@nat@width\fi}
\def\maxheight{\ifdim\Gin@nat@height>\textheight\textheight\else\Gin@nat@height\fi}
\makeatother
\setkeys{Gin}{width=\maxwidth,height=\maxheight,keepaspectratio}
\makeatletter
\def\fps@figure{htbp}
\makeatother
\providecommand{\tightlist}{%
  \setlength{\itemsep}{0pt}\setlength{\parskip}{0pt}}
  \usepackage{fancyhdr}
  \usepackage{algorithm,algorithmic}
  \usepackage{bm}
  \usepackage[margin=1in]{geometry}
  \usepackage{colortbl}

\newcommand{\cH}{\mathcal{H}}

\newcommand{\cD}{\mathcal{D}}

\newcommand{\E}{\mathbb{E}}

\DeclareMathOperator*{\sspan}{span}
\DeclareMathOperator{\mean}{mean}

\ifLuaTeX
  \usepackage{selnolig}  
\fi
\usepackage[numbers]{natbib}
\IfFileExists{bookmark.sty}{\usepackage{bookmark}}{\usepackage{hyperref}}
\IfFileExists{xurl.sty}{\usepackage{xurl}}{} 
\hypersetup{
  pdftitle={: Towards a Unified Implementation of Regression Monte Carlo Algorithms},
  pdfauthor={Mike Ludkovski},
  hidelinks,
  pdfcreator={LaTeX via pandoc}}

\title{\texttt{mlOSP}: Towards a Unified Implementation of Regression Monte Carlo Algorithms}
\author{Mike Ludkovski\footnote{Department of Statistics and Applied Probability, University of California, Santa Barbara, 93106-3110. Work partially supported by NSF DMS-1821240. I also thank Xiong Lyu and Heming Chen for their contributions to the R package. The underlying source file was knitted from a RMarkdown document via pandoc. \href{mailto:ludkovski@pstat.ucsb.edu}{\nolinkurl{ludkovski@pstat.ucsb.edu}}}}
\date{October 1, 2022}

\begin{document}
\maketitle
\begin{abstract}
We introduce \texttt{mlOSP}, a computational template for Machine Learning for Optimal Stopping Problems. The template is implemented in the \texttt{R} statistical environment and publicly available via a \texttt{GitHub} repository. \texttt{mlOSP} presents a unified numerical implementation of Regression Monte Carlo (RMC) approaches to optimal stopping, providing a state-of-the-art, open-source, reproducible and transparent platform. Highlighting its modular nature, we present multiple novel variants of RMC algorithms, especially in terms of constructing simulation designs for training the regressors, as well as in terms of machine learning regression modules. Furthermore, \texttt{mlOSP} nests most of the existing RMC schemes, allowing for a consistent and verifiable benchmarking of extant algorithms. The article contains extensive \texttt{R} code snippets and figures, and serves as a vignette to the underlying software package.
\end{abstract}

\hypertarget{introduction}{%
\section{Introduction}\label{introduction}}

Numerical resolution of optimal stopping problems has been an active area of research for more than two decades. Originally investigated in the context of American option pricing, it has since metamorphosed into a field unto itself, with wide-ranging applications and dozens of proposed approaches.
A major strand that is increasingly dominating the subject is simulation-based methods rooted in the Monte Carlo paradigm. Developed in the late 1990s by Longstaff \& Schwartz \citep{LS} and Tsitsiklis \& van Roy \citep{TsitsiklisVanRoy} this framework remains without an agreed-upon name; I shall refer to it as Regression Monte Carlo (RMC). The main feature of RMC is its marriage of a probabilistic approach, namely simulation of the underlying stochastic state dynamics, and statistical tools for approximating the quantities of interest: the value and/or continuation functions, and the stopping region. This combination of simulation and statistics brings scalability, flexibility in terms of underlying model assumptions, and a vast arsenal of potential implementations. These benefits have translated into excellent performance which has made RMC popular both in the academic and practitioner quantitative finance communities. Indeed, in the opinion of the author, these developments can claim to be the most successful numerical strategy that emerged from Financial Mathematics. RMC has by now percolated down into standard Masters-level curriculum, and is increasingly utilized in cognate engineering and mathematical sciences domains.

Despite hundreds of journal publications presenting and analyzing variants of RMC (see e.g.~the surveys \citep{Kohler10review, NadarajahSecomandi17} and the monograph by \citet{BelomestnyBook}), there remains a dearth of user-friendly benchmarks or unified overviews of the algorithms. In particular, to my knowledge, there is no \texttt{R} (or other common programming environments, such as \texttt{Python}) package for RMC. Small stand-alone versions exist in \citep{LSMonteCarlo, rlsm} but are essentially unknown by the research community; see also \citep{StOpt} that has a broader scope and is primarily written in C++. One reason for this gap is that many research articles tend to explore one narrow aspect of RMC and then illustrate their contributions on a few idiosyncratic examples. As a result, it is difficult to independently evaluate the relative pros and cons or to explore the merit of new ideas.

In the present article, I offer an algorithmic template for RMC, dubbed \textbf{mlOSP} (Machine Learning for Optimal Stopping Problems), coupled with its implementation within the \texttt{R} programming language. This endeavor is meant to be a ``living'' project, offering a platform to centralize and standardize new RMC approaches as they are proposed. The associated \texttt{\{mlOSP\}} library \citep{mlOSP} was first released in late 2020; by September 2022 it is in its Version 1.2 incarnation, and is expected to have further (hopefully regular) updates.

\hypertarget{contributions}{%
\subsection{Contributions}\label{contributions}}

The \textbf{mlOSP} template detailed below offers a unified description of RMC via the underlying statistical concepts.
The aim of \textbf{mlOSP} is a \emph{generic} RMC template that emphasizes the building blocks, rather than specific approaches. This perspective therefore tries to \emph{nest} as many of the numerous existing approaches as possible, shedding light on the differences and the similarities between them.
To specify \textbf{mlOSP}, I utilize statistical terminology (in contrast to financial or probabilistic lingo). In particular, I place RMC in the context of modern statistical machine learning, highlighting links to Design of Experiments, Stochastic Simulation, and Sequential Learning literatures.

By providing a templated perspective and tying disparate extant concepts, \textbf{mlOSP} allows for numerous new twists, extensions, and variants to ``classical'' RMC. To this end, in this article and in the respective \texttt{R} code I discuss and showcase the following new RMC flavors:

\begin{enumerate}
\def\labelenumi{\roman{enumi})}
\item
  Space-filling training designs (multiple variants in Section 4.1);
\item
  Variable simulation design size \(N_k\) across time-steps (Section 4.1);
\item
  Adaptive simulation designs with varying batch sizes generated sequentially (Section 4.3);
\item
  Cross-validated kernel regression emulators (Section 5.1);
\item
  Relevance Vector Machine emulators (Section 5.1);
\item
  Heteroskedastic and local Gaussian Process emulators (Section 5.2).
\end{enumerate}

As a further contribution, I extend \textbf{mlOSP} to the context of multiple-stopping problems, designing its analogue for valuing Swing options (Section 7.1). To my knowledge, this is the first paper that applies concepts of simulation design and advanced statistical emulators to swing option pricing.
This generalization hints at further possibilities of building upon \textbf{mlOSP} to develop an interlocking collection of computational tools for decision-making under uncertainty.

This article moreover contributes to the literature by providing a comprehensive \emph{benchmarking} of different RMC schemes across solver types and problem instances. By providing fully reproducible results, I aim to initiate a common collection of testbeds (in the spirit of say the UCI ML repository \url{http://archive.ics.uci.edu/ml}) that allow for a transparent, apples-to-apples comparison and can be easily augmented by other researchers.

\hypertarget{organization-of-the-article}{%
\subsection{Organization of the article}\label{organization-of-the-article}}

The eponymous \texttt{R} package \texttt{\{mlOSP\}} \citep{mlOSP} is a central companion to this article. Accordingly, I added many \texttt{R} code snippets throughout, providing a ``User Guide'' to \texttt{\{mlOSP\}}. Conversely, the code is used to illustrate and interpret the discussion. Appendices provide further information about the \texttt{R} implementation.

Section 2 presents the \textbf{mlOSP} template, emphasizing the three blocks of the RMC framework: stochastic simulation, training design, and statistical approximation. These pieces are modularized and can be fully mixed-and-matched within the core backward dynamic programming loop. Section 3 provides an illustration of this fundamental concepts in \texttt{\{mlOSP\}} serving as a short vignette. Sections 4 and 5 explore the two key parts of the template: simulation designs and regression emulators.
Section 6 benchmarks some of the implemented schemes, running 10 \texttt{\{mlOSP\}} solvers on 9 instances of Bermudan options in 1-5 dimensions. The solvers range from piecewise linear regression, to RVM kernel regression, to Gaussian process emulators. It also includes additional experimental results related to scheme stability and fine-tuning various algorithm ingredients.
Section 7 discusses extensions: Section 7.1 showcases a generalization to handle multiple-stopping problems and Section 7.2 goes through the steps to define a new OSP instance within the package. Section 8 concludes.

\hypertarget{the-mlosp-template}{%
\section{The mlOSP template}\label{the-mlosp-template}}

\hypertarget{regression-monte-carlo-methodology}{%
\subsection{Regression Monte Carlo methodology}\label{regression-monte-carlo-methodology}}

An optimal stopping problem (OSP) is described through two main ingredients: the state process and the reward function. I shall use \(X= (X(t))\) to denote the state process, assumed to be a stochastic process indexed by the time \(t\) and taking values in \({\cal X} \subseteq \mathbb{R}^d\). The reward function \(h(t,x): [0,T] \times \cal{X} \to \mathbb{R}\) represents the net present value (in dollars) of stopping at time \(t\) in state \(x\), where the notation emphasizes the common possibility of the reward depending on time, e.g.~due to discounting. I assume the standard probabilistic structure of \((\Omega, {\cal F}, ({\cal F}_t), \mathbb{P})\), where \(X\) is adapted to the filtration \(({\cal F}_t)\), and \(h(t,\cdot) \in L^2(\mathbb{P})\).

We seek the decision rule \(\tau\), a \emph{stopping time} to maximize expected reward until the horizon \(T\):
\begin{align}\mathbb{E} \left[ h(\tau,X(\tau)) \right] \rightarrow \max! \end{align}
To this end, we wish to evaluate the \textbf{value function} \(V : [0,T] \times \cal{X} \to \mathbb{R}\),
\begin{align}
V(t,x) := \sup_{\tau \in \mathfrak{S}_t} \mathbb{E} \left[ h(\tau, X(\tau)) |\; X(t) = x \right],
\end{align}
where \(\mathfrak{S}_t\) is the collection of all \(({\cal F}_t)\)-stopping times bigger than \(t\) and less than or equal to \(T\).

The state \((X(t))\) is typically assumed to satisfy a Stochastic Differential Equation of Îto type,
\begin{align}\label{eq:sde}
dX(t) = \mu(X(t)) \, dt + \sigma(X(t))\, dW(t),
\end{align}
where \((W(t))\) is a (multi-dimensional) Brownian motion and the drift \(\mu(\cdot)\) and volatility \(\sigma(\cdot)\) are smooth enough to yield a unique strong solution to \eqref{eq:sde}. While utilized in all presented case studies, the structure of \eqref{eq:sde} is not essential beyond its Markov property, and other stochastic processes \(X\), for example jump-diffusions, can be treated interchangeably. Note that while I use \(\mathbb{P},\mathbb{E}\) to denote the probability and expectation operators, financially we are working under the risk-neutral ``\(\mathbb{Q}\)''-measure that is the relevant one for contingent claim valuation.

For the remainder of the section I adopt the discrete-time paradigm, where decisions are made at \(K\) pre-specified instances \(t_0=0 < t_1 < \ldots < t_k < t_{k+1} < \ldots < t_K = T\), where typically \(t_k = k\Delta t\) for a given discretization step \(\Delta t\). Henceforth, I index time steps by \(k\) and work with \({\cal T} = (t_k)_{k=0}^K\).

It is most intuitive to think of optimal stopping as dynamic decision making. At each exercise step \(k\), the controller must decide whether to stop (0) or continue (1), which within a Markovian structure is encoded via the action strategy \({\cal A}=(A_{0:K}(\cdot))\) with each \(A_k(x) \in \{ 0,1 \}\). The action map \(A_k\) gives rise to the stopping region
\[ {\cal S}_k := \{ x : A_k(x) = 0\} \subseteq {\cal X}, \]
where the decision is to stop, and in parallel defines the corresponding first hitting time \(\tau_{A_{k:K}} := \inf \{ \ell \ge k :\, A_\ell(X(\ell)) = 0 \} \wedge K\) which is an optimal exercise time after \(k\).
Hence, solving an OSP is equivalent to classifying each pair \((x \in {\cal X}, t_k\in {\cal T})\) into \({\cal S}_k\) or its complement the continuation set. Recursively, the action set \(A_k\) is characterized as
\begin{align}
\label{eq:A} A_k(x) = 1 \quad \Longleftrightarrow \quad \E \left[ h(\tau_{A_{k+1:K}}, X({\tau_{A_{k+1:K}}})) | \, X(k) = x \right] > h(k,x),
\end{align}
i.e.~one should continue if the expected \emph{reward-to-go} on the left of \eqref{eq:A} dominates the immediate payoff on the right. Denoting the step-ahead conditional expectation of the value function by
\begin{align}\label{eq:q} q(k,x) := \mathbb{E} \left[ V(k+1, X({k+1})) | X(k) =x  \right]
\end{align}
and using the Dynamic Programming principle, we can equivalently write
\(A_k(x) = 1 \; \Longleftrightarrow \; q(k,x) > h(k,x)\)
because we have \(V(k+1,X(k+1)) = \E\left[ h(\tau_{A_{k+1:K}}, X({\tau_{ A_{k+1:K}}})) | X(k+1) \right]\). The \(q\)-value \(q(k,\cdot)\) is also known as the continuation value.

The above logic yields a recursive algorithm to construct approximate action maps \(\widehat{A}_k\)'s through iterating on either \eqref{eq:q} or \eqref{eq:A}. Thus, the RMC framework generates functional approximations of the continuation values \(\hat{q}(k,\cdot)\) in order to build \(\widehat{A}_k(\cdot)\). This is initialized with \(\widehat{V}(K,\cdot) \equiv h(K,\cdot)\) and proceeds with the following loop:

For \(k=K-1,\ldots,1\) repeat:

\begin{enumerate}
\def\labelenumi{\roman{enumi})}
\item
  Learn the \(q\)-value \(\hat{q}(k,\cdot)\);
\item
  Set \(\widehat{A}_k(\cdot) := 1_{\{ \hat{q}(k,\cdot) > h(k,\cdot) \} }\) and
  \(\widehat{V}(k,\cdot) := \max \bigl( \hat{q}(k,\cdot), h(k,\cdot) \bigr)\).
\end{enumerate}

The loop makes it clear that RMC hinges on a sequence of probabilistic function approximation tasks that are recursively defined. Note that the principal step i) of learning the continuation function can be represented in multiple ways:

-- of approximating
\(x\mapsto {\E}\left[ \widehat{V} \bigl(k+1, X(k+1) \bigr) \big| \, X(k) = x \right]\);

-- of approximating
\(x\mapsto \E\left[ h \bigl(\tau_{\widehat{A}_{k+1:K}}, X({\tau_{ \widehat{A}_{k+1:K}}}) \bigr) \big|\, X(k) = x\right]\);

-- by picking some \(1 \le w \le K-k\) and approximating
\[x \mapsto \E\left[ h \bigl(\tau_{\widehat{A}_{k+1:k+w}}, X({\tau_{ \widehat{A}_{k+1:k+w}}}) \bigr) 1_{\{\tau_{ \widehat{A}_{k+1:k+w}} < k+w\}} + \widehat{V}(k+w,X(k+w))1_{\{\tau_{ \widehat{A}_{k+1:k+w}} = k+w\}} \big|\, X(k) = x\right].
\]
Above, the look-ahead horizon \(w \in\{ 1,\ldots, K-k\}\) allows to combine both pathwise rewards based on \(\tau_{\widehat{A}_{k+1:k+w}}\,\) and the approximate value function \(\widehat{V}(k+w,\cdot)\) \(w\)-steps into the future \citep{EgloffKohlerTodorovic07}. These choices are not numerically identical, because \(\widehat{V}(k+1, x) \neq \E\left[ h(\tau_{\widehat{A}_{k+1:K}}, X({\tau_{ \widehat{A}_{k+1:K}}})) |\, X(k+1)=x \right]\) due to the approximation error.

As a final step, once all the action sets are computed they induce the expected reward
\begin{align}\label{eq:tildeV}
\widetilde{V}(0,x; \widehat{A}_{0:K}) := \E\left[ h \bigl(\tau_{\widehat{A}_{0:K}}, X({\tau_{ \widehat{A}_{0:K}}}) \bigr) \big|\, X(0) = x\right].
\end{align}
The gap between \(\tilde{V}(0,x)\) and \(V(0,x)\) (which is the maximal possible expected reward) is the performance metric for evaluating \(\widehat{A}_{0:K}\).

\hypertarget{workflow}{%
\subsection{Workflow}\label{workflow}}

The key step requiring numerical approximation in the above loop is learning the continuation functions \(q(k,\cdot)\). In the RMC paradigm it is handled by re-interpreting conditional expectation as the mean response within a stochastic model. Thus, given an input \(x\) and fixing \(k\), there is a generative model \(x \mapsto q(k,x)\) which is not directly known but accessible through a pathwise reward \emph{simulator} \(x \mapsto Y(x)\), where \(Y(x)\) is a random variable with mean \(\E[ Y(x)] = q(k,x)\) and finite variance. The aim is then to \emph{predict} the mean output of this simulator for an arbitrary \(x\). This learning task is resolved by running the simulator many times and then utilizing a statistical model to capture the observed input-output relationship. It can be broken further down into three sub-problems:

\begin{enumerate}
\def\labelenumi{\arabic{enumi}.}
\item
  Defining the stochastic simulator;
\item
  Defining the simulation design;
\item
  Defining the regression model.
\end{enumerate}

Recasting in the machine learning terminology, for sub-step 1 we need to define a \emph{Simulation Device} that accepts an input \(x \in \cal{X}\) (the state at time \(k\)) and returns a random sample \(Y(x)\), which is a random realization of the pathwise reward starting at \((k,x)\) such that \(\mathbb{E}[Y(x)] = q(k,x)\). We have already discussed multiple versions of such simulators, cf.~\eqref{eq:A}-\eqref{eq:q}. In sub-step 2, we then need to decide which collection of \(x\)'s should be applied as a training set. After selecting such \emph{Simulation Design} of some size \(N\), \(x^{1:N}\), we collect the simulation outputs \(y^{1:N} = Y(x^{1:N})\) and reconstruct the model
\begin{align}\label{eq:stat-model}
Y(x) = f(x) + \epsilon(x), \qquad \E[ \epsilon(x)] = 0, \quad\mathbb{V}ar(\epsilon(x)) = \sigma^2(x),
\end{align}
where the inferred \emph{Emulator} \(f(x) \equiv \hat{q}(k,x)\) is the approximate continuation value.

The above 3 sub-steps constitute one iteration of the overall loop; the aggregate RMC is
a \textbf{sequence} of tasks, indexed by \(k\). With that in mind, the following remarks are helpful to guide the implementation:

\begin{itemize}
\item
  The RMC tasks are recursive, the sub-problem at step \(k\) is linked to the previous simulators/emulators at steps \(\ell > k\). Therefore, approximation errors will back-propagate.
\item
  While the tasks are inter-related, since they are solved one-by-one, there is a large scope for modularization, adaptation, etc., to be utilized---the solvers need not be identical across \(k\).
\item
  When deciding how to train \(\hat{q}\), there is no ``data'' per se and the controller is fully in charge of selecting what simulations to run. Judicious choice of how to do so is an important criterion of numerical efficiency.
\item
  In classical machine learning tasks, there is a well-defined loss function that quantifies the quality of the constructed approximation \(\hat{q}\). In OSP, this loss function is highly implicit; ultimately we judge algorithm performance in terms of the quality of the overall \(\widehat{A}_{0:K}\) in \eqref{eq:tildeV}.
\item
  The stochasticity in RMC is deeply embedded in all the pathwise simulators and the dependence across time-steps. Understanding the sources of this stochasticity is critical for maximizing performance.
\item
  The ``error term'' \(\epsilon(x)\) in \eqref{eq:stat-model} is simulation noise coming from the \(Y\)-simulator. As a result, its statistical properties tend to be quite complex and non-Gaussian. For example, in the common scenario where the reward has a lower bound of zero \(h(t,x) \ge 0\), \(\epsilon(x)\) has a mixed distribution with a point mass.
\item
  Like in any statistical model, we may distinguish between a training set used to construct the emulators---the functional approximations \(\hat{q}\)'s, and a subsequent test set used to evaluate the quality of the resulting exercise rule \(\hat{\tau}\).
\end{itemize}

\hypertarget{illustration-with-a-toy-example}{%
\subsection{\texorpdfstring{Illustration with a Toy Example \label{sec:toy}}{Illustration with a Toy Example }}\label{illustration-with-a-toy-example}}

To illustrate the recursive construction of \(\hat{q}\)'s, we proceed with a toy example. We take the point of view of a buyer of a Bermudan Put. The contract has an expiration date \(T\), a strike \(\cal K\), and exercise frequency \(\Delta t\) (such as daily).
The state \(X(t)\) is interpreted as the share price of the underlying asset at date \(t\), with state space \(\cal{X} = \mathbb{R}_+\) and is assumed to follow Geometric Brownian motion with constant coefficients \(r,\sigma\).
Taking into account the time-value of money, the reward function at time \(t\) is
\[
  h_{Put}(t,x) = e^{-r t} ({\cal K}-x)_+,
\]
where \(r\) is the constant risk-free interest rate. With this setup, the stopping set \(\mathcal{S}_k\) is known as the \emph{exercise region}. For the Black-Scholes Bermudan Put it is known that \(\mathcal{S}_k = [0,\bar{s}_k]\), i.e.~one should stop as soon as the asset price drops below the exercise thresholds \(\bar{s}_k\).

\begin{figure}
\centering
\includegraphics{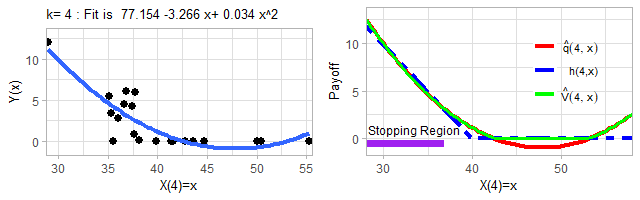}
\caption{\label{fig:unnamed-chunk-2}\label{fig:quad-reg-4} Bermudan Put toy example at \(k=4\). Left: Quadratic fit for the continuation value \(q(k,\cdot)\). Right: Fitted continuation value \(\hat{q}(k,\cdot)\), payoff \(h(k,\cdot)\), value function \(\widehat{V}(4,\cdot)\) and the stopping region \({\cal S}_4\).}
\end{figure}

We assume that \({\cal K}=X(0)=40\), \(T=0.5\) and there are \(K=5\) periods, \(\Delta t= 0.1\), and illustrate how the RMC steps i)-ii)-iii) might be implemented at \(k=4\). Because we are at the penultimate step, the simulation device is simply a one-step-ahead sampling of the terminal payoff \(h_{Put}(K,\cdot)\), i.e.\\
\begin{align}\label{eq:sim-device}
Y(x) := e^{-5 r \Delta t} \cdot \left[{\cal K} - x \cdot \exp \left( (r-0.5\sigma^2)\Delta t + \sigma \sqrt{\Delta t} \cdot \xi \right) \right]_+,
\end{align}
where \(\xi \sim {\cal N}(0,1)\) is the Brownian shock. Note that \(\E[ Y(x)] = q(4,x)\): since \(V(5,x) = h(5,x)\), the \(q\)-value at \(k=4\) is \(q(4,x) := \mathbb{E} \left[ e^{-5r \Delta t}({\cal K}-X(5))_+ | X(4) = x \right]\), which is indeed the expectation of \(Y(x)\). For the Simulation Design I draw \[x^n \sim LogNorm \left( 40\exp[(r-0.5\sigma^2)4\Delta t], \sigma \sqrt{4 \Delta t}\right), \qquad n=1,\ldots,N\] i.i.d., interpreted as the realized values at \(k=4\) obtained by simulating paths \(x^n(0:K)\) of \(X\) starting with the fixed initial share price \(X(0)=40\). Specifically, I take \(N=20\) inputs \(x^{1:N}(k)\) at step \(4\) (note that the randomness in \(x^n\) is superfluous to the algorithm and in particular independent of the \(\xi\)'s in \(Y(x)\)). The corresponding \(y^{1:N}(k)\), \(y^n(k) \sim Y(x^n)\) are then interpreted as \emph{realized} future payoffs at \(K=5\) if one waits one more period. Because the \(\xi\)'s inside \(Y(x)\) are sampled in an i.i.d. manner, the \(y^n\)'s are independent draws from the conditional payoff distribution given \(X(4)\). With a path-based perspective, these payoffs are simply \(e^{-r T}({\cal K} -x^{1:N}(5))_+\) across independent paths of \(X\). The left panel of Figure \ref{fig:quad-reg-4} visualizes the resulting scatter plot of these input-output pairs \(x^{1:N}(4), y^{1:N}(4)\) for \(r=0.05, \sigma=0.2\).

\begin{Shaded}
\begin{Highlighting}[]
\NormalTok{lsModel }\OtherTok{\textless{}{-}} \FunctionTok{list}\NormalTok{(}\AttributeTok{dim=}\DecValTok{1}\NormalTok{,}\AttributeTok{r=}\FloatTok{0.05}\NormalTok{,}\AttributeTok{sigma=}\FloatTok{0.2}\NormalTok{,}\AttributeTok{div=}\DecValTok{0}\NormalTok{,}\AttributeTok{dt=}\FloatTok{0.1}\NormalTok{,}\AttributeTok{T=}\FloatTok{0.5}\NormalTok{)}
\NormalTok{rmc }\OtherTok{\textless{}{-}} \FunctionTok{list}\NormalTok{(); paths }\OtherTok{\textless{}{-}} \FunctionTok{list}\NormalTok{(); payoff }\OtherTok{\textless{}{-}} \FunctionTok{list}\NormalTok{()}
\NormalTok{x0 }\OtherTok{\textless{}{-}} \DecValTok{40}\NormalTok{;  Np }\OtherTok{\textless{}{-}} \DecValTok{20}\NormalTok{; Nt }\OtherTok{\textless{}{-}} \DecValTok{5}\NormalTok{; }\FunctionTok{set.seed}\NormalTok{(}\DecValTok{262}\NormalTok{)}
\NormalTok{paths[[}\DecValTok{1}\NormalTok{]] }\OtherTok{\textless{}{-}} \FunctionTok{sim.gbm}\NormalTok{( }\FunctionTok{matrix}\NormalTok{(}\FunctionTok{rep}\NormalTok{(x0, Np), }\AttributeTok{nrow=}\NormalTok{Np,}\AttributeTok{byrow=}\NormalTok{T), lsModel, lsModel}\SpecialCharTok{$}\NormalTok{dt)}
\ControlFlowTok{for}\NormalTok{ (j }\ControlFlowTok{in} \DecValTok{2}\SpecialCharTok{:}\NormalTok{(Nt}\SpecialCharTok{+}\DecValTok{1}\NormalTok{))  }\CommentTok{\# Save paths of X: paths[[j]] = X( j \textbackslash{}Delta t)}
\NormalTok{    paths[[j]] }\OtherTok{\textless{}{-}} \FunctionTok{sim.gbm}\NormalTok{( paths[[j}\DecValTok{{-}1}\NormalTok{]], lsModel, lsModel}\SpecialCharTok{$}\NormalTok{dt)}
\end{Highlighting}
\end{Shaded}

To infer the continuation value function \(q(4, \cdot)\) we apply regression, specifically let me fit a quadratic function of the input \(x\) to predict the expected value of \(Y(x)\). In other words I \emph{postulate} that \(\hat{q}(4,x) = \beta_0 + \beta_1 x + \beta_2 x^2\) or
\begin{align}\label{eq:quadratic-y}
Y(x) = \beta_0 + \beta_1 x + \beta_2 x^2 + \epsilon,
\end{align}
where the regression coefficients \(\beta_{0:2}\) parametrize the unknown continuation function \(q(4, \cdot)\) we are after, and \(\epsilon \sim \mathcal{N}(0,\sigma_\epsilon^2)\) is assumed to be Gaussian. As is standard, we fit \(\beta\)'s via Ordinary Least Squares, i.e.~by minimizing a quadratic penalty function which is the residual sum of squares between observed and predicted values. In \texttt{R} this is achieved by the \texttt{lm} command as follows:

\begin{Shaded}
\begin{Highlighting}[]
\NormalTok{payoff[[}\DecValTok{4}\NormalTok{]] }\OtherTok{\textless{}{-}} \FunctionTok{exp}\NormalTok{(}\SpecialCharTok{{-}}\FloatTok{0.05}\SpecialCharTok{*}\FloatTok{0.5}\NormalTok{)}\SpecialCharTok{*}\FunctionTok{pmax}\NormalTok{(}\DecValTok{40}\SpecialCharTok{{-}}\NormalTok{paths[[}\DecValTok{5}\NormalTok{]],}\DecValTok{0}\NormalTok{)  }\CommentTok{\# outputs Y(x). Input is paths[[4]]}
\NormalTok{rmc[[}\DecValTok{4}\NormalTok{]] }\OtherTok{\textless{}{-}} \FunctionTok{lm}\NormalTok{(y }\SpecialCharTok{\textasciitilde{}} \FunctionTok{poly}\NormalTok{(x,}\DecValTok{2}\NormalTok{,}\AttributeTok{raw=}\ConstantTok{TRUE}\NormalTok{), }\FunctionTok{data.frame}\NormalTok{(}\AttributeTok{y=}\NormalTok{ payoff[[}\DecValTok{4}\NormalTok{]], }\AttributeTok{x =}\NormalTok{ paths[[}\DecValTok{4}\NormalTok{]]))}
\end{Highlighting}
\end{Shaded}

\emph{Remark}: It is instructive to compare the assumed \eqref{eq:quadratic-y} with the exact simulation device in \eqref{eq:sim-device}. One of the core concepts of RMC is to abstract away from the ``innards'' of \eqref{eq:sim-device} and substitute with a statistical treatment as in \eqref{eq:quadratic-y}.

The blue line in the right panel of Figure \ref{fig:quad-reg-4} is our resulting estimate of expected payoff at the terminal time step \(K=5\) given \(X(4) = x\). Arithmetically, we have
\[\mathbb{E}[ e^{-r T}({\cal K}-X(5))_+ | X(4) =x ] \simeq \hat{q}(4,x) := 77.154  -3.266 x+ 0.034 x^2.\]
So for example \(\hat{q}(4,40)= 1.15\) and \(\hat{q}(4,35)\) = 4.674. Proceeding with step ii) of the RMC loop, we may conclude that \(\widehat{A}_4(40)=1\) (continue at \(X(4)=40\)) and \(A_4(35)=0\) (stop at \(X(4)=35\), since \(\hat{q}(4,35)<h(4,35)=e^{-0.4 \cdot 0.05}(40-35)_+ = 4.900\)).
Clearly \(\hat{q}(4,\cdot)\) is an approximation: the conditional expectation is not \emph{actually} quadratic, but since we do not know what it is, we simply \emph{model} it via that fitted quadratic function.

Note that the output of the regression is \emph{not} any particular prediction, but the \emph{object} \(\hat{q}(4,\cdot)\), so that the corresponding decision rule \(\widehat{A}_4(\cdot)\) is implicit. In fact, RMC never requires explicitly characterizing \(\widehat{A}_k(\cdot)\); all we need is to be able to evaluate \(\widehat{A}_k(x)\) at any inputted \(x\). In Figure \ref{fig:quad-reg-4} I solve for the stopping boundary \(\bar{s}_4\) by searching at which threshold stopping becomes more profitable than continuing and shading the resulting stopping region \({\cal S}_4\). Obtaining the exercise threshold \(\bar{s}_4\) is purely ``cosmetic'' and plays no role in the solver.

The above completes one step of the RMC recursion, defining \(\widehat{V}(4,\cdot)\) and (implicitly) \(\widehat{A}_4(\cdot)\). We can then continue by repeating the above logic over \(k=3,2,1\) and using \eqref{eq:q}, i.e.~simulating the one-step-ahead value function to obtain \(Y\)'s. Figure \ref{fig:tvr-panels} in Appendix B illustrates this sequential estimation of \(\hat{q}\)'s, utilizing \(N=20\) training inputs sampled from a log-normal distribution at each \(k\), and then constructing quadratic fits.

The above choices of the simulation design (and its size) and regression model are arbitrary and for motivation only. In the next section, they are generalized and scaled up into a generic template that is the main object of study for the remainder of the paper.

\hypertarget{dynamic-emulation-template}{%
\subsection{Dynamic Emulation Template}\label{dynamic-emulation-template}}

Before presenting the template in Algorithm \ref{alg:1} below, several conceptual adjustments are necessary. First, in order to abstract from the payoff function specifics, \textbf{mlOSP} operates with the \emph{timing value}, \[T(k,x) := q(k,x) - h(k,x).\] Exercising is optimal when the timing value is negative \(A_k(x) = 1 \Leftrightarrow T(k,x) \ge 0\); the stopping boundary is the zero contour of \(T(k,\cdot)\). So instead of learning \(q(k,\cdot)\), \textbf{mlOSP} learns \(T(k,\cdot)\), solely a numerical rather than conceptual change. Given \(\widehat{T}(k,x)\) one can recover the value function (and hence the option price) via
\(\hat{V}(k,x)= h(k, x) + \max \bigl(0, \widehat{T}(k, x) \bigr)\).

Second, the abstract specification of the timing value emulator \(\widehat{T}(k,\cdot)\) is as the empirical minimizer in some function space \({\cal H}_k\) of the mean squared error from the observations,
\begin{align}\label{eq:H-min}
\widehat{T}(k,\cdot) := \arg \min_{f \in {\cal H}_k} \sum_{n=1}^{N_k} (f(x^n(k)) - y^n)^2.
\end{align}
For instance, in the linear model as above, \({\cal H}_k = \sspan(B_1, \ldots, B_{R-1})\) are the linear combinations of the basis functions \(B_r(\cdot)\) and \eqref{eq:H-min} is a finite-dimensional optimization problem in terms of the \(R\) respective coefficients \(\vec\beta\), with \(\widehat{T}(k,x) = \beta_0 + \sum_{r=1}^{R-1} \beta_r B_r(x)\). Note that \(\widehat{T}(k,\cdot)\) is a statistical model; it is viewed as an object (rather than say a vector of numbers) and passed as a ``function pointer'' to the pathwise reward simulator in subsequent steps. The latter simulator in turn applies a \textbf{predict} method, asking \(\widehat{T}(k,\cdot)\) to furnish a predicted timing value at arbitrary \(x\).

\emph{Remark:} In \texttt{\{mlOSP\}} implementation, the emulator is employed only when the decision is non-trivial, i.e.~in the in-the-money region \({\cal X}_{in} := \{ x : h(k,x) > 0\}\). This insight dates back to \citet{LS} who noted that learning \(\widehat{T}(k,x)\) is only necessary when \(h(k,x) >0\); otherwise it is clear that \(T(k,x) > 0\) and hence we may set \(\widehat{A}_k(x) = 1\). Thus, the simulation design \(\cD_k\) is restricted to be in \({\cal X}_{in}\).

Finally, I index everything by the time steps \(k=1,\ldots,K\), with corresponding \(t_k = k\Delta t\) by default, and the following notation:

\begin{itemize}
\item
  \(N_k\): number of training inputs at step \(k\) (can again vary across \(k\));
\item
  \({\cal D}_k\): simulation design, i.e.~the collection of training inputs \(x^{(k),1:N_k}\) at step \(k\); \(|{\cal D}_k| = N_k\)
\item
  \({\cal H}_k\): functional approximation space where \(\widehat{T}(k,\cdot)\) is searched within;
\item
  \(y^{1:N_k}_k\) pathwise samples of timing value used as the responses in the regression model.
\end{itemize}

Equipped with above, Algorithm \ref{alg:1} presents the \textbf{mlOSP} template, emphasizing the ``moving parts'' of the simulation designs \(\mathcal{D}_{k}\) and fitting \(\hat{T}(k,\cdot)\). These two building blocks are explored in Sections 3 and 4 below.

\begin{algorithm}
\begin{algorithmic}[1]
\REQUIRE{ $K=T/\Delta t$ (time steps), $(N_k)$ (simulation budget per step)}, $w$ (path lookahead)
\FOR{$ k=K - 1, \ldots, 0$ }
\STATE Generate training design $\mathcal{D}_{k} := (x^{(k),1:N_k}({k}))$ of size $N_{k}$
\STATE Generate $w$-step paths $x^{(k),n}({k}) \mapsto x^{(k),n}(s)$ for $n=1,\ldots,N_{k}$, $s=k+1,\ldots,(k+w) \wedge K$
\STATE Generate pathwise forward stopping rule $\tau^n_{k} = \min\{ s \ge k+1 : \widehat{T}(s, x^{(k),n}(s)) < 0 \} \wedge (k+w) \wedge K$
\STATE Generate pathwise timing value {\small $$ y^n_{k+1} = \left\{ \begin{aligned}  & h\bigl(\tau^n_{k},x^{(k),n}({\tau}^n_{k}) \bigr) - h \bigl(k,x^{(k),n}({k}) \bigr) \hspace{2.1in} \text{if} \quad \tau^n_{k} < k+w;\\
&\widehat{T}\bigl(k+w, x^{(k),n}({k+w}) \bigr)+h\bigl(k+w, x^{(k),n}({k+w}) \bigr)- h \bigl(k,x^{(k),n}({k}) \bigr) \quad \text{if} \quad\tau^n_{k} = (k+w) \wedge K. \end{aligned}\right.
$$}
\STATE Fit $\widehat{T}(k,\cdot) \leftarrow \arg \min_{f(\cdot) \in \mathcal{H}_k} \sum_{n=1}^{N_k} |f(x^{(k),n}(k)) - y_{k+1}^n |^2$
\ENDFOR
\STATE Return fitted objects $\{ \widehat{T}(k,\cdot) \}_{k=0}^{K-1}$
\end{algorithmic}
\caption{Dynamic Emulation Algorithm: the \textbf{mlOSP} template. \label{alg:1}}
\end{algorithm}

The last piece of Algorithm \ref{alg:1} that remains to be clarified concerns the simulation device and the respective \emph{lookahead parameter} \(w\) that determines how long are the forward trajectories.
The two most common approaches are TvR (for Tsitsiklis-van Roy \citep{TsitsiklisVanRoy}) and LS (for Longstaff-Schwartz \citep{LS}).
In the TvR variant \(w=1\) and the simulator generates one-step-ahead paths, resulting in the regression of \(\widehat{T}(k+1,X(k+1))\) against \(X(k)\). In the LS variant, \(w=K-k\), and one regresses the pathwise timing values \(y^n_{k}\) against \(X(k)\).

\hypertarget{expected-payoff}{%
\subsubsection{Expected Payoff}\label{expected-payoff}}

The output of Algorithm \ref{alg:1} is the approximate decision rules \(\widehat{A}_k(\cdot) = 1_{\{ \widehat{T}_k(\cdot) < 0\}}\). These are functions, taking as input the system state and returning the action to take (stop or continue). Additional computation is needed to evaluate the corresponding \emph{expected payoff} \(\tilde{V}\) of the stopping rule \(\tau_{\widehat{A}_{0:K-1}}\) in \eqref{eq:tildeV}.
To do so, one again employs Monte Carlo simulation, approximating with a sample average on a set of fresh forward test scenarios \(x^{1:N'}(k), k=1,\ldots,K\), \(x^{n'}(0) = X(0)\),
\begin{align}\label{eq:out-of-sample}
\check{V}(0,X(0)) = \frac{1}{N'} \sum_{n'=1}^{N'} h \left(\tau^{n'}, x^{n'}({\tau^{n'}}) \right),
\end{align}
where \(\tau^n := \min \{ k \ge 0 : x^n(k) \in \widehat{\mathcal{S}}_k \}\) is the pathwise stopping time.
Eqn.~\eqref{eq:out-of-sample} also yields a confidence interval for the expected payoff by using the empirical standard deviation of \(h(\tau^{n'}, x^{n'}_{\tau^{n'}})\)'s. Because \(\check{V}(0; X(0))\) is based on out-of-sample test scenarios, it is an unbiased estimator of \(\tilde{V}\) and hence for large enough test sets it will yield a lower bound on the true optimal expected reward, \(\E[ \check{V}(0,X(0))] = \tilde{V}(0,X(0);\widehat{A}_{0:K}) < V(0,X(0))\). The interpretation is that the simulation design \({\cal D}_k\) defines a training set, while \(x^{1:N'}(k)\) forms the test set. The latter is not only important to obtain unbiased estimates of expected reward from the rule \(\tau_{\widehat{A}_{0:K}}\), but also to enable an apples-to-apples comparison between multiple solvers which should be run on a fixed test set of \(X\)-paths.

\hypertarget{longstaff-schwartz-scheme}{%
\subsubsection{Longstaff-Schwartz Scheme}\label{longstaff-schwartz-scheme}}

In the LS variant of \citep{LS}, the training sets are constructed from a single global simulation of the \(X\)-process, so that \(x^{(k),n}(\ell) = x^{(\ell),n}(\ell)\) and the forward paths that yield \(Y\)'s are \emph{re-used} across time-steps. This requires storing all those paths in memory, but reduces the need to keep generating new shocks \(\xi\)'s underlying \(y^n(k)\). Moreover, it provides access to an \emph{in-sample} estimator \(\hat{V}(0,X(0))\) which is the average payoff on the training paths. As is well known, \(\hat{V}\) suffers from look-ahead bias and therefore cannot be reliably compared to the true option price. Heuristically, it has been observed that the in-sample estimator that is available in the global-path design approach tends to give \emph{upper} bounds, and hence can be used to roughly ``sandwich'' the final estimate of the option price between the lower bound of the test set and the upper bound of the training set, ``\(V(0,X(0)) \in [ \check{V}(0,X(0)), \hat{V}(0,X(0))]\)''.

\emph{Remark}: our RMC formulation returns the action map \(\widehat{A}_k(\cdot)\) globally, or at least anywhere within the neighborhood of the training sets \({\cal D}_k\). As such, during forward simulation one may evaluate the expected payoff with arbitrary initial condition \({X}(0)\). Since the LS scheme is based on starting all training paths from a fixed \({X}(0)\) (i.e.~a degenerate \({\cal D}_0\)) it is sensitive to testing with a different initial condition.

\hypertarget{what-is-random}{%
\subsubsection{What is Random?}\label{what-is-random}}

The core of RMC is its probabilistic (hence Monte Carlo) nature. Indeed, Algorithm \ref{alg:1} is random and re-running it would yield a different set of functional approximations \(\widehat{T}(k,\cdot)\) at each \(k\), and ultimately a different \(\check{V}(0,{X}(0))\). Why does this happen exactly?

In the \textbf{mlOSP} template, the immediate place where random samples are generated is during the forward simulation of trajectories (step 3 of Algorithm \ref{alg:1}) \(x^{(k),1:N}(k:k+w)\). In other words, the ``y''s in the regression indeed have a random component to them. Secondly, in the LS variant the simulation design is random, i.e.~the ``x''s in \({\cal D}_k\) (step 2) are random. Namely, the training set is based on simulated trajectories \((X^{1:N}(k))\) which vary across algorithm runs. Third, many machine learning emulators involve nonlinear optimization during the fitting step 6, and the low-level optimizers frequently invoke randomization. For example, neural network and Gaussian process emulators rely on stochastic gradient descent and/or genetic optimization and thus return not the true global minimizer but a randomly found local minimum in \({\cal H}_k\). Fourth, the test collection of paths \(X^{1:N'}\) is random.

To sum up, \(\check{V}(0,X(0))\) is a random variable, and so are all the intermediate solution outputs, such as \(\widehat{A}_k(\cdot)\) or \(\widehat{T}(k,\cdot)\). Indeed, due to the coupling between \(\widehat{T}(k,\cdot)\) and \(\{\widehat{T}(\ell,\cdot), \ell>k\}\), as soon as \(\widehat{T}(k,\cdot)\) has some probabilistic aspect to it, so will all preceding-in-time emulators. Similarly, the \emph{budget} of the solver in terms of its running time and number of uniform variates generated is also non-constant across runs. With so many random pieces, scheme stability is important. The standard way to measure stability is through sampling variance, i.e.~the variance of \(\check{V}(0,X(0))\) across independent algorithm runs, \emph{holding the test set fixed}. More efficient RMC versions would be expected to yield lower sampling variance, exhibiting less dependence on the particular training set. Statistically, this dependence is primarily driven by the sensitivity of the regression emulator to the realizations in \(y^{1:N}_k\), and thus it is preferable to ``regularize'' the regression architecture to avoid sensitivity to outliers, or other statistical mis-specifications, cf.~Section 5.

\hypertarget{getting-started-with-mlosp}{%
\section{\texorpdfstring{Getting Started with \texttt{\{mlOSP\}}}{Getting Started with \{mlOSP\}}}\label{getting-started-with-mlosp}}

The \texttt{\{mlOSP\}} package \citep{mlOSP} implements multiple versions of Algorithm \ref{alg:1}. The following user guide highlights its key aspects. I have strived to make it fully reproducible, with the full underlying \texttt{R} code available on GitHub. Since the schemes are intrinsically based on generating random outputs from the underlying stochastic simulator, I fix the random number generator (RNG) seeds. Depending on the particular machine and \texttt{R} version, these seeds might nevertheless lead to different results.

The workflow pipeline in \texttt{\{mlOSP\}} consists of:

\begin{enumerate}
\def\labelenumi{(\roman{enumi})}
\item
  Defining the \texttt{model}, which is a list of (a) parameters that determine the dynamics of \((X(t))\) in \eqref{eq:sde}; (b) the payoff function \(h(t,x)\); and (c) the tuning parameters determining the regression emulator specification.
\item
  Calling the appropriate top-level solver. The solvers are indexed by the underlying type of \emph{simulation design}. They also use a top-level \texttt{method} argument that selects from a collection of implemented regression modules. Otherwise, all other parameters are passed through the above \texttt{model} argument. The solver returns a collection of fitted emulators---an \texttt{R} list containing the respective regression object of \(\widehat{T}(k,\cdot)\) for each time step \(k\), plus a few diagnostics;
\item
  Evaluating the obtained fitted emulators through an out-of-sample forward simulator via \texttt{forward.sim.policy} that evaluates \eqref{eq:tildeV}. The latter is a top-level function that can work with any of the implemented regression objects. Alternatively, one may also \emph{visualize} the emulators through a few provided plotting functions.
\end{enumerate}

The aim of the above architecture is to maximize modularity and to simplify construction of user-defined OSP instances, such as additional system dynamics or bespoke payoffs. The latter piece is illustrated in Section 7.2.

The following example illustrates this workflow. We consider a 1-D Bermudan Put with payoff \(e^{-r t}({\cal K}-x)_+\) where the underlying dynamics are given by Geometric Brownian Motion (GBM)
\[ dX(t) = (r-\delta) X(t) dt + \sigma X(t) dW(t), \qquad X(0) = x_0,\]
with scalar parameters \(r, \delta,\sigma,x_0\). Thus, \(X(t_k)\) can be simulated exactly by sampling from the respective log-normal distribution.
In the model specification below we have \(r=0.06, \delta=0, T=1, \sigma=0.2\) and the Put strike is \({\cal K}=40\). Exercising the option is possible \(K=25\) times before expiration, i.e., \(\Delta t = 0.04\). To implement the above OSP instance just requires defining a named list with the respective parameters (plus a few more for the regression model specified below):

\begin{Shaded}
\begin{Highlighting}[]
\NormalTok{put1d.model }\OtherTok{\textless{}{-}} \FunctionTok{c}\NormalTok{(}\AttributeTok{K=}\DecValTok{40}\NormalTok{, }\AttributeTok{payoff.func=}\NormalTok{put.payoff,  }\CommentTok{\# payoff function}
            \AttributeTok{x0=}\DecValTok{40}\NormalTok{,}\AttributeTok{sigma=}\FloatTok{0.2}\NormalTok{,}\AttributeTok{r=}\FloatTok{0.06}\NormalTok{,}\AttributeTok{div=}\DecValTok{0}\NormalTok{,}\AttributeTok{T=}\DecValTok{1}\NormalTok{,}\AttributeTok{dt=}\FloatTok{0.04}\NormalTok{,}\AttributeTok{dim=}\DecValTok{1}\NormalTok{, }\AttributeTok{sim.func=}\NormalTok{sim.gbm,}
            \AttributeTok{km.cov=}\DecValTok{4}\NormalTok{,}\AttributeTok{km.var=}\DecValTok{1}\NormalTok{,}\AttributeTok{kernel.family=}\StringTok{"matern5\_2"}\NormalTok{,  }\CommentTok{\# GP emulator params}
            \AttributeTok{pilot.nsims=}\DecValTok{0}\NormalTok{,}\AttributeTok{batch.nrep=}\DecValTok{200}\NormalTok{,}\AttributeTok{N=}\DecValTok{25}\NormalTok{)}
\end{Highlighting}
\end{Shaded}

As a representative solver, I utilize \texttt{osp.fixed.design} that asks the user to specify the simulation design directly. To select a particular regression approach, \texttt{osp.fixed.design} has a \texttt{method} field which can take a large range of regression methods. Specifics of each regression are controlled through additional \texttt{model} fields. As example, the code snippet below employs the \texttt{\{DiceKriging\}} Gaussian Process (GP) emulator with fixed hyperparameters and a constant prior mean, selected through \texttt{method="km"}. The kernel family is Matérn-5/2, with hyperparameters specified via \texttt{km.cov}, \texttt{km.var} above. For the simulation design (\texttt{input.domain} field) I take \(\{16,17,\ldots, 40\}\) with 200 replications (\texttt{batch.nrep}) per site, yielding \(N=|{\cal D}|=200 \cdot 25\), see Section 4.1. The replications are treated using the SK method \citep{ankenman2010stochastic}, pre-averaging the replicated outputs before training the GP.

\begin{Shaded}
\begin{Highlighting}[]
\NormalTok{train.grid}\FloatTok{.1}\NormalTok{d }\OtherTok{\textless{}{-}} \FunctionTok{seq}\NormalTok{(}\DecValTok{16}\NormalTok{, }\DecValTok{40}\NormalTok{, }\AttributeTok{len=}\DecValTok{25}\NormalTok{)  }\CommentTok{\# simulation design}
\NormalTok{km.fit }\OtherTok{\textless{}{-}} \FunctionTok{osp.fixed.design}\NormalTok{(put1d.model,}\AttributeTok{input.domain=}\NormalTok{train.grid}\FloatTok{.1}\NormalTok{d, }\AttributeTok{method=}\StringTok{"km"}\NormalTok{)}
\end{Highlighting}
\end{Shaded}

Note that no output is printed: the produced object \texttt{km.fit} contains an array of 24 (one for each time step, except at maturity) fitted GP models, but does not yet contain the estimate of the option price \({V}(0,X(0))\). Indeed, we have not defined any test set, and consequently are momentarily postponing the computation of \(\check{V}(0,X(0))\).

\hypertarget{comparing-solvers}{%
\subsection{Comparing solvers}\label{comparing-solvers}}

\texttt{\{mlOSP\}} has several solvers and allows straightforward swapping of the different pieces of the template. Below, for example, I consider the LS scheme, implemented in the \textbf{osp.prob.design} solver, where the simulation design is based on forward \(X\)-paths. I moreover replace the regression module with a smoothing cubic spline (\textbf{smooth.spline}; see \citep{Kohler08spline} for a discussion of using splines for RMC). The latter requires specifying the number of knots \texttt{model\$nk}. Only 2 lines of code are necessary to make all these modifications and obtain an alternative solution of the same OSP:

\begin{Shaded}
\begin{Highlighting}[]
\NormalTok{put1d.model}\SpecialCharTok{$}\NormalTok{nk}\OtherTok{=}\DecValTok{20}  \CommentTok{\# number of knots for the smoothing spline}
\NormalTok{spl.fit }\OtherTok{\textless{}{-}} \FunctionTok{osp.prob.design}\NormalTok{(}\AttributeTok{N=}\DecValTok{30000}\NormalTok{,put1d.model,}\AttributeTok{method=}\StringTok{"spline"}\NormalTok{)  }\CommentTok{\# 30000 training paths}
\end{Highlighting}
\end{Shaded}

Again, there is no visible output; \texttt{spl.fit} now contains a list of 24 fitted spline objects that are each parametrized by 20 (number of chosen knots) cubic spline coefficients.

\begin{figure}

{\centering \includegraphics[trim={0 0.5cm 0 0},clip]{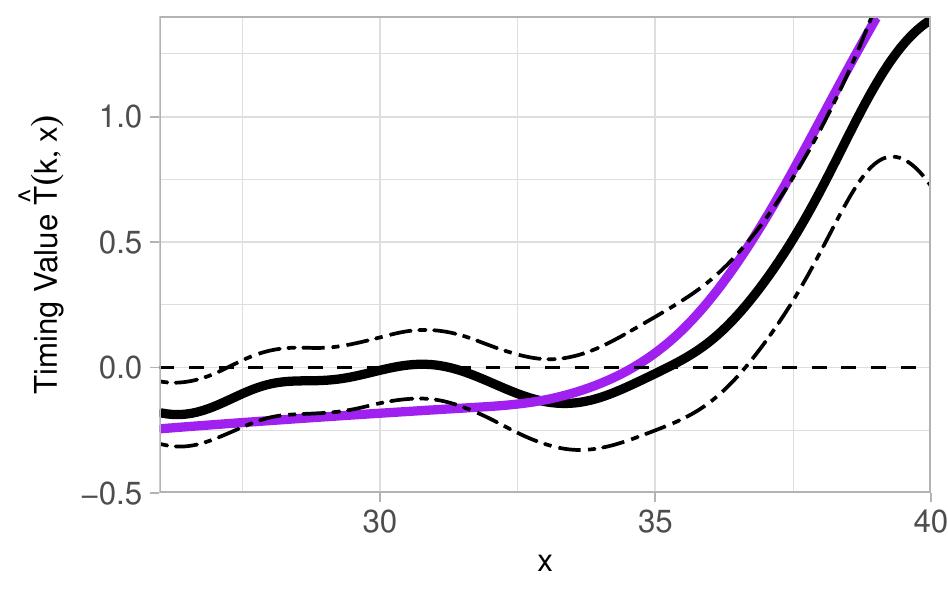}

}

\caption{\label{fig:1d}Timing value of a Bermudan Put based on GP emulator (black) and a Smoothing Spline emulator (purple) at $k=10$ ($t=0.4$). I also display  the uncertainty quantification regarding the GP fit of $\widehat{T}(k,\cdot)$ (the dashed $95\%$ band).}\label{fig:Fitted-GP}
\end{figure}

Figure \ref{fig:1d} visualizes the fitted timing value from one time step. To do so, I predict \(\widehat{T}(k,x)\) based on a fitted emulator (at \(t=10\Delta t=0.4\)) over a collection of test locations. In the Figure, this is done for both of the above solvers (GP-km and Spline), moreover we also display the 95\% credible intervals of the GP emulator for \(\widehat{T}(k,\cdot)\). Keep in mind that the exact \emph{shape} of \(\widehat{T}(k,\cdot)\) is irrelevant in \textbf{mlOSP}, all that matters is the implied \(\widehat{A}_k\) which is the zero level set of the timing value, i.e.~determined by the \emph{sign} of \(\widehat{T}(k,\cdot)\). Therefore, the two regression methods yield quite similar exercise rules, although they do differ (e.g.~at \(x=35\) the spline-based rule is to continue, while the GP-based one is to stop and exercise the Put).
Since asymptotically both solvers should recover the optimal rule, their difference can be attributed to training errors. As such, uncertainty quantification of \(\widehat{T}(k,\cdot)\) is a useful diagnostic to assess the perceived accuracy of \(\widehat{\cal A}_k\). In Figure \ref{fig:1d} the displayed uncertainty band shows that the GP emulator has low confidence about the right action to take for nearly all \(x \le 37\), since zero is inside the 95\% credible band and therefore the positivity of \(\widehat{T}(k,\cdot)\) in that region is not statistically significant.

\texttt{\{mlOSP\}} is designed to be dimension-agnostic, so that building a multi-dimensional model follows the exact same steps. For example, the two-line snippet below defines a 2D model with Geometric Brownian motion dynamics for the two assets \(X_1, X_2\) and a basket average Put payoff
\(h_{\text{avePut}}(t,x) = e^{-r t}({\cal K} - (x_1+x_2)/2)_+, x \in {\mathbb{R}}^2_+.\) The two assets are assumed to be uncorrelated with identical dynamics. I continue to use a strike of \({\cal K}=40\) and at-the-money (ATM) initial condition \(X(0)=(40,40)\).

\begin{Shaded}
\begin{Highlighting}[]
\NormalTok{model2d }\OtherTok{\textless{}{-}} \FunctionTok{list}\NormalTok{(}\AttributeTok{K=}\DecValTok{40}\NormalTok{,}\AttributeTok{x0=}\FunctionTok{rep}\NormalTok{(}\DecValTok{40}\NormalTok{,}\DecValTok{2}\NormalTok{),}\AttributeTok{sigma=}\FunctionTok{rep}\NormalTok{(}\FloatTok{0.2}\NormalTok{,}\DecValTok{2}\NormalTok{),}\AttributeTok{r=}\FloatTok{0.06}\NormalTok{,}\AttributeTok{div=}\DecValTok{0}\NormalTok{,}
                      \AttributeTok{T=}\DecValTok{1}\NormalTok{,}\AttributeTok{dt=}\FloatTok{0.04}\NormalTok{,}\AttributeTok{dim=}\DecValTok{2}\NormalTok{,}\AttributeTok{sim.func=}\NormalTok{sim.gbm, }\AttributeTok{payoff.func=}\NormalTok{put.payoff)}
\end{Highlighting}
\end{Shaded}

With the model defined, we can use the same solvers as above, with exactly the same syntax. To illustrate a different regression module, let us employ a linear model, \texttt{method="lm"}. To do so, we need to first define the basis functions \(B_r(\cdot)\) that are passed to the \texttt{model\$bases} parameter. Below I select polynomial bases of degree \(\le 2\), \({\cal H}_k = \sspan \{x_1, x_1^2, x_2, x_2^2, x_1 \cdot x_2\}\); by default the \texttt{R} \texttt{lm} model also includes the constant term, so there are a total of 6 regression coefficients \(\vec{\beta}\).

\begin{Shaded}
\begin{Highlighting}[]
\NormalTok{bas22 }\OtherTok{\textless{}{-}} \ControlFlowTok{function}\NormalTok{(x) }\FunctionTok{return}\NormalTok{(}\FunctionTok{cbind}\NormalTok{(x[,}\DecValTok{1}\NormalTok{],x[,}\DecValTok{1}\NormalTok{]}\SpecialCharTok{\^{}}\DecValTok{2}\NormalTok{,x[,}\DecValTok{2}\NormalTok{],x[,}\DecValTok{2}\NormalTok{]}\SpecialCharTok{\^{}}\DecValTok{2}\NormalTok{,x[,}\DecValTok{1}\NormalTok{]}\SpecialCharTok{*}\NormalTok{x[,}\DecValTok{2}\NormalTok{]))}
\NormalTok{model2d}\SpecialCharTok{$}\NormalTok{bases }\OtherTok{\textless{}{-}}\NormalTok{ bas22}
\NormalTok{prob.lm }\OtherTok{\textless{}{-}} \FunctionTok{osp.prob.design}\NormalTok{(}\DecValTok{15000}\NormalTok{,model2d, }\AttributeTok{method=}\StringTok{"lm"}\NormalTok{)  }\CommentTok{\# Train with 15,000 paths}
\end{Highlighting}
\end{Shaded}

For comparison, I re-solve with a GP emulator that has a space-filling training design of \(N_{unique}=150\) sites replicated with batches of \(N_{rep}=100\) each for a total of \(N=15,000\) simulations again.

\begin{Shaded}
\begin{Highlighting}[]
\NormalTok{model2d}\SpecialCharTok{$}\NormalTok{N }\OtherTok{\textless{}{-}} \DecValTok{150}  \CommentTok{\# N\_unique }
\NormalTok{model2d}\SpecialCharTok{$}\NormalTok{kernel.family }\OtherTok{\textless{}{-}} \StringTok{"gauss"} \CommentTok{\# squared{-}exponential kernel}
\NormalTok{model2d}\SpecialCharTok{$}\NormalTok{batch.nrep }\OtherTok{\textless{}{-}} \DecValTok{100}
\NormalTok{model2d}\SpecialCharTok{$}\NormalTok{pilot.nsims }\OtherTok{\textless{}{-}} \DecValTok{0}

\NormalTok{sob150 }\OtherTok{\textless{}{-}} \FunctionTok{sobol}\NormalTok{(}\DecValTok{276}\NormalTok{, }\AttributeTok{d=}\DecValTok{2}\NormalTok{) }\CommentTok{\# Sobol space{-}filling sequence}
\CommentTok{\# triangular approximation domain for the simulation design}
\NormalTok{sob150 }\OtherTok{\textless{}{-}}\NormalTok{ sob150[ }\FunctionTok{which}\NormalTok{( sob150[,}\DecValTok{1}\NormalTok{] }\SpecialCharTok{+}\NormalTok{ sob150[,}\DecValTok{2}\NormalTok{] }\SpecialCharTok{\textless{}=} \DecValTok{1}\NormalTok{) ,]  }
\NormalTok{sob150 }\OtherTok{\textless{}{-}} \DecValTok{25}\SpecialCharTok{+}\DecValTok{30}\SpecialCharTok{*}\NormalTok{sob150  }\CommentTok{\# Lower{-}left triangle in [25,55]x[25,55], see Fig }

\NormalTok{sob.km }\OtherTok{\textless{}{-}} \FunctionTok{osp.fixed.design}\NormalTok{(model2d,}\AttributeTok{input.domain=}\NormalTok{sob150, }\AttributeTok{method=}\StringTok{"trainkm"}\NormalTok{)}
\end{Highlighting}
\end{Shaded}

\begin{figure}
{\centering \includegraphics[trim={0 0.1in 0 0},clip]{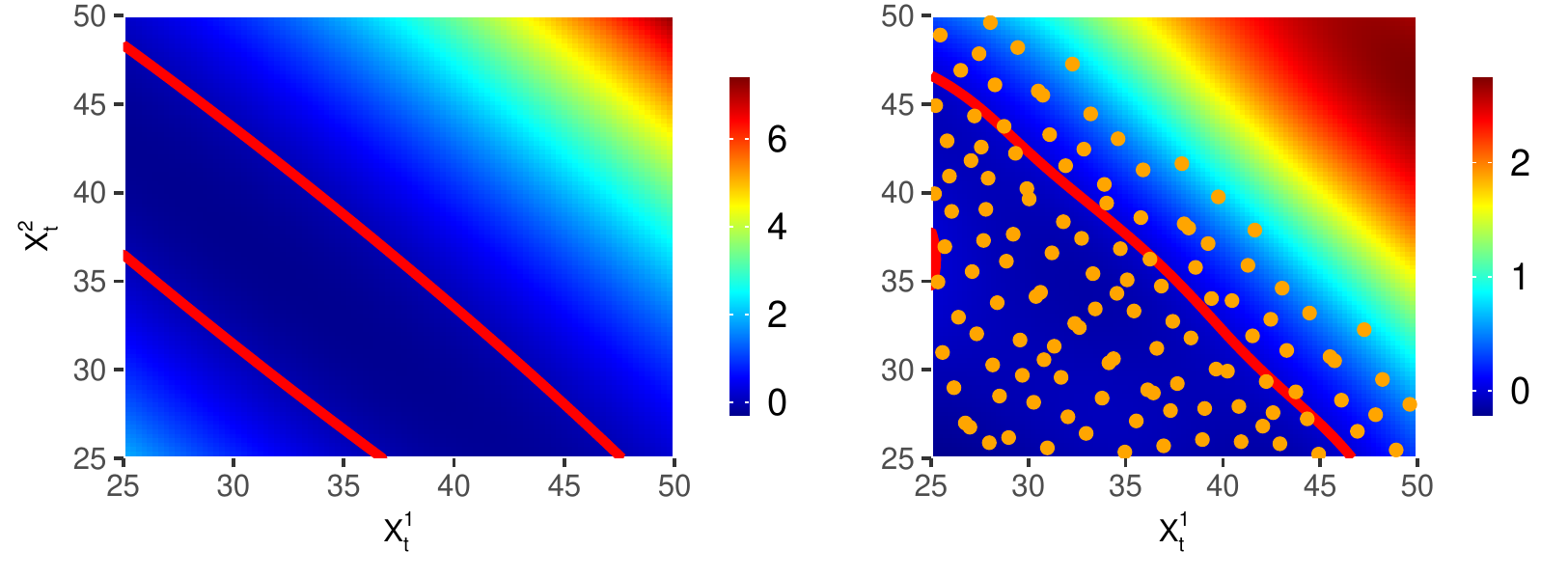}}
\caption{\label{fig:plt-surf}Timing Value of a 2D Basket Put at $k=15$ ($t=0.6$). The red contour shows the boundary of the stopping set (bottom-left corner). The colors indicate the value of $\widehat{T}(k,x)$.}\label{fig:contour-plt}
\end{figure}

Figure \ref{fig:plt-surf} visualizes the estimated stopping boundary from the above 2 solvers using the \texttt{plt.2d.surf} function from the package. It shows an image plot of the emulator \(\widehat{T}(k,\cdot)\) at a single time-step \(k\). Recall that the stopping region is the level set where the timing value is negative, indicated with the red contours that delineate the exercise boundary. In the left panel, the exercise boundary is a parabola because \(\widehat{T}(k,\cdot)\) is modeled as a quadratic. On the right panel, the exercise boundary has a much more complex shape since that \(\widehat{T}(k,\cdot)\) has many more degrees of freedom. For the latter case, Figure \ref{fig:plt-surf} also displays the underlying training design of \(N_{unique}=150\) sites.

Appendix C lists all the simulation, payoff and solver functions available in \texttt{\{mlOSP\}}.

\hypertarget{out-of-sample-tests}{%
\subsection{Out-of-sample Tests}\label{out-of-sample-tests}}

By default, the solvers only create the functional representations of the timing/value function and do not return any explicit estimates of the option price or stopping rule. This reflects the strict separation between training and testing, as well as the fact that RMC builds a global estimate of \(\widehat{T}(k,\cdot)\) and hence can be used to obtain a range of option prices (e.g.~by changing \(X(0)\)).

The following code snippet builds an out-of-sample database of \(X\)-paths. This is done iteratively by employing the underlying simulator, already saved under the \texttt{model2d\$sim.func} field. The latter is interpreted as the function to generate a vector of \(X(k+1)\)'s conditional on \(X(k)\).
By applying \texttt{payoff.func} to the final vector (which stores values of \(X(T)\)) we can get an estimate of the respective European option price \(\mathbb{E}[ h(T,X(T))]\). By calling \texttt{forward.sim.policy} with a previously saved \texttt{\{mlOSP\}} object we then obtain a collection of \(h(\tau^n, X^n({\tau^n})), n=1,\ldots,\) that can be averaged to obtain a \(\check{V}(0,X(0))\) as in \eqref{eq:out-of-sample}.

\begin{Shaded}
\begin{Highlighting}[]
\NormalTok{nSims}\FloatTok{.2}\NormalTok{d }\OtherTok{\textless{}{-}} \DecValTok{40000}  \CommentTok{\# use N\textquotesingle{}=40,000 fresh trajectories}
\NormalTok{nSteps}\FloatTok{.2}\NormalTok{d }\OtherTok{\textless{}{-}} \DecValTok{25}
\FunctionTok{set.seed}\NormalTok{(}\DecValTok{102}\NormalTok{)}
\NormalTok{test}\FloatTok{.2}\NormalTok{d }\OtherTok{\textless{}{-}} \FunctionTok{list}\NormalTok{()  }\CommentTok{\# store a database of forward trajectories}
\NormalTok{test}\FloatTok{.2}\NormalTok{d[[}\DecValTok{1}\NormalTok{]] }\OtherTok{\textless{}{-}}\NormalTok{ model2d}\SpecialCharTok{$}\FunctionTok{sim.func}\NormalTok{( }\FunctionTok{matrix}\NormalTok{(}\FunctionTok{rep}\NormalTok{(model2d}\SpecialCharTok{$}\NormalTok{x0, nSims}\FloatTok{.2}\NormalTok{d),}\AttributeTok{nrow=}\NormalTok{nSims}\FloatTok{.2}\NormalTok{d,}\AttributeTok{byrow=}\NormalTok{T), }
\NormalTok{                               model2d, model2d}\SpecialCharTok{$}\NormalTok{dt)}
\ControlFlowTok{for}\NormalTok{ (i }\ControlFlowTok{in} \DecValTok{2}\SpecialCharTok{:}\NormalTok{(nSteps}\FloatTok{.2}\NormalTok{d}\SpecialCharTok{+}\DecValTok{1}\NormalTok{))  }\CommentTok{\# generate forward trajectories}
\NormalTok{   test}\FloatTok{.2}\NormalTok{d [[i]] }\OtherTok{\textless{}{-}}\NormalTok{ model2d}\SpecialCharTok{$}\FunctionTok{sim.func}\NormalTok{( test}\FloatTok{.2}\NormalTok{d [[i}\DecValTok{{-}1}\NormalTok{]], model2d, model2d}\SpecialCharTok{$}\NormalTok{dt)}
\NormalTok{oos.lm }\OtherTok{\textless{}{-}} \FunctionTok{forward.sim.policy}\NormalTok{( test}\FloatTok{.2}\NormalTok{d, nSteps}\FloatTok{.2}\NormalTok{d, prob.lm}\SpecialCharTok{$}\NormalTok{fit, model2d)  }\CommentTok{\# prob.lm payoff}
\NormalTok{oos.km }\OtherTok{\textless{}{-}} \FunctionTok{forward.sim.policy}\NormalTok{( test}\FloatTok{.2}\NormalTok{d, nSteps}\FloatTok{.2}\NormalTok{d, sob.km}\SpecialCharTok{$}\NormalTok{fit,  model2d)  }\CommentTok{\# sob.km payoff}
\FunctionTok{cat}\NormalTok{( }\FunctionTok{paste}\NormalTok{(}\StringTok{\textquotesingle{}Price estimates\textquotesingle{}}\NormalTok{, }\FunctionTok{round}\NormalTok{(}\FunctionTok{mean}\NormalTok{(oos.lm}\SpecialCharTok{$}\NormalTok{payoff),}\DecValTok{3}\NormalTok{), }\FunctionTok{round}\NormalTok{(}\FunctionTok{mean}\NormalTok{(oos.km}\SpecialCharTok{$}\NormalTok{payoff),}\DecValTok{3}\NormalTok{)))}
\end{Highlighting}
\end{Shaded}

\begin{verbatim}
## Price estimates 1.446 1.44
\end{verbatim}

\begin{Shaded}
\begin{Highlighting}[]
\CommentTok{\# check: estimated European option value}
\FunctionTok{cat}\NormalTok{(}\FunctionTok{mean}\NormalTok{( }\FunctionTok{exp}\NormalTok{(}\SpecialCharTok{{-}}\NormalTok{model2d}\SpecialCharTok{$}\NormalTok{r}\SpecialCharTok{*}\NormalTok{model2d}\SpecialCharTok{$}\NormalTok{T)}\SpecialCharTok{*}\NormalTok{model2d}\SpecialCharTok{$}\FunctionTok{payoff.func}\NormalTok{(test}\FloatTok{.2}\NormalTok{d[[nSteps}\FloatTok{.2}\NormalTok{d]], model2d)))}
\end{Highlighting}
\end{Shaded}

\begin{verbatim}
## 1.214149
\end{verbatim}

Since we evaluated both estimators on the same set of test paths, we can conclude that the LM-Poly emulator leads to a better approximation \(\check{V}^{LM}(0,x_0)=\) 1.446 \(>\) 1.44 \(=\check{V}^{GP}(0,x_0)\) than the GP-km one. The reported European Put estimate of 1.214 can be used as a control variate to adjust \(\check{V}\) since it is based on the same test paths and so we expect that \(\check{V}^{EU}(0,x_0) - \E\left[ h(T,X(T)) | X(0)=x_0 \right] \simeq \check{V}(0,x_0) - \E \left[ h(\hat{\tau}, X({\hat{\tau}})) | X(0)=x_0 \right]\).

The \texttt{osp.prob.design} LS solver can evaluate in parallel the in-sample estimator \(\hat{V}(0,X(0))\) and the out-of-sample \(\check{V}(0,X(0))\) by splitting the total simulation budget between training and testing paths. This is controlled by the \texttt{subset} parameter. For example, I re-run the linear model from above but now using 15,000 training inputs and 15,000 test paths:

\begin{Shaded}
\begin{Highlighting}[]
\NormalTok{ls.lm }\OtherTok{\textless{}{-}} \FunctionTok{osp.prob.design}\NormalTok{(}\DecValTok{30000}\NormalTok{,model2d,}\AttributeTok{method=}\StringTok{"lm"}\NormalTok{,}\AttributeTok{subset=}\DecValTok{1}\SpecialCharTok{:}\DecValTok{15000}\NormalTok{)}
\end{Highlighting}
\end{Shaded}

\begin{verbatim}
## [1] "In-sample price estimate 1.462; and out-of-sample: 1.446"
\end{verbatim}

The returned price estimates are consistent with the previous discussion that \(\hat{V}\) (in-sample) is biased high and \(\check{V}\) (out-of-sample) is biased low compared to the true \(V(0,X(0))\).

\hypertarget{training-designs}{%
\section{Training Designs}\label{training-designs}}

The simulation designs \({\cal D}_k\) determine the training domains of the regression emulators. While the size \(N_k = |{\cal D}_k|\) of the training set obviously plays a major role in how accurate the approximation will end up, the \emph{geometry} of \({\cal D}_k\) also has a significant impact.
To represent the design geometry I consider the respective \emph{training density} \(p_{\cal D}(k,\cdot)\), with \(x^n(k)\)'s viewed as samples from that target density
\begin{align}
x^n(k) \sim p_{\cal D}(k, \cdot).
\end{align}
In a randomized simulation design, this is precisely how samples are generated. For instance, the LS scheme \citep{LS} constructs \(x^n\)'s by sampling from the conditional density of \(X(k) | X(0)\); thus in the case where the dynamics are GBM, \(p_{\cal D}\) is log-normal and concentrates on regions reachable by \(X(k)\) starting from \(X(0)\). Uniform target densities, \(p_{\cal D} \equiv \text{Unif}_{\tilde{\cal X}}\) for a given bounded input domain \(\tilde{\cal X}\), are common because they correspond to space-filling simulation designs. The latter capture the intuition of learning through exploration, i.e.~sampling a diverse set of \(x\)'s in order to observe the corresponding \(y\)'s. However, note that a Uniform \(p_{\cal D}\) requires the user to specify the supporting bounding region \(\tilde{\cal X}\), something not needed when \(p_{\cal D}\) is log-normal.

Alternatively, one can generate deterministic representations of \(p_{\cal D}(k,\cdot)\). For example, one can replace sampling from a uniform \(p_{\cal D}\) with placing \(x^n(k)\)'s on a lattice (i.e.~a grid which is a discrete-uniform target density). Yet another way is to mimic a Uniform \(p_{\cal D}\) by using a Quasi Monte Carlo (QMC) sequence.

\textbf{Replication.} Conventionally, a training design of size \(N\) consists of \(N\) unique sites \(x^{1:N}\). In contrast, in a replicated design, all (some) sites appear multiple times. In a most common \emph{batched} design, we have \(N_{unique}\) distinct sites, the so-called macro-design, and each unique \(x^n\) is then repeated \(N_{rep}\) times, so that
\begin{align}\label{eq:rep-design}
 {\cal D} = \Bigl\{ \underbrace{x^{1}, x^{1}, \ldots, x^{1}}_{N_{rep} \text{ times}}, x^{2},  \ldots, \underbrace{x^{N_{unique}}, \ldots, x^{N_{unique}}}_{N_{rep} \text{ times}} \Bigr\},
\end{align}
where the superscripts now index \emph{unique} inputs and the total simulation budget is \(|\cD| = N_{unique} \times N_{rep}\). The corresponding simulator outputs are denoted as \(y^{1,1},y^{1,2},\ldots,y^{n,i},\ldots, y^{N_{unique},N_{rep}}\).

A replicated design allows to pre-average the corresponding \(y\)-values, \(\bar{y}^n := \frac{1}{N_{rep}} \sum_{i=1}^{N_{rep}} y^{n,i}\), and then to call the regression module on the reduced dataset \((x^{1:N_{unique}},\bar{y}^{1:N_{unique}})\). This is especially relevant for non-parametric regressions, where such pre-averaging greatly cuts down on the regression overhead. For example, kernel-type regressions have cubic runtime complexity and orders-of-magnitude savings are achieved by training them on only \(N_{unique} = N/N_{rep}\) samples rather than \(N\). Thus, replication is a must to be able to tractably implement kernel regression in a moderately-high dimensional problem where thousands of samples are needed \(N \gg 10^3\). Since pre-averaging involves loss of information, an additional approximation error would be incurred.

Aspects of the simulation design that can be varied include randomized or deterministic; replicated or not; sequential or one-shot. For example, the classical LS scheme has a randomized, adaptive, non-replicated design.
In the subsections below I describe the various design options that can be specified within the \textbf{mlOSP} template.
In \texttt{\{mlOSP\}} these choices are controlled through the selection of the top-level solvers, as well as \texttt{model} parameters. Most of these ideas are new for RMC training and to my knowledge have not appeared elsewhere except for a tangential mention in \citet{Lu18}. In aggregate they offer a lot of latitude for finetuning RMC.

\hypertarget{space-filling-designs}{%
\subsection{Space Filling Designs}\label{space-filling-designs}}

A space-filling design \({\cal D}\) aims to ``uniformly'' cover the domain of approximation \(\tilde{\cal X}\). Space-filling simulation designs are implemented within the \texttt{osp.fixed.design} solver and primarily controlled through the \texttt{input.domain} parameter which determines the domain of approximation \(\tilde{\cal X}\). Below I illustrate how these are used on the 2D Basket Put case study.

\begin{figure}

{\centering \includegraphics[trim=0 0.3in 0 0]{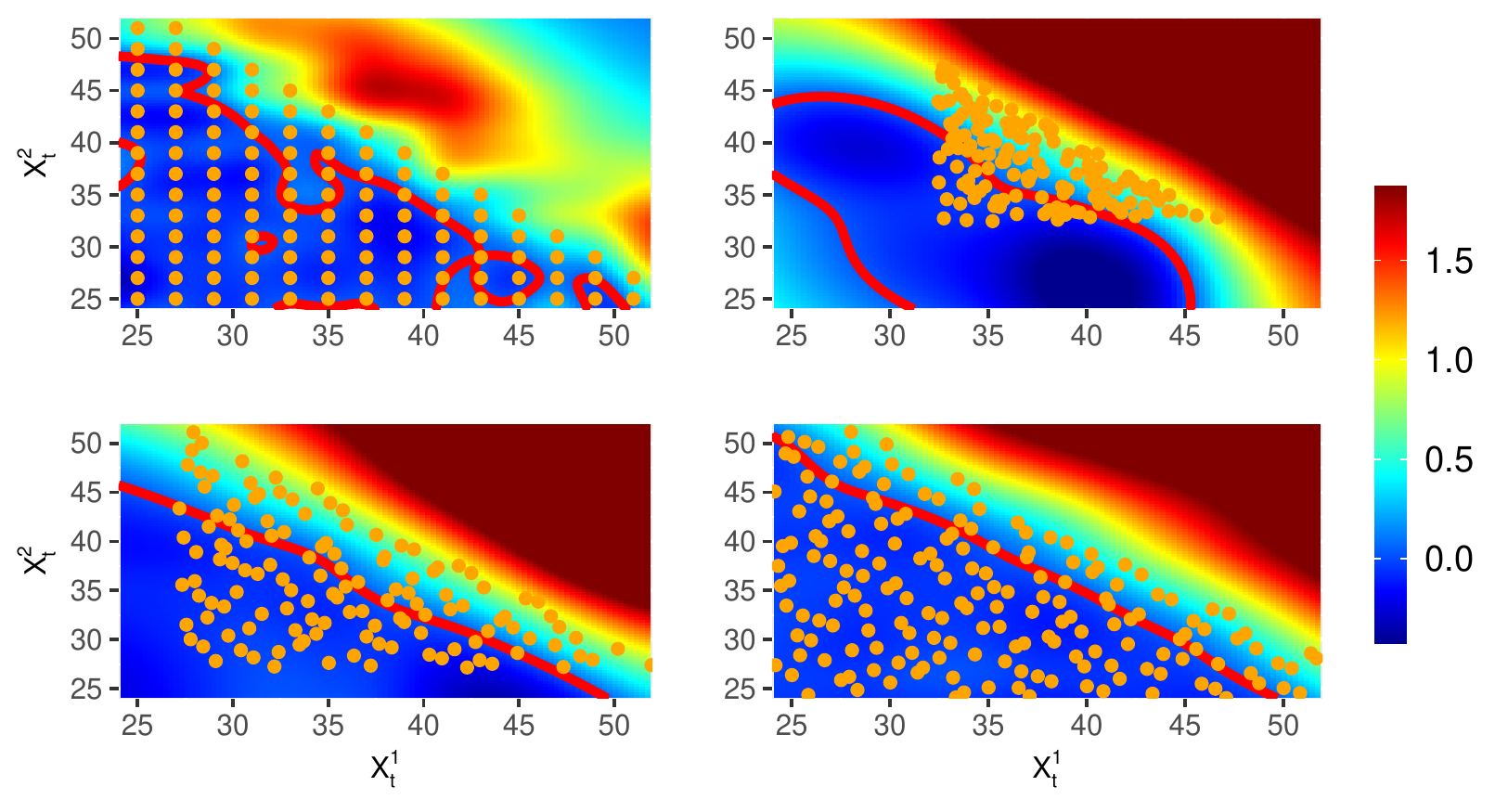}
}
\caption{\label{fig:design-comp} GP-km emulators with different space-filling simulation designs. Top row: lattice (left) and Latin Hypercube Sampling (right)  designs at $k=10$. Bottom row: Halton QMC sequence with time-varying design size: $k=10$ ($t=0.4$ left) and $k=20$ ($t=0.8$ right).  As $k$ increases the input domain $\tilde{\cal X}_k$ grows.}\label{fig:design-plot}
\end{figure}

\begin{enumerate}
\def\labelenumi{(\roman{enumi})}
\tightlist
\item
  One may directly specify a simulation design \({\cal D}\) to be used as-is. For example, \({\cal D}\) can be chosen to be a space-filling lattice. The next code snippet generates a fixed training lattice over a triangular \(\tilde{\cal X}\) (namely the lower-left triangle since we only train in-the-money). Based on model parameters a good domain of approximation is \(\{ (x_1, x_2) : 25 \le x_i \le 55, x_1+x_2 \le 80\}\), cf.~Figure \ref{fig:design-comp} top-left. I batch each of the resulting \(N_{unique}=136\) inputs with \(N_{rep}=100\) replications/site for a total simulation budget of \(N=13,600\). For the regression module I utilize a GP-km Matérn-5/2 emulator.
\end{enumerate}

\begin{Shaded}
\begin{Highlighting}[]
\NormalTok{lattice136 }\OtherTok{\textless{}{-}} \FunctionTok{as.matrix}\NormalTok{(}\FunctionTok{expand.grid}\NormalTok{( }\FunctionTok{seq}\NormalTok{(}\DecValTok{25}\NormalTok{,}\DecValTok{55}\NormalTok{,}\AttributeTok{len=}\DecValTok{16}\NormalTok{), }\FunctionTok{seq}\NormalTok{(}\DecValTok{25}\NormalTok{,}\DecValTok{55}\NormalTok{,}\AttributeTok{len=}\DecValTok{16}\NormalTok{)))}
\NormalTok{lattice136 }\OtherTok{\textless{}{-}}\NormalTok{ lattice136[ }\FunctionTok{which}\NormalTok{( lattice136[,}\DecValTok{1}\NormalTok{] }\SpecialCharTok{+}\NormalTok{ lattice136[,}\DecValTok{2}\NormalTok{] }\SpecialCharTok{\textless{}=} \DecValTok{80}\NormalTok{) ,]}
\NormalTok{model2d}\SpecialCharTok{$}\NormalTok{batch.nrep }\OtherTok{\textless{}{-}} \DecValTok{100}
\NormalTok{put2d.lattice }\OtherTok{\textless{}{-}} \FunctionTok{osp.fixed.design}\NormalTok{(model2d,}\AttributeTok{input.dom=}\NormalTok{lattice136, }\AttributeTok{method=}\StringTok{"km"}\NormalTok{)}
\end{Highlighting}
\end{Shaded}

\begin{enumerate}
\def\labelenumi{(\roman{enumi})}
\setcounter{enumi}{1}
\tightlist
\item
  A bit more generally, one can specify a hyper-rectangular \(\tilde{\cal X}\) and then space-fill
  by employing QMC sequences to place \(x^{1:N_k}({k})\). The \texttt{\{randtoolbox\}} package provides several space-filling methods, such as Sobol sequences, Halton sequences, etc. In \texttt{\{mlOSP\}} this is achieved by providing the ranges for each coordinate \(x_i\) (activated if \texttt{length(input.domain)==2*model\$dim}) and specifying \texttt{model\$qmc.method}. The implementation then automatically clips \(\cD_k\) to the in-the-money region.
\end{enumerate}

The first two approaches above select the same \(\tilde{\cal X}\) across the time-steps. We next consider time-varying domains of interest. Since the variance of \(X(k)\) grows in \(k\), the training region should also be larger for later steps.

\begin{enumerate}
\def\labelenumi{(\roman{enumi})}
\setcounter{enumi}{2}
\tightlist
\item
  To adaptively construct an expanding \(\tilde{\cal X}_k\), \texttt{\{mlOSP\}} offers an automated space-filling procedure based on \emph{pilot simulations}. The pilot simulations are forward trajectories \(\tilde{x}^{1:N_{pilot}}(0\!:\! K)\) from the given initial condition \(X(0)\) (\(N_{pilot}\) set via \texttt{pilot.nsims}). An adaptive hyper-rectangular \(\tilde{\cal X}_k\) is then obtained in terms of the coordinate-wise \emph{quantiles} of \(\tilde{x}^{1:N_{pilot}}(k)\), with the latter acting as ``scaffolding'' to determine an appropriate size of the bounding box. Given \(\tilde{\cal X}_k\), it can be space-filled as before. The motivation for the pilot trajectories is to absolve the end-user from having to directly specify the range of \(\tilde{\cal X}_k\).
\end{enumerate}

To illustrate, I use Latin Hypercube Sampling (LHS, a randomized space filling method, as implemented in the \texttt{\{tgp\}} package, selected by \texttt{qmc.method=NULL}) to space-fill a rectangle between the 4th and 96th percentiles (selected via \texttt{input.domain=0.04}) of the \texttt{pilot.nsim=1000} pilot trajectories at each time step. The resulting \(\tilde{\cal X}_k\) organically expands in \(k\), cf.~Figure \ref{fig:design-comp} top-right.

\begin{Shaded}
\begin{Highlighting}[]
\NormalTok{model2d}\SpecialCharTok{$}\NormalTok{pilot.nsims }\OtherTok{\textless{}{-}} \DecValTok{1000}
\NormalTok{model2d}\SpecialCharTok{$}\NormalTok{qmc.method }\OtherTok{\textless{}{-}} \ConstantTok{NULL}  \CommentTok{\# Use LHS to space{-}fill}
\NormalTok{model2d}\SpecialCharTok{$}\NormalTok{N }\OtherTok{\textless{}{-}} \DecValTok{400}  \CommentTok{\# number of inputs to propose. Only in{-}the{-}money ones are kept}
\NormalTok{put2d.lhsAdaptive}\OtherTok{\textless{}{-}} \FunctionTok{osp.fixed.design}\NormalTok{(model2d,}\AttributeTok{input.dom=}\FloatTok{0.04}\NormalTok{, }\AttributeTok{method=}\StringTok{"trainkm"}\NormalTok{)}
\end{Highlighting}
\end{Shaded}

In this setting, \({\cal D}_k\) is randomized and since we only keep in-the-money inputs, the resulting \(N_k\) is non-constant. For example, in the above run I obtain \(N_k=\) 142
at \(k=12\), \(N_{11}=\) 139 and \(N_{10}=\) 146.

\emph{Remark:} A non-uniform space-filling approach is proposed in \citep{GoudenegeZanette19} and consists in using QMC sampling to draw the \(\epsilon\)'s in the log-normal representation of \(x^n \sim X(k)|X(0)\). This keeps a log-normal target training density \(p_{\cal D}\) but simultaneously space-fills the (unbounded) region of interest.

In option (iii) above, \({\cal D}_k\) is generated separately for each \(k\). As such, one can straightforwardly also vary the design size \(N_k\). One reason why variable \(N_k\) is useful is to reflect the growing volume of \(\tilde{\cal X}_k\) based on the pilot trajectories. Indeed, learning \(A_k(\cdot)\) over a larger domain necessarily requires more simulation effort. Moreover, backpropagation of approximation errors implies that the algorithm is more sensitive to errors in the middle/end time-steps than in the early ones. While conventional RMC schemes use a constant \(N\), this restriction is superfluous for constructing a training set \(\cD_k\).

In \texttt{\{mlOSP\}} this feature is achieved by making the \texttt{model\$N} parameter a vector. Further note that all the space-filling methods can handle arbitrary \texttt{\$N\_k\$} in contrast to naive lattice designs that force \(N_k\) to be a product of marginal lattice lengths.

In the code snippet below I use the full range of the pilot scenarios to select \(\tilde{\cal X}_k\) (achieved by setting \texttt{input.domain=-1}) together with space-filling based on the Halton QMC sequence and a varying \texttt{model\$N}. Note that the latter is not equal to \(N_k\) since the algorithm internally drops all out-of-the-money inputs.

\begin{Shaded}
\begin{Highlighting}[]
\NormalTok{model2d}\SpecialCharTok{$}\NormalTok{qmc.method }\OtherTok{\textless{}{-}}\NormalTok{ randtoolbox}\SpecialCharTok{::}\NormalTok{halton}
\NormalTok{model2d}\SpecialCharTok{$}\NormalTok{N }\OtherTok{\textless{}{-}} \FunctionTok{c}\NormalTok{(}\FunctionTok{rep}\NormalTok{(}\DecValTok{300}\NormalTok{,}\DecValTok{8}\NormalTok{), }\FunctionTok{rep}\NormalTok{(}\DecValTok{500}\NormalTok{,}\DecValTok{8}\NormalTok{), }\FunctionTok{rep}\NormalTok{(}\DecValTok{800}\NormalTok{,}\DecValTok{8}\NormalTok{)) }\CommentTok{\# design size across time{-}steps  }
\NormalTok{put2d.haltonRange }\OtherTok{\textless{}{-}} \FunctionTok{osp.fixed.design}\NormalTok{(model2d,}\AttributeTok{input.dom=}\SpecialCharTok{{-}}\DecValTok{1}\NormalTok{, }\AttributeTok{method=}\StringTok{"km"}\NormalTok{)}
\end{Highlighting}
\end{Shaded}

\begin{enumerate}
\def\labelenumi{(\roman{enumi})}
\setcounter{enumi}{3}
\tightlist
\item
  For completeness, I mention briefly one further option supported by \texttt{\{mlOSP\}}. If \texttt{input.domain} is omitted entirely, \texttt{osp.fixed.design} creates a path-based \({\cal D}_k = \tilde{x}^{(0),0:N}(k)\) directly from the pilot paths. This choice still embeds the replication aspect via \texttt{batch.nrep}, but otherwise is equivalent to \texttt{osp.prob.design}. Thus, \texttt{input.domain=NULL} generates a probabilistic, replicated design.
\end{enumerate}

Figure \ref{fig:design-comp} compares training designs at \(k=10\) from option (i) with Lattice space-filling (\(N_{10}=\) 120); option (iii) with LHS space-filling (\(N_{10}=\) 146), and time-varying \(N_k\) using Halton-sequence space-filling. For the latter, I visualize both \({\cal D}_k\) at \(k=10\) with \(N_{10}=\) 132 and \({\cal D}_{20}\) (\(t=0.8\)), where \(N_{20}=\) 206, but the training region is much larger too.

I end this section by comparing the above three solvers head-to-head in order to assess how the training domain affects the ultimate option price \(\check{V}(0,X(0))\):

\begin{Shaded}
\begin{Highlighting}[]
\NormalTok{oos}\FloatTok{.1} \OtherTok{\textless{}{-}} \FunctionTok{forward.sim.policy}\NormalTok{(test}\FloatTok{.2}\NormalTok{d, nSteps}\FloatTok{.2}\NormalTok{d, put2d.lattice}\SpecialCharTok{$}\NormalTok{fit,  model2d)}
\NormalTok{oos}\FloatTok{.2} \OtherTok{\textless{}{-}} \FunctionTok{forward.sim.policy}\NormalTok{(test}\FloatTok{.2}\NormalTok{d, nSteps}\FloatTok{.2}\NormalTok{d, put2d.lhsAdaptive}\SpecialCharTok{$}\NormalTok{fit,  model2d)}
\NormalTok{oos}\FloatTok{.3} \OtherTok{\textless{}{-}} \FunctionTok{forward.sim.policy}\NormalTok{(test}\FloatTok{.2}\NormalTok{d, nSteps}\FloatTok{.2}\NormalTok{d, put2d.haltonRange}\SpecialCharTok{$}\NormalTok{fit,  model2d)}
\FunctionTok{print}\NormalTok{( }\FunctionTok{round}\NormalTok{(}\FunctionTok{c}\NormalTok{(}\FunctionTok{mean}\NormalTok{(oos}\FloatTok{.1}\SpecialCharTok{$}\NormalTok{payoff), }\FunctionTok{mean}\NormalTok{(oos}\FloatTok{.2}\SpecialCharTok{$}\NormalTok{payoff), }\FunctionTok{mean}\NormalTok{(oos}\FloatTok{.3}\SpecialCharTok{$}\NormalTok{payoff)), }\DecValTok{3}\NormalTok{) )}
\end{Highlighting}
\end{Shaded}

\begin{verbatim}
## [1] 1.432 1.448 1.445
\end{verbatim}

The main take-away is that there is not much impact from the specifics of the space-filling, but the range and density of \(\tilde{\cal X}_k\) plays a significant role. Above, the hard-coded lattice design does noticeably worse, because its fixed domain of approximation and number of inputs are not adjusting to the time-dependent region of interest of the optimal stopping problem. Thus, I strongly advocate adaptive construction of \(\tilde{\cal X}_k\).

\hypertarget{sequential-designs}{%
\subsection{Sequential Designs}\label{sequential-designs}}

The quality of the training design \({\cal D}_k\) is linked to how well it supports the learning of \(\widehat{T}(k,\cdot)\). As discussed in \citep{Lu18}, it is most efficient to select inputs around the exercise boundary, i.e.~the region where the correct decision rule is most unclear. Of course, the exercise boundary is unknown a priori, motivating a sequential construction
of \({\cal D}_k\), gradually adding training samples \(x^n(k)\) indexed by \(n\). The goal of such adaptive \({\cal D}_k\) is to improve simulation efficiency through targeted placement of \(x^{n}(k)\)'s to maximize the learning of the timing value. Such \emph{active learning} methods are common in machine learning applications.

The \textbf{osp.seq.design} solver implements active learning of the exercise region through greedily optimizing an \emph{acquisition function} \(x \mapsto {\cal I}_n(x)\) that is a proxy for the information gain for the respective input location \(x\). The selection of \(x^{n}(k)\) is done one-by-one by maximizing \({\cal I}_n(\cdot)\). Currently, five different acquisition functions are implemented, specified via the \textbf{ei.func} parameter: ``smcu'', ``tmse'', ``sur'', ``csur'', ``amcu'', cf. \citep{binois2016practical} and \citep{Lu18}. Because all these acquisition function \({\cal I}_n(x)\) rely on the posterior uncertainty of \(\widehat{T}^{(n)}(k,\cdot)\), they require working with a GP-type emulator, namely \texttt{method} being one of ``km'', ``trainkm'', ``hetgp'' or ``homtp''.

\begin{figure}
\centering
\includegraphics[trim=0 0.2in 0 0]{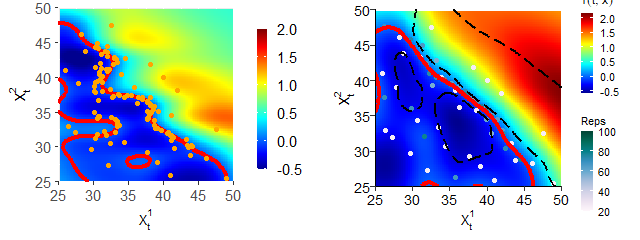}
\caption{\label{fig:adsa-plot}\label{fig:adsa}Left: Sequential design with sMCU. Right: Adaptive batching with ADSA. Both designs are at \(k=15\) (\(t=0.6\)) for the 2D Basket Put example. Replication counts \(r^n(k)\) are input-dependent and color coded in grayscale. The surface plots are color-coded according to \(\widehat{T}(k,\cdot)\).}
\end{figure}

The left panel of Figure \ref{fig:adsa} illustrates a sequential design constructed using the straddle Maximum Contour Uncertainty \texttt{sMCU} acquisition heuristic. The resulting design \(\cD_k\) at time step \(k=15\) (i.e.~\(t=0.6\)) has a total of \(N_{unique}=120\) unique training sites. Compared to the space-filling designs in Figure \ref{fig:design-comp}, \textbf{osp.seq.design} yields a highly non-uniform placement of \(x^n(k)\)'s,
placing them \emph{around the stopping boundary}. The different heuristics above vary in how aggressively they do this, known as the exploitation-exploration trade-off.
This adaptive behavior improves upon the space-filling designs that will often have many simulation sites far from the stopping boundary (e.g.~deep in-the-money) and which are therefore not informative about the optimal stopping rule. See \citet{Lu18} for a detailed analysis of sequential designs in RMC, including the trade-off between number of unique inputs \(N_{unique}\) and replication count \(N_{rep}\) (keeping \(N=N_{unique} \cdot N_{rep}\) fixed).

To set up the \textbf{osp.seq.design} solver, several additional ingredients must be specified. First, I initialize the experimental design of a given size \(n_0\) and provide the respective locations (taken from a space-filling QMC sequence). Second, the number of sequential rounds is specified via \texttt{seq.design.size} parameter which determines the final number of unique inputs. Thus, a total of \texttt{seq.design.size-init.size} rounds are run, during each of which \(x \mapsto {\cal I}_n(x)\) is maximized to pick the next unique input. To collect more samples, \textbf{osp.seq.design} relies on replication, i.e.~a simulation design of the form \eqref{eq:rep-design}. Third, I specify the details of the acquisition heuristic.
In the example below I start with \(n_0=30\) space-filling design sites \(x^{1:n_0}(k)\) (\texttt{init.size=30}), and add an additional 90 with \(N_{rep}=25\) replications each, for a total \(|\cD_k| = 120 \cdot 25 = 3000\). The sMCU acquisition function has another parameter \texttt{ucb.gamma} which is on the order of 1. Larger \texttt{ucb.gamma} emphasizes expoitation over exploration.

\begin{Shaded}
\begin{Highlighting}[]
\NormalTok{model2d}\SpecialCharTok{$}\NormalTok{init.size }\OtherTok{\textless{}{-}} \DecValTok{30}   \CommentTok{\# initial design size}
\NormalTok{sob30 }\OtherTok{\textless{}{-}}\NormalTok{ randtoolbox}\SpecialCharTok{::}\FunctionTok{sobol}\NormalTok{(}\DecValTok{55}\NormalTok{, }\AttributeTok{d=}\DecValTok{2}\NormalTok{)  }\CommentTok{\# build a Sobol space{-}filling design to initialize}
\NormalTok{sob30 }\OtherTok{\textless{}{-}}\NormalTok{ sob30[ }\FunctionTok{which}\NormalTok{( sob30[,}\DecValTok{1}\NormalTok{] }\SpecialCharTok{+}\NormalTok{ sob30[,}\DecValTok{2}\NormalTok{] }\SpecialCharTok{\textless{}=} \DecValTok{1}\NormalTok{) ,]  }
\NormalTok{sob30 }\OtherTok{\textless{}{-}} \DecValTok{25}\SpecialCharTok{+}\DecValTok{30}\SpecialCharTok{*}\NormalTok{sob30}
\NormalTok{model2d}\SpecialCharTok{$}\NormalTok{init.grid }\OtherTok{\textless{}{-}}\NormalTok{ sob30}

\NormalTok{model2d}\SpecialCharTok{$}\NormalTok{batch.nrep }\OtherTok{\textless{}{-}} \DecValTok{25}  \CommentTok{\# N\_rep}
\NormalTok{model2d}\SpecialCharTok{$}\NormalTok{seq.design.size }\OtherTok{\textless{}{-}} \DecValTok{120}  \CommentTok{\# final design size {-}{-} a total of 3000 simulations}
\NormalTok{model2d}\SpecialCharTok{$}\NormalTok{ei.func }\OtherTok{\textless{}{-}} \StringTok{"smcu"}  \CommentTok{\# straddle maximum contour uncertainty}
\NormalTok{model2d}\SpecialCharTok{$}\NormalTok{ucb.gamma }\OtherTok{\textless{}{-}} \DecValTok{1}  \CommentTok{\# sMCU parameter}

\NormalTok{model2d}\SpecialCharTok{$}\NormalTok{kernel.family }\OtherTok{\textless{}{-}} \StringTok{"Matern5\_2"}
\NormalTok{model2d}\SpecialCharTok{$}\NormalTok{update.freq }\OtherTok{\textless{}{-}} \DecValTok{5}      \CommentTok{\# frequency of re{-}fitting GP hyperparameters}

\NormalTok{put2d.mcu.gp }\OtherTok{\textless{}{-}} \FunctionTok{osp.seq.design}\NormalTok{(model2d, }\AttributeTok{method=}\StringTok{"hetgp"}\NormalTok{)}
\NormalTok{oos.mcu.gp }\OtherTok{\textless{}{-}} \FunctionTok{forward.sim.policy}\NormalTok{( test}\FloatTok{.2}\NormalTok{d, nSteps}\FloatTok{.2}\NormalTok{d, put2d.mcu.gp}\SpecialCharTok{$}\NormalTok{fit, model2d)}
\end{Highlighting}
\end{Shaded}

The primary take-away is that \textbf{osp.seq.design} is highly efficient in terms of keeping \(N\) low, but is rather slow due to shifting the work from running simulations to updating the emulator. Indeed, since each round has nontrivial overhead in optimizing oer \({\cal I}_n(\cdot)\), I do not recommend that users run it with more than 150 or so sequential rounds. The gains from sequential design would be higher in models where simulation is expensive (e.g.~where very small simulation steps are needed to generate \(X(k+1)|X(k)\)).

\hypertarget{deep-dive-adaptive-batching}{%
\subsection{Deep Dive: Adaptive Batching}\label{deep-dive-adaptive-batching}}

Conceptually, active learning is intended to favor locations close to the exercise boundary where the correct decision rule is hardest to resolve. As a result, sequential designs will increasingly concentrate, i.e.~the added \(x^n(k)\)'s cluster as \(n\) grows, see Figure \ref{fig:adsa} left. \emph{Adaptive batching} takes advantage of this by gradually increasing the replication level, in effect replacing clusters of \(x^n\)'s with a single replicated input. This allows to reduce the number of unique inputs \(N_{unique}\) and speeds up the construction of the sequential design. In an adaptively batched design, the constant \(N_{rep}\) in \eqref{eq:rep-design} is replaced with input-dependent replication counts \(r^n(k)\):
\begin{align}
 {\cal D}_k = \Bigl\{ \underbrace{x^{1}, x^{1}, \ldots, x^{1}}_{r^1(k) \text{ times}}, \underbrace{x^{2}, \ldots}_{r^2(k) \text{ times}}, x^{3}, \ldots, \ldots, x^{{N}_{unique}} \Bigr\},
\end{align}
where the algorithm now specifies both the unique inputs \(x^1(k),\ldots\) and the respective \(r^1(k), r^2(k),\ldots\). This idea was explored in detail in \citet{Lyu20} that proposed several strategies to construct \(r^n(k)\) sequentially and is implemented in the \texttt{osp.seq.batch.design} function. The latter works with a GP-based emulator and includes a choice of several \emph{batching heuristics} that control how \(r^n(k)\) is obtained: \texttt{fb} (Fixed Batching); \texttt{mlb}: Multi-Level Batching;
\texttt{rb}: Ratchet Batching; \texttt{absur}: Adaptively Batched SUR; \texttt{adsa}: Adaptive Batching Design with Stepwise Allocation; \texttt{ddsa}: Deterministic Batching Design with Stepwise Allocation.

To implement adaptive batching one needs to specify the batch heuristic via \texttt{batch.heuristic} and the sequential design acquisition function \({\cal I}_n(\cdot)\) as in Section \ref{sequential-designs} via \texttt{ei.func}. Below I apply the Adaptive Design with Sequential Allocation (ADSA) scheme. ADSA relies on the AMCU acquisition function and at each batching round either adds a new training input \(x^{n+1}(k)\) \emph{or} allocates additional simulations to existing \(x^{1:n}(k)\).
Below I employ ADSA with a heteroskedastic GP emulator (method set to \texttt{hetgp}). The other settings match the \textbf{osp.seq.design} solver from the previous section to allow for the best comparison.

\begin{Shaded}
\begin{Highlighting}[]
\NormalTok{model2d}\SpecialCharTok{$}\NormalTok{total.budget }\OtherTok{\textless{}{-}} \DecValTok{3000}  \CommentTok{\# total simulation budget N}

\FunctionTok{set.seed}\NormalTok{(}\DecValTok{110}\NormalTok{)}
\NormalTok{model2d}\SpecialCharTok{$}\NormalTok{batch.heuristic }\OtherTok{\textless{}{-}} \StringTok{\textquotesingle{}adsa\textquotesingle{}}  \CommentTok{\# ADSA with AMCU acquisition function}
\NormalTok{model2d}\SpecialCharTok{$}\NormalTok{ei.func }\OtherTok{\textless{}{-}} \StringTok{\textquotesingle{}amcu\textquotesingle{}}
\NormalTok{model2d}\SpecialCharTok{$}\NormalTok{cand.len }\OtherTok{\textless{}{-}} \DecValTok{1000}
\NormalTok{put2d.adsa }\OtherTok{\textless{}{-}} \FunctionTok{osp.seq.batch.design}\NormalTok{(model2d, }\AttributeTok{method=}\StringTok{"hetgp"}\NormalTok{)}
\end{Highlighting}
\end{Shaded}

At \(k=15\) ADSA yields \(\cD_{15}\) with \(N_{unique}=\) 65 unique inputs and with replication counts ranging up to \(r^{n}(k)=79\), see right panel of Figure \ref{fig:adsa}. Note that the geometry of \(|\cD_k|\) is very similar to that in Figure \ref{fig:adsa} left, but \(N_{unique}\) is noticeably lower. Thus, adaptive batching allows to reduce the number of sequential design rounds and the associated computational overhead, running multiple times faster than a comparable \textbf{osp.seq.design} solver. The respective contract prices are 1.4453 for sMCU and 1.4478 for ADSA which are both excellent, however the running times of 7.46 and 1.7 minutes respectively, show that the ADSA-based solver is more than 4 times faster. \texttt{osp.seq.batch.design} yields the most (so far) time-efficient GP-based solver in dimensions higher than \(d > 2\).

\hypertarget{regression-emulators}{%
\section{Regression Emulators}\label{regression-emulators}}

The selection of the regression module is the most well known aspect of RMC.
Over the past two decades, numerous proposals have been put forth beyond the original idea of ordinary least squares regression with user-specified basis functions. Indeed, there are literally hundreds of potential tools to fit a \(\widehat{T}(k,\cdot)\) or equivalently \(\hat{V}(k,\cdot)\), coming from the worlds of statistical and/or machine learning. With \texttt{\{mlOSP\}} I take advantage of the standardized \texttt{R} regression API to implement more than a dozen such emulators. At its core, a regression module is based on the generic \texttt{fit} and \texttt{predict} methods and is otherwise fully interchangeable in terms of learning \(\widehat{T}(k,\cdot)\). This means that it is possible to, say, substitute a neural network emulator with a random forest one, keeping all other aspects of the scheme exactly the same, and only changing a couple lines of code. In \texttt{\{mlOSP\}} this is enabled through the \texttt{method} field of the respective solver. The implementation allows for a straightforward swapping of regression methods, facilitating comparison and experimentation.

Before presenting new ideas for emulating \(\widehat{T}(k,\cdot)\), below is a summary of the best-known extant proposals and their corresponding \texttt{\{mlOSP\}} implementation if available:

\begin{itemize}
\item
  Piecewise regression with adaptive sub-grids by \citet[\textbf{osp.probDesign.piecewisebw} solver, see below]{BouchardWarin10};
\item
  Regularized regression, such as LASSO, by \citet{Kohler12lasso};
\item
  Kernel regression by \citet{Belomestny11}. \texttt{method=rvm} implements relevance vector machine regression, with the radial basis function kernel as one option.
\item
  Gaussian Process regression by Goudenege et al.~\citep{GoudenegeZanette19, GoudenegeZanette20} and the author \citep{Lu18}, see Section 5.3;
\item
  Neural nets by \citet{Kohler10nn}. \texttt{method=nnet} implements a single-layer neural network with a linear activation function.
\item
  Dynamic trees by \citet{GL13}. \texttt{method=dynatree} implements dynamic trees using \texttt{\{dynaTree\}} package, representing \(\widehat{T}(k,\cdot)\) via a piecewise regression with adaptively generated spatial partitions. The hierarchical partitioning is similar to random forest, but utilizes a different Bayesian-inspired mechanism.
\end{itemize}

Appendix C.2 lists all the available choices, including further tools like Multivariate Adaptive Regression Splines, Local Linear Regression (LOESS), and smoothing splines.

\hypertarget{a-potpourri-of-mlosp-regressions}{%
\subsection{\texorpdfstring{A potpourri of \texttt{\{mlOSP\}} regressions}{A potpourri of \{mlOSP\} regressions}}\label{a-potpourri-of-mlosp-regressions}}

As a way to showcase new variants of RMC emulators, I illustrate them on a fixed multidimensional OSP, ending with an apples-to-apples horse race of their performance. To this end, consider a 3D OSP instance of a max-Call option \[h_{\text{maxCall}}(t,x) := e^{-r t} \left(\max_{i=1,\ldots, d} X_i - {\cal K} \right)_+\]
where the underlying assets follow i.i.d.~GBMs. Following \citep[(MS'04), Table 2 p.~1230]{Broadie} I take \(\Delta t = 1/3\) with \(T=3\), i.e.~\(K=9\) exercise dates. I further take an OTM initial condition \(X(0) = (90,90,90)\) with strike \({\cal K}=100\). The true price of this max-Call is about \(V(0,{X}(0))=11.25\) under continuous exercise optionality.

\begin{Shaded}
\begin{Highlighting}[]
\NormalTok{modelBrGl3d }\OtherTok{\textless{}{-}} \FunctionTok{list}\NormalTok{(}\AttributeTok{K=}\DecValTok{100}\NormalTok{, }\AttributeTok{r=}\FloatTok{0.05}\NormalTok{, }\AttributeTok{div=}\FloatTok{0.1}\NormalTok{, }\AttributeTok{sigma=}\FunctionTok{rep}\NormalTok{(}\FloatTok{0.2}\NormalTok{,}\DecValTok{3}\NormalTok{),}\AttributeTok{T=}\DecValTok{3}\NormalTok{, }\AttributeTok{dt=}\DecValTok{1}\SpecialCharTok{/}\DecValTok{3}\NormalTok{,}
  \AttributeTok{x0=}\FunctionTok{rep}\NormalTok{(}\DecValTok{90}\NormalTok{,}\DecValTok{3}\NormalTok{),}\AttributeTok{dim=}\DecValTok{3}\NormalTok{, }\AttributeTok{sim.func=}\NormalTok{sim.gbm,}\AttributeTok{payoff.func=}\NormalTok{maxi.call.payoff)}
\end{Highlighting}
\end{Shaded}

To allow comparisons between solvers, I generate a shared test set of 20,000 out-of-sample trajectories and report the resulting mean payoff on that same, fixed \texttt{test.3d}.

I start with the random forest (RF) regression emulator. \citet{TompaidisYang13} investigated the use of regression trees for RMC, but I could not find a prior reference for RF's. RF regression generates a piecewise constant fit obtained as an ensemble estimator from a collection of hierarchical partition trees. Even though the fit is discontinuous, RF's are known to be among the most robust (in terms of scalability, resistance to non-Gaussianity, etc.) emulators and perform excellently in many statistical learning contexts.
To specify a RF emulator requires inputting the number of trees (\texttt{rf.ntree}) and the maximum size of each tree terminal node (\texttt{rf.maxnode}) that are passed to the \texttt{\{randomForest\}} package. The rest is all taken care of by \texttt{\{mlOSP\}} and the previously defined \texttt{model}. RF is not very fast, but can be scaled to \(N > 10^6\) inputs and is agnostic (in terms of coding effort) to problem dimension \(d\). Below I use \(N=10^5\) training paths.

\begin{Shaded}
\begin{Highlighting}[]
\NormalTok{modelBrGl3d}\SpecialCharTok{$}\NormalTok{rf.ntree }\OtherTok{=} \DecValTok{200}  \CommentTok{\# random forest parameters}
\NormalTok{modelBrGl3d}\SpecialCharTok{$}\NormalTok{rf.maxnode}\OtherTok{=}\DecValTok{200}
\NormalTok{call3d.rf }\OtherTok{\textless{}{-}} \FunctionTok{osp.prob.design}\NormalTok{(}\DecValTok{100000}\NormalTok{,modelBrGl3d,}\AttributeTok{method=}\StringTok{"randomforest"}\NormalTok{)}
\end{Highlighting}
\end{Shaded}

Another general-purpose emulator is neural networks (NN). The \texttt{\{nnet\}} package builds a neural network for \(\widehat{T}(k,\cdot)\) with a single hidden layer specified via the number of neurons \texttt{nn.nodes} (40 below) and the linear activation function. It provides a streamlined interface that is agnostic to \(d\). More advanced versions, such as those in \texttt{\{keras\}} tend to require much more fine-tuning.

\begin{Shaded}
\begin{Highlighting}[]
\NormalTok{modelBrGl3d}\SpecialCharTok{$}\NormalTok{nn.nodes }\OtherTok{\textless{}{-}} \DecValTok{40}
\NormalTok{call3d.nnet }\OtherTok{\textless{}{-}} \FunctionTok{osp.prob.design}\NormalTok{(}\AttributeTok{N=}\DecValTok{100000}\NormalTok{,modelBrGl3d, }\AttributeTok{method=}\StringTok{"nnet"}\NormalTok{)}
\end{Highlighting}
\end{Shaded}

Both RF and NN provide non-parametric fits which are attractive in the sense of universal approximation---zero asymptotic regression error as data and model complexity go to infinity. The oldest and best-understood non-parametric emulators are furnished by kernel regression. \texttt{\{mlOSP\}} provides a couple of different variants. First, I propose to use Relevance Vector Machine (RVM) emulators from the package \texttt{\{kernlab\}}. The underlying \texttt{rvm.kernel} kernel defaults to the Gaussian radial basis functions (\texttt{rbfdot}). RVM is computationally intensive in the number of kernel functions to use and directly using thousands of \(x^n(k)\) is too slow. To this end, I propose to combine RVM with a replicated design. The example below uses 800 unique inputs and total simulation budget of \(N = 20,000 = 800 \cdot 25\), i.e.~\(N_{rep}=25\) replicates per site. This time, for variety's sake, I pick a space-filling design:

\begin{Shaded}
\begin{Highlighting}[]
\NormalTok{modelBrGl3d}\SpecialCharTok{$}\NormalTok{N }\OtherTok{\textless{}{-}} \DecValTok{800}  \CommentTok{\# N\_unique}
\NormalTok{modelBrGl3d}\SpecialCharTok{$}\NormalTok{batch.nrep }\OtherTok{\textless{}{-}} \DecValTok{25} \CommentTok{\# N\_rep}
\NormalTok{modelBrGl3d}\SpecialCharTok{$}\NormalTok{pilot.nsims }\OtherTok{\textless{}{-}} \DecValTok{1000}
\NormalTok{lhs.rect }\OtherTok{\textless{}{-}} \FunctionTok{matrix}\NormalTok{(}\DecValTok{0}\NormalTok{, }\AttributeTok{nrow=}\DecValTok{3}\NormalTok{, }\AttributeTok{ncol=}\DecValTok{2}\NormalTok{) }\CommentTok{\# domain of approximation}
\NormalTok{lhs.rect[}\DecValTok{1}\NormalTok{,] }\OtherTok{\textless{}{-}}\NormalTok{ lhs.rect[}\DecValTok{2}\NormalTok{,] }\OtherTok{\textless{}{-}}\NormalTok{ lhs.rect[}\DecValTok{3}\NormalTok{,] }\OtherTok{\textless{}{-}} \FunctionTok{c}\NormalTok{(}\DecValTok{50}\NormalTok{,}\DecValTok{150}\NormalTok{)}
\NormalTok{modelBrGl3d}\SpecialCharTok{$}\NormalTok{qmc.method }\OtherTok{\textless{}{-}}\NormalTok{ randtoolbox}\SpecialCharTok{::}\NormalTok{sobol  }\CommentTok{\# space{-}filling using QMC sequence}
\NormalTok{call3d.lhsFixed.rvm }\OtherTok{\textless{}{-}} \FunctionTok{osp.fixed.design}\NormalTok{(modelBrGl3d,}\AttributeTok{input.domain=}\NormalTok{lhs.rect, }\AttributeTok{method=}\StringTok{"rvm"}\NormalTok{)}
\end{Highlighting}
\end{Shaded}

An alternative kernel regression package is \texttt{\{npreg\}}. Below I use it with a Epanechnikov order-4 kernel and local-constant regression \texttt{np.regtype=lc}. The bandwidth is estimated using least squares cross-validation (default \texttt{npreg} option).

\begin{Shaded}
\begin{Highlighting}[]
\NormalTok{modelBrGl3d}\SpecialCharTok{$}\NormalTok{np.kertype }\OtherTok{\textless{}{-}} \StringTok{"epanechnikov"}
\NormalTok{modelBrGl3d}\SpecialCharTok{$}\NormalTok{np.kerorder }\OtherTok{\textless{}{-}} \DecValTok{4}\NormalTok{; modelBrGl3d}\SpecialCharTok{$}\NormalTok{np.regtype }\OtherTok{\textless{}{-}} \StringTok{"lc"}
\NormalTok{call3d.sobFixed.np }\OtherTok{\textless{}{-}} \FunctionTok{osp.fixed.design}\NormalTok{(modelBrGl3d,}\AttributeTok{input.domain=}\NormalTok{lhs.rect, }\AttributeTok{method=}\StringTok{"npreg"}\NormalTok{)}
\end{Highlighting}
\end{Shaded}

As a final emulation idea, consider multivariate adaptive regression splines (MARS) as implemented in the \texttt{\{earth\}} package.
MARS provides adaptive feature selection using a forward-backward selection algorithm. I set the degree to be \texttt{earth.deg=2}, so that bases consist of linear/quadratic hinge functions, and allow up to \texttt{earth.nk=200} hinges. Finally I also set the backward fit threshold \texttt{earth.thresh}. I take this opportunity to also illustrate the TvR scheme which relies on the one-step-ahead value function \(\hat{V}(k+1,\cdot)\) instead of \(h({\tau}_{\widehat{A}_{k+1:K}}, \cdot)\) during the regression. This is available via the top-level solver \texttt{osp.tvr} that otherwise follows the exact same syntax as \texttt{osp.prob.design} and so can be mixed and matched with any of the above regression methods.

\begin{Shaded}
\begin{Highlighting}[]
\NormalTok{earthParams }\OtherTok{\textless{}{-}} \FunctionTok{c}\NormalTok{(}\AttributeTok{earth.deg=}\DecValTok{2}\NormalTok{,}\AttributeTok{earth.nk=}\DecValTok{200}\NormalTok{,}\AttributeTok{earth.thresh=}\FloatTok{1E{-}8}\NormalTok{)  }\CommentTok{\# passed to \{earth\}}
\NormalTok{call3d.tvr.mars }\OtherTok{\textless{}{-}} \FunctionTok{osp.tvr}\NormalTok{(}\AttributeTok{N=}\DecValTok{100000}\NormalTok{, }\FunctionTok{c}\NormalTok{(modelBrGl3d,earthParams), }\AttributeTok{method=}\StringTok{"earth"}\NormalTok{)}
\end{Highlighting}
\end{Shaded}

To my knowledge, most of the above choices are new, or at least partially new. Many more can be added.
After having built all these models one can do horse racing on a fixed out-of-sample set of scenarios. A typical call to do so is

\begin{Shaded}
\begin{Highlighting}[]
\NormalTok{oos.tvr.mars }\OtherTok{\textless{}{-}} \FunctionTok{forward.sim.policy}\NormalTok{( test}\FloatTok{.3}\NormalTok{d,nSteps}\FloatTok{.3}\NormalTok{d,call3d.tvr.mars}\SpecialCharTok{$}\NormalTok{fit, modelBrGl3d)}
\end{Highlighting}
\end{Shaded}

I repeat this for 25 times each in order to record also the sampling standard error of each \(\check{V}(0,X(0))\) estimator, which is a measure of the scheme's stability. The above sampling error reflects sensitivity to the training data, keeping all model tuning parameters and the test set fixed.

\begin{longtable}[]{@{}lrrr@{}}
\caption{\label{tab:table-regrs}\label{tbl:2dput}Comparison of 3D max-Call solvers with 5 different regression emulators. Running times based on a Surface Studio laptop with 32GB memory and Intel Core i-7 3.3GHZ processor.}\tabularnewline
\toprule()
Emulator & Mean Price & Std. Error & Time (secs) \\
\midrule()
\endfirsthead
\toprule()
Emulator & Mean Price & Std. Error & Time (secs) \\
\midrule()
\endhead
RF & 11.138 & 0.0058 & 34.5 \\
NNet & 11.152 & 0.0071 & 349.0 \\
npreg & 10.738 & 0.0095 & 62.9 \\
RVM & 11.151 & 0.0078 & 35.2 \\
MARS & 10.932 & 0.0041 & 51.6 \\
\bottomrule()
\end{longtable}

According to Table \ref{tbl:2dput}, RVM and NNet emulators perform best, yielding the highest out-of-sample payoffs. At the other end, npreg and MARS perform quite poorly. However, looking at running times, NNet is in fact seen to be the slowest of the bunch, taking more than 10x extra time compared to RVM, which is the fastest. In terms of standard errors, all emulators are quite close to another, with \texttt{npreg} (which performs worst) being the most stable. We note that direct comparisons are not straightforward since the different solvers utilized different training sets, i.e.~\(N\) is method-specific. In Section 6.3 below, I carry out additional fine-tuning of the LM and NNet solvers.

\hypertarget{deep-dive-specifying-bases-for-a-linear-model}{%
\subsection{Deep Dive: Specifying bases for a linear model}\label{deep-dive-specifying-bases-for-a-linear-model}}

The most extensively studied RMC approach follows the classical regression paradigm of a linear model with explicitly specified bases \(B_1(\cdot), \ldots, B_{R-1}(\cdot)\). This means that the approximation space \(\cH_k\) has \(R\) degrees of freedom and \(\widehat{T}(k,\cdot) \in \sspan( B_1, \ldots, B_{R-1})\). From a statistical perspective, any set of linearly independent bases will do, although for theoretical analysis one often picks special (orthogonal) basis families, such as Hermite polynomials. In \texttt{\{mlOSP\}} selecting \texttt{method=lm} allows the user to specify any collection of bases, giving complete transparency on constructing \(\cH_k\).

It is well known that linear models are prone to overfitting, in part due to their sensitivity to the Gaussian homoskedastic noise assumption that is strongly violated in RMC. The skewed distribution of \(\epsilon(x)\) and state-dependent simulation variance make the resulting fit biased and less stable compared to regularized regression approaches, such as MARS or RF. To avoid overfitting, large training sets are needed, illustrating one of the trade-offs that must be considered for RMC implementations: namely the trade-off between speed (\texttt{lm} is effectively the fastest possible regression emulator) and memory (very large \texttt{N} required in order not to overfit). The memory constraint precludes scalability of \texttt{method=lm} to high dimensions whereby the needed number of bases grows quickly.

As an illustration, below I provide definitions of polynomial bases of degree up to 2 (\texttt{bas2}) and up to degree 3 (\texttt{bas3}) for the above 3D max-Call example, using \(N=3 \cdot 10^5\) (300 thousand) paths. In the latter case, I also append the payoff \(h(k,x)\) to the set of bases.
The first expression below takes \(\cH_k = \sspan(1, x_1, x_1^2, x_2, x_2^2, x_1 \cdot x_2, x_3, x_3^2, x_3 \cdot x_2, x_3 \cdot x_1)\) and similarly for degree-3 polynomials.

\begin{Shaded}
\begin{Highlighting}[]
\CommentTok{\# polynomials of degree \textless{}= 2}
\NormalTok{bas2 }\OtherTok{\textless{}{-}} \ControlFlowTok{function}\NormalTok{(x) }\FunctionTok{return}\NormalTok{(}\FunctionTok{cbind}\NormalTok{(x[,}\DecValTok{1}\NormalTok{],x[,}\DecValTok{1}\NormalTok{]}\SpecialCharTok{\^{}}\DecValTok{2}\NormalTok{,x[,}\DecValTok{2}\NormalTok{],x[,}\DecValTok{2}\NormalTok{]}\SpecialCharTok{\^{}}\DecValTok{2}\NormalTok{,x[,}\DecValTok{1}\NormalTok{]}\SpecialCharTok{*}\NormalTok{x[,}\DecValTok{2}\NormalTok{],x[,}\DecValTok{3}\NormalTok{],x[,}\DecValTok{3}\NormalTok{]}\SpecialCharTok{\^{}}\DecValTok{2}\NormalTok{,}
\NormalTok{                                 x[,}\DecValTok{3}\NormalTok{]}\SpecialCharTok{*}\NormalTok{x[,}\DecValTok{2}\NormalTok{],x[,}\DecValTok{1}\NormalTok{]}\SpecialCharTok{*}\NormalTok{x[,}\DecValTok{3}\NormalTok{]))}
\CommentTok{\# polynomials up to degree 3 + the payoff}
\NormalTok{bas3 }\OtherTok{\textless{}{-}} \ControlFlowTok{function}\NormalTok{(x) }\FunctionTok{return}\NormalTok{(}\FunctionTok{cbind}\NormalTok{(x[,}\DecValTok{1}\NormalTok{],x[,}\DecValTok{1}\NormalTok{]}\SpecialCharTok{\^{}}\DecValTok{2}\NormalTok{,x[,}\DecValTok{2}\NormalTok{],x[,}\DecValTok{2}\NormalTok{]}\SpecialCharTok{\^{}}\DecValTok{2}\NormalTok{,x[,}\DecValTok{1}\NormalTok{]}\SpecialCharTok{*}\NormalTok{x[,}\DecValTok{2}\NormalTok{],x[,}\DecValTok{3}\NormalTok{],x[,}\DecValTok{3}\NormalTok{]}\SpecialCharTok{\^{}}\DecValTok{2}\NormalTok{,}
\NormalTok{                                 x[,}\DecValTok{3}\NormalTok{]}\SpecialCharTok{*}\NormalTok{x[,}\DecValTok{2}\NormalTok{],x[,}\DecValTok{1}\NormalTok{]}\SpecialCharTok{*}\NormalTok{x[,}\DecValTok{3}\NormalTok{], x[,}\DecValTok{1}\NormalTok{]}\SpecialCharTok{\^{}}\DecValTok{3}\NormalTok{,x[,}\DecValTok{2}\NormalTok{]}\SpecialCharTok{\^{}}\DecValTok{3}\NormalTok{,x[,}\DecValTok{3}\NormalTok{]}\SpecialCharTok{\^{}}\DecValTok{3}\NormalTok{,}
\NormalTok{                             x[,}\DecValTok{1}\NormalTok{]}\SpecialCharTok{\^{}}\DecValTok{2}\SpecialCharTok{*}\NormalTok{x[,}\DecValTok{2}\NormalTok{],x[,}\DecValTok{1}\NormalTok{]}\SpecialCharTok{\^{}}\DecValTok{2}\SpecialCharTok{*}\NormalTok{x[,}\DecValTok{3}\NormalTok{],x[,}\DecValTok{2}\NormalTok{]}\SpecialCharTok{\^{}}\DecValTok{2}\SpecialCharTok{*}\NormalTok{x[,}\DecValTok{1}\NormalTok{],x[,}\DecValTok{2}\NormalTok{]}\SpecialCharTok{\^{}}\DecValTok{2}\SpecialCharTok{*}\NormalTok{x[,}\DecValTok{3}\NormalTok{],}
\NormalTok{                             x[,}\DecValTok{3}\NormalTok{]}\SpecialCharTok{\^{}}\DecValTok{2}\SpecialCharTok{*}\NormalTok{x[,}\DecValTok{1}\NormalTok{],x[,}\DecValTok{3}\NormalTok{]}\SpecialCharTok{\^{}}\DecValTok{2}\SpecialCharTok{*}\NormalTok{x[,}\DecValTok{2}\NormalTok{],x[,}\DecValTok{1}\NormalTok{]}\SpecialCharTok{*}\NormalTok{x[,}\DecValTok{2}\NormalTok{]}\SpecialCharTok{*}\NormalTok{x[,}\DecValTok{3}\NormalTok{],}
                             \FunctionTok{maxi.call.payoff}\NormalTok{(x,modelBrGl3d)) )  }\CommentTok{\# include the payoff}

\NormalTok{modelBrGl3d}\SpecialCharTok{$}\NormalTok{bases }\OtherTok{\textless{}{-}}\NormalTok{ bas2  }\CommentTok{\# 10 coefficients to fit}
\NormalTok{lm.run2 }\OtherTok{\textless{}{-}} \FunctionTok{osp.prob.design}\NormalTok{(}\DecValTok{300000}\NormalTok{,modelBrGl3d,}\AttributeTok{method=}\StringTok{"lm"}\NormalTok{)}
\NormalTok{oos.lm2 }\OtherTok{\textless{}{-}} \FunctionTok{forward.sim.policy}\NormalTok{(test}\FloatTok{.3}\NormalTok{d, nSteps}\FloatTok{.3}\NormalTok{d, lm.run2}\SpecialCharTok{$}\NormalTok{fit, modelBrGl3d)}
\NormalTok{modelBrGl3d}\SpecialCharTok{$}\NormalTok{bases }\OtherTok{\textless{}{-}}\NormalTok{ bas3  }\CommentTok{\# 21 coefficients to fit}
\NormalTok{lm.run3 }\OtherTok{\textless{}{-}} \FunctionTok{osp.prob.design}\NormalTok{(}\DecValTok{300000}\NormalTok{,modelBrGl3d,}\AttributeTok{method=}\StringTok{"lm"}\NormalTok{)}
\NormalTok{oos.lm3 }\OtherTok{\textless{}{-}} \FunctionTok{forward.sim.policy}\NormalTok{(test}\FloatTok{.3}\NormalTok{d, nSteps}\FloatTok{.3}\NormalTok{d, lm.run3}\SpecialCharTok{$}\NormalTok{fit, modelBrGl3d)}
\end{Highlighting}
\end{Shaded}

The second estimator is found to be significantly better, yielding \(\check{V}(0,X(0)) =\) 11.2239 compared to 10.7213 for the first one. In other words, having 21 bases is better than having only 10. At the same time it is also 2.5 times slower (11.21 vs 3.999 seconds).

One may straightforwardly apply more sophisticated bases, for example based on \emph{order statistics} of \(X(k)\) which is a common trick in the context of a max-Call payoff \citep{BroadieCao08}. Below I first sort \(x\) as \(x_{(1)} \ge x_{(2)} \ge x_{(3)}\) and then work with the 10-dimensional \(\cH_k = \sspan(1, x_{(1)}, x_{(1)}^2, x_{(1)}^3, x_{(1)}^4, x_{(2)}, x_{(2)}^2, x_{(3)}, x_{(1)}\cdot x_{(2)}, x_{(1)} \cdot x_{(3)})\).

\begin{Shaded}
\begin{Highlighting}[]
\NormalTok{modelBrGl3d}\SpecialCharTok{$}\NormalTok{bases }\OtherTok{\textless{}{-}}\ControlFlowTok{function}\NormalTok{(x) \{}
\NormalTok{  sortx }\OtherTok{\textless{}{-}} \FunctionTok{t}\NormalTok{(}\FunctionTok{apply}\NormalTok{(x, }\DecValTok{1}\NormalTok{, sort, }\AttributeTok{decreasing =} \ConstantTok{TRUE}\NormalTok{))  }\CommentTok{\# sort coordinates in decreasing order}
  \FunctionTok{return}\NormalTok{(}\FunctionTok{cbind}\NormalTok{(sortx[,}\DecValTok{1}\NormalTok{],sortx[,}\DecValTok{1}\NormalTok{]}\SpecialCharTok{\^{}}\DecValTok{2}\NormalTok{, sortx[,}\DecValTok{1}\NormalTok{]}\SpecialCharTok{\^{}}\DecValTok{3}\NormalTok{,sortx[,}\DecValTok{1}\NormalTok{]}\SpecialCharTok{\^{}}\DecValTok{4}\NormalTok{, sortx[,}\DecValTok{2}\NormalTok{], }
\NormalTok{               sortx[,}\DecValTok{2}\NormalTok{]}\SpecialCharTok{\^{}}\DecValTok{2}\NormalTok{, sortx[,}\DecValTok{3}\NormalTok{], sortx[,}\DecValTok{1}\NormalTok{]}\SpecialCharTok{*}\NormalTok{sortx[,}\DecValTok{2}\NormalTok{], sortx[,}\DecValTok{1}\NormalTok{]}\SpecialCharTok{*}\NormalTok{sortx[,}\DecValTok{3}\NormalTok{] ))}
\NormalTok{\}}
\NormalTok{lm.run4 }\OtherTok{\textless{}{-}} \FunctionTok{osp.prob.design}\NormalTok{(}\DecValTok{300000}\NormalTok{,modelBrGl3d,}\AttributeTok{method=}\StringTok{"lm"}\NormalTok{)}
\end{Highlighting}
\end{Shaded}

\begin{Shaded}
\begin{Highlighting}[]
\NormalTok{oos.lm.sorted }\OtherTok{\textless{}{-}} \FunctionTok{forward.sim.policy}\NormalTok{(test}\FloatTok{.3}\NormalTok{d, nSteps}\FloatTok{.3}\NormalTok{d, lm.run4}\SpecialCharTok{$}\NormalTok{fit, modelBrGl3d)}
\FunctionTok{print}\NormalTok{(}\FunctionTok{mean}\NormalTok{(oos.lm.sorted}\SpecialCharTok{$}\NormalTok{payoff))}
\end{Highlighting}
\end{Shaded}

\begin{verbatim}
## [1] 11.27393
\end{verbatim}

The sorted coordinates better convey information about \({\cal S}_k\) than the unsorted ones, yielding the highest option price \(\check{V}(0,X(0))\) among all solvers presented. The above examples allude to the limitless scope for customizing \(\cH_k\) that is possible with \textbf{mlOSP}.

Another approach within the linear model regression framework is to build a piecewise-linear fit for \(\widehat{T}(k,\cdot)\) that is defined in terms of partitions of the state space. The Bouchard-Warin algorithm \citep{BouchardWarin10}
adaptively picks the regression sub-domains based on an equi-probable partition of \(\cD_k\). Thus, the sub-domains are empirically selected, typically with a fixed number of partitions in each coordinate of \(X\). The
\texttt{osp.probDesign.piecewisebw} solver implements the above, generating \(\cD_k\) from forward trajectories like in the LS scheme (cf.~\texttt{osp.prob.design}). In the example below, I take \texttt{nBins=5} bins per coordinate. With \(d=3\) and \(N=3 \cdot 10^5\) training samples, this implies having \(5^3=125\) sub-domains with 2400 trajectories in each. The overall \(\widehat{T}(k,\cdot)\) then has 500 coefficients to be fitted, based on running 125 degree-1 \texttt{lm} models across the partitions.

\begin{Shaded}
\begin{Highlighting}[]
\NormalTok{modelBrGl3d}\SpecialCharTok{$}\NormalTok{nBins }\OtherTok{\textless{}{-}} \DecValTok{5}
\NormalTok{bw.run }\OtherTok{\textless{}{-}} \FunctionTok{osp.probDesign.piecewisebw}\NormalTok{(}\DecValTok{300000}\NormalTok{,modelBrGl3d,}\AttributeTok{tst.paths=}\NormalTok{test}\FloatTok{.3}\NormalTok{d)}
\end{Highlighting}
\end{Shaded}

\begin{verbatim}
## In-sample estimated price:  11.245  and out-of-sample  11.107
\end{verbatim}

\hypertarget{variants-of-gp-emulators}{%
\subsection{Variants of GP Emulators}\label{variants-of-gp-emulators}}

Gaussian Process emulators for RMC were originally proposed in \citet{Lu18} and also investigated in \citet{GoudenegeZanette19}. One of their benefits is uncertainty quantification of the resulting fit, enabling the use of sequential design via \texttt{osp.seq.design} and \texttt{osp.seq.batch.design}. Another advantage is their expressivity, i.e.~ability to fit complex input-output relations based on just a few training inputs. Thanks to these properties, GP emulators can provide highly accurate fits even with juts a few dozen (well-placed) training \(x^n(k)\). At the same time, GPs have a cubic computational complexity in design size, and hence cannot directly handle more than a couple thousand unique inputs. Replicated designs offer one way out of this challenge; other solutions could be local GPs or sparse GPs.

The classical GP emulator assumes homoskedastic Gaussian simulation noise in \eqref{eq:stat-model} and learns its magnitude as part of the MLE fitting. However, this is a poor assumption for RMC because the variance in pathwise rewards is highly state-dependent, i.e.~heteroskedastic. Indeed, deep out-of-the-money the conditional variance of the payoff is negligible, whereas it is very high deep in-the-money. A partial solution to account for this is to combine a replicated design with Stochastic Kriging (SK) \citep{ankenman2010stochastic}. SK estimates \(\sigma^2(x)\) empirically via the classical MC variance estimator based on the batch of the \(N_{rep}\) pathwise timing values originating at \(x^n\):
\[ \hat{\sigma}^2(x^n) = \frac{1}{N_{rep}-1} \sum_{i=1}^{N_{rep}} (y^{n,i} - \bar{y}^n)^2.\]
SK does require a large batch size \(N_{rep}\) to be reliable; in practice we find that \(N_{rep} \gg 20\) is necessary.

In \texttt{\{mlOSP\}} I implement the following types of GP emulators:

\begin{itemize}
\item
  \texttt{km} based on user-specified hyperparameters \texttt{km.var}, \texttt{km.cov}. The GP is fit via the \texttt{\{DiceKriging\}} \citep{kmPackage-R} package that offers several choices for \texttt{kernel.family} and utilizes SK. Thus, the GP-km emulator treats noise variance as known (rather than a parameter to be estimated) and uses \(\hat{\sigma}^2(x^n)\) as a proxy for the true \(\sigma^2(x^n)\).
\item
  \texttt{trainkm} based on MLE-optimized GP hyperparameters as implemented in \texttt{\{DiceKriging\}} and SK;
\item
  \texttt{homgp} which is essentially the same as \texttt{trainkm} but with a different MLE optimizer from the \texttt{\{hetGP\}} package. Also uses SK;
\item
  \texttt{homtp}: homoskedastic \(t\)-Process regression also implemented in \texttt{\{hetGP\}};
\item
  \texttt{lagp} based on local approximate GP regression proposed in \citet{GramacyApley13} and implemented in \texttt{\{laGP\}};
\item
  \texttt{hetgp} heteroskedastic Gaussian process regression based on the eponymous \texttt{\{hetGP\}} package from \citet{binois2016practical};
\end{itemize}

The latter two choices are new in the RMC literature, and are proposed here for the first time due to their relevance for \texttt{mlOSP}, namely handling heteroskedasticity and spatial non-stationarity. The \texttt{hetGP} framework directly learns \(\sigma^2(\cdot)\) via a second spatial model that is jointly inferred with the model for the mean response. This has been recently implemented \citep{binois2016practical} in the \texttt{\{hetGP\}} package and available in \texttt{\{mlOSP\}} via \texttt{method="hetgp"}.

\begin{Shaded}
\begin{Highlighting}[]
\NormalTok{modelBrGl3d}\SpecialCharTok{$}\NormalTok{N }\OtherTok{\textless{}{-}} \DecValTok{500}
\NormalTok{modelBrGl3d}\SpecialCharTok{$}\NormalTok{batch.nrep }\OtherTok{\textless{}{-}} \DecValTok{60}
\NormalTok{modelBrGl3d}\SpecialCharTok{$}\NormalTok{kernel.family }\OtherTok{\textless{}{-}} \StringTok{"Matern5\_2"} \CommentTok{\# different naming compared to km }
\NormalTok{put3d.hetgp }\OtherTok{\textless{}{-}} \FunctionTok{osp.fixed.design}\NormalTok{(modelBrGl3d,}\AttributeTok{input.dom=}\FloatTok{0.02}\NormalTok{, }\AttributeTok{method=}\StringTok{"hetgp"}\NormalTok{)}
\NormalTok{oos.hetgp }\OtherTok{\textless{}{-}} \FunctionTok{forward.sim.policy}\NormalTok{(test}\FloatTok{.3}\NormalTok{d, nSteps}\FloatTok{.3}\NormalTok{d, put3d.hetgp}\SpecialCharTok{$}\NormalTok{fit, modelBrGl3d)}
\FunctionTok{cat}\NormalTok{(}\FunctionTok{round}\NormalTok{(}\FunctionTok{mean}\NormalTok{(oos.hetgp}\SpecialCharTok{$}\NormalTok{payoff),}\DecValTok{3}\NormalTok{) )}
\end{Highlighting}
\end{Shaded}

\begin{verbatim}
## 11.114
\end{verbatim}

Another way to achieve better scalability is to replace a single GP emulator that assumes a global correlation/noise structure with a local GP fit, analogous to replacing a global linear model with LOESS regression. To this end, I leverage the \texttt{laGP} framework which adopts a prediction-focused approach that builds a new \emph{sparse} GP every time a \(\widehat{T}(k,x)\) is needed. Those GPs are local (akin to nearest neighbors), allowing to capture spatial non-stationarity and heteroskedasticity. To illustrate, I utilize \texttt{lagp} with a non-replicated design of 5000 unique inputs which would be prohibitively expensive with \texttt{km}. The variant below constructs local fits with 40 inputs, selected according to the \texttt{alcray} criterion.

\begin{Shaded}
\begin{Highlighting}[]
\NormalTok{modelBrGl3d}\SpecialCharTok{$}\NormalTok{lagp.type}\OtherTok{=}\StringTok{"alcray"}
\NormalTok{modelBrGl3d}\SpecialCharTok{$}\NormalTok{lagp.end}\OtherTok{=}\DecValTok{40}
\NormalTok{put3d.lagp }\OtherTok{\textless{}{-}} \FunctionTok{osp.prob.design}\NormalTok{(}\DecValTok{7500}\NormalTok{,modelBrGl3d,}\AttributeTok{method=}\StringTok{"lagp"}\NormalTok{,}\AttributeTok{subset=}\DecValTok{1}\SpecialCharTok{:}\DecValTok{2500}\NormalTok{)}
\end{Highlighting}
\end{Shaded}

\begin{verbatim}
## [1] "In-sample price estimate 12.224; and out-of-sample: 10.584"
\end{verbatim}

\hypertarget{benchmarks}{%
\section{Benchmarks}\label{benchmarks}}

A key motivation for building the \textbf{mlOSP} template was to create transparent and verifiable benchmarks for RMC algorithms. While many published works contain detailed comparisons between a particular RMC version and competing approaches, these results are necessarily limited in scope and are often very difficult to reproduce. The numerous nuances that inevitably crop up when implementing RMC make such comparisons fraught, leading to a lack of consensus on what strategies are more efficient. Some of the challenges inherent to benchmarking are:

\begin{itemize}
\item
  RMC algorithms tend to have a slew of tuning parameters that significantly affect performance (from the number of simulations, to the precise choice of the regression specification);
\item
  One must define a complete problem instance, i.e.~the payoff function, state dynamics, initial condition, etc. While some test cases have appeared repeatedly in numerous articles, there is no agreed-upon portfolio of test instances. Many algorithms are highly sensitive to the dimensionality of the problem, the geometry of the value function, the initial condition, etc., requiring a diverse set of tests to reveal all their pros and cons.
\item
  There are multiple criteria one could utilize to compare solution quality across algorithms, including accuracy at a fixed simulation budget, accuracy at a fixed computation time, running time at fixed budget.
\item
  The above metrics are heavily affected by the particular environment, including the programming language (e.g.~\texttt{R} vs.~\texttt{C++}), hardware setup, operating system, etc.
\item
  A key practical goal of benchmarking is to assess scalability of an RMC scheme, i.e.~its ability to work well for a wide range of OSP instances. However, scalability is again a multi-faceted concept, related to simulation budget, running time, memory requirements, and of course accuracy, that all change nonlinearly as problems become more complex.
\end{itemize}

Due to all the above, the only way to create a ``level playing field'' for the different algorithms is to bring them all under one roof, coded side-by-side within a transparent computing environment. This is precisely what is achieved in \texttt{\{mlOSP\}}. In the companion software appendix, posted also on GitHub, I provide the RMarkdown code that can be run by any reader or end-user who downloads the \texttt{\{mlOSP\}} package to fully reproduce our results, figures and tables. To my knowledge, this is the most comprehensive verifiable set of RMC benchmarks (the StochOpt \citep{StOpt} also contains some benchmarks but primarily addresses more general stochastic control problems). The \texttt{R} code also makes completely transparent the chosen set of fully specified problem instances, which would be useful for other researchers in the future.

\emph{Remark}: the preprint \citet{herreraTeichmann} provides a related set of benchmarks for high-dimensional OSP instances with \(d \ge 5\).

\hypertarget{benchmarked-mlosp-instances}{%
\subsection{\texorpdfstring{Benchmarked \texttt{\{mlOSP\}} instances}{Benchmarked \{mlOSP\} instances}}\label{benchmarked-mlosp-instances}}

Below I present a list of 9 models and 10 solvers. All of them have appeared in previous articles. The OSP instances span a range of case studies (see Table \ref{tbl:osp-instances} as well as Appendix D for a full specification) in terms of:

\begin{itemize}
\item
  Problem dimension \(d\): 1D, 2D, 3D and 5D;
\item
  Number of time-steps \(K\) from 9 to 50;
\item
  Underlying dynamics, including Geometric Brownian motion and stochastic volatility;
\item
  Option payoffs \(h(t,x)\), including Puts, basket average Puts, max-Calls;
\item
  Problem geometry: both symmetric settings with i.i.d.~assets (coordinates of \(X(k)\)), as well as asymmetric/correlated asset dynamics.
\end{itemize}

\begin{table}[ht]

\caption{\label{tab:kable-info}\label{tbl:osp-instances}Benchmarked OSP Instances}
\centering
\resizebox{\linewidth}{!}{\begin{tabular}[t]{lrrlp{0.8in}p{3.25in}l}
\toprule
Model & Dim & Steps & Payoff & Dynamics & Notes & Ref\\
\midrule
\cellcolor{gray!6}{M1} & \cellcolor{gray!6}{1} & \cellcolor{gray!6}{25} & \cellcolor{gray!6}{Put} & \cellcolor{gray!6}{GBM} & \cellcolor{gray!6}{Classic 1D at-the-money Put} & \cellcolor{gray!6}{\cite{LS}}\\
M2 & 1 & 25 & Put & GBM & Same as M1 but out-of-the-money &  \cite{LS}\\
\cellcolor{gray!6}{M3} & \cellcolor{gray!6}{2} & \cellcolor{gray!6}{25} & \cellcolor{gray!6}{Ave Put} & \cellcolor{gray!6}{GBM} & \cellcolor{gray!6}{2D symmetric in-the-money basket Put} & \cellcolor{gray!6}{\cite{LS}} \\
M4 & 2 & 9 & Max Call & GBM & 2D symmetric in-the-money max-Call & \cite{LS}\\
\cellcolor{gray!6}{M5} & \cellcolor{gray!6}{2} & \cellcolor{gray!6}{50} & \cellcolor{gray!6}{SV Put} & \cellcolor{gray!6}{SV Heston} & \cellcolor{gray!6}{Put in a Stochastic volatility model, in-the-money} & \cellcolor{gray!6}{\cite{Rambharat11} }\\
\addlinespace
M6 & 3 & 9 & Max Call & GBM & 3D symmetric out-of-the-money max-Call & \cite{BroadieCao08}\\
\cellcolor{gray!6}{M7} & \cellcolor{gray!6}{5} & \cellcolor{gray!6}{9} & \cellcolor{gray!6}{Max Call} & \cellcolor{gray!6}{GBM} & \cellcolor{gray!6}{5D symmetric at-the-money max-Call} & \cellcolor{gray!6}{\cite{BroadieCao08}}\\
M8 & 5 & 9 & Max Call & GBM & 5D Asymmetric out-of-the-money max-Call & \cite{BroadieCao08}\\
\cellcolor{gray!6}{M9} & \cellcolor{gray!6}{5} & \cellcolor{gray!6}{20} & \cellcolor{gray!6}{Ave Put} & \cellcolor{gray!6}{GBM Cor} & \cellcolor{gray!6}{5D asymmetric correlated basket Put} & \cellcolor{gray!6}{\cite{Lelong19}}\\
\bottomrule
\end{tabular}}
\end{table}

The resulting benchmarks are reproducible (all random number seeds fixed and provided) and utilize shared out-of-sample test sets which are available upon request from the author, comprising more than 100Mb of data.

\hypertarget{benchmark-results}{%
\subsection{Benchmark Results}\label{benchmark-results}}

\begin{table}[ht]

\caption{\label{tab:solver-table}\label{tbl:bench-prices}Benchmarked Bermudan option prices. For each instance all methods share a common test set.}
\centering
\resizebox{\linewidth}{!}{
\begin{tabular}[t]{lrrrrrrrrrr}
\toprule
Model & S1-LM & S2-RF & S3-MARS & S4-TvR & S5-NNet & S6-BW & S7-GP & S8-ADSA & S9-Seq & S10-hetGP\\
\midrule
\cellcolor{gray!6}{M1} & \cellcolor{gray!6}{2.07} & \cellcolor{gray!6}{2.24} & \cellcolor{gray!6}{2.30} & \cellcolor{gray!6}{2.20} & \cellcolor{gray!6}{2.31} & \cellcolor{gray!6}{2.30} & \cellcolor{gray!6}{2.31} & \cellcolor{gray!6}{2.31} & \cellcolor{gray!6}{2.31} & \cellcolor{gray!6}{2.29}\\
M2 & 1.01 & 1.04 & 1.09 & 1.07 & 1.09 & 1.10 & 1.10 & 1.10 & 1.10 & 1.09\\
\cellcolor{gray!6}{M3} & \cellcolor{gray!6}{1.23} & \cellcolor{gray!6}{1.33} & \cellcolor{gray!6}{1.45} & \cellcolor{gray!6}{1.37} & \cellcolor{gray!6}{1.45} & \cellcolor{gray!6}{1.44} & \cellcolor{gray!6}{1.46} & \cellcolor{gray!6}{1.43} & \cellcolor{gray!6}{1.44} & \cellcolor{gray!6}{1.44}\\
M4 & 21.48 & 21.21 & 21.39 & 20.21 & 21.44 & 21.36 & 21.41 & 21.30 & 21.32 & 21.31\\
\cellcolor{gray!6}{M5} & \cellcolor{gray!6}{16.43} & \cellcolor{gray!6}{16.44} & \cellcolor{gray!6}{15.99} & \cellcolor{gray!6}{16.40} & \cellcolor{gray!6}{16.35} & \cellcolor{gray!6}{15.97} & \cellcolor{gray!6}{16.41} & \cellcolor{gray!6}{10.00} & \cellcolor{gray!6}{16.39} & \cellcolor{gray!6}{16.17}\\
M6 & 11.15 & 11.04 & 11.13 & 10.85 & 11.05 & 11.01 & 11.06 & 11.15 & 11.12 & 11.10\\
\cellcolor{gray!6}{M7} & \cellcolor{gray!6}{25.84} & \cellcolor{gray!6}{25.34} & \cellcolor{gray!6}{25.32} & \cellcolor{gray!6}{25.10} & \cellcolor{gray!6}{25.31} & \cellcolor{gray!6}{25.00} & \cellcolor{gray!6}{25.44} & \cellcolor{gray!6}{25.12} & \cellcolor{gray!6}{25.31} & \cellcolor{gray!6}{25.37}\\
M8 & 11.81 & 11.51 & 11.65 & 11.74 & 11.69 & 11.53 & 11.39 & 11.60 & 11.60 & 11.65\\
\cellcolor{gray!6}{M9} & \cellcolor{gray!6}{2.71} & \cellcolor{gray!6}{3.90} & \cellcolor{gray!6}{4.10} & \cellcolor{gray!6}{3.50} & \cellcolor{gray!6}{4.15} & \cellcolor{gray!6}{3.95} & \cellcolor{gray!6}{4.11} & \cellcolor{gray!6}{4.12} & \cellcolor{gray!6}{4.10} & \cellcolor{gray!6}{4.13}\\
\bottomrule
\end{tabular}}
\end{table}

Table \ref{tbl:bench-prices} presents the obtained option prices across the 9 OSP instances M1-M9 and the solvers S1-S10, see full description in Appendix A. I emphasize that the wide range of tuning parameters of each scheme makes comparisons fraught. For example, just for the linear model emulator, there have been dozens of proposals on how to pick the basis functions and the benchmarked S1-LM is not claimed to be ideal in any way. Moreover, for a more fair assessment, all solvers have some heuristics for picking their parameters across the 9 instances, for example the scaling of number of paths in terms of the problem dimension, the scaling of the number of trees for the RF solver, etc. These scalings are rule-based and thus inherently sub-optimal for any particular OSP instance.

Therefore, rather than providing authoritative statements about relative performance, I make a few general observations. First, no single scheme clearly dominates in terms of accuracy. This might be expected and reflects the diversity of our problem instances. In 1D (instances M1-M2), the best performing are the GP methods (S7-S9) which are arguably an overkill for such a straightforward setting, where even PDE methods would work well. S3-MARS and S5-NNet also work very well. In 2D (M3-M5) the situation is similar, except that LM-poly also becomes competitive (and runs extremely fast). In 3D (M6), S1-LM wins on accuracy, followed still by S5-NNet and S3-MARS. In 5D (M7-M9), the best performing is S3-MARS and S8-ADSA. S1-LM does great for Max-Calls in M7-M8 but is terrible for the basket Put in M9.

A few model-solver combinations perform poorly. For instance, M1 with S1-LM is more than 10\% below the best benchmark, which is very poor but not surprising since in that case we are attempting to capture the timing value function with just a cubic fit. Another poor choice is S10-hetGP for M7.

Table \ref{tbl:bench-times} lists the running times of each solver-model combination. Of note, M5 has 50 time periods and therefore takes the longest to solve. We observe an extreme range of the runtimes---from 2 to 4400 seconds. Multiple aspects come together to create this effect, from the choice of \(N\) (varies across solvers) to the regression overhead, to the number of time steps \(K\). The solvers with simpler emulators, like S1-LM and S3-MARS run fastest, while complex GP emulators can be more than 100 times slower. The top-performing S5-NNet is also among the slowest.

\begin{table}

\caption{\label{tab:solver-time}\label{tbl:bench-times}Benchmarked Algorithm Running Times (secs)}
\centering
\resizebox{\linewidth}{!}{
\begin{tabular}[t]{lrrrrrrrrrr}
\toprule
Model & S1-LM & S2-RF & S3-MARS & S4-TvR & S5-NNet & S6-BW & S7-GP & S8-ADSA & S9-Seq & S10-hetGP\\
\midrule
\cellcolor{gray!6}{M1} & \cellcolor{gray!6}{14.5} & \cellcolor{gray!6}{77.9} & \cellcolor{gray!6}{14.4} & \cellcolor{gray!6}{25.2} & \cellcolor{gray!6}{81.7} & \cellcolor{gray!6}{14.1} & \cellcolor{gray!6}{62.5} & \cellcolor{gray!6}{196.9} & \cellcolor{gray!6}{206.7} & \cellcolor{gray!6}{9.2}\\
M2 & 13.6 & 50.6 & 10.5 & 24.7 & 42.3 & 17.0 & 24.6 & 196.7 & 208.4 & 9.4\\
\cellcolor{gray!6}{M3} & \cellcolor{gray!6}{15.2} & \cellcolor{gray!6}{78.7} & \cellcolor{gray!6}{37.3} & \cellcolor{gray!6}{114.4} & \cellcolor{gray!6}{334.7} & \cellcolor{gray!6}{13.4} & \cellcolor{gray!6}{43.6} & \cellcolor{gray!6}{234.1} & \cellcolor{gray!6}{422.7} & \cellcolor{gray!6}{20.5}\\
M4 & 2.0 & 37.3 & 24.5 & 35.4 & 210.3 & 2.0 & 53.7 & 160.1 & 521.8 & 24.4\\
\cellcolor{gray!6}{M5} & \cellcolor{gray!6}{1.9} & \cellcolor{gray!6}{249.7} & \cellcolor{gray!6}{165.5} & \cellcolor{gray!6}{197.9} & \cellcolor{gray!6}{1731.6} & \cellcolor{gray!6}{8.2} & \cellcolor{gray!6}{1465.4} & \cellcolor{gray!6}{573.6} & \cellcolor{gray!6}{577.4} & \cellcolor{gray!6}{302.9}\\
M6 & 3.9 & 87.0 & 24.1 & 54.4 & 732.1 & 4.6 & 323.1 & 375.4 & 2023.8 & 120.3\\
\cellcolor{gray!6}{M7} & \cellcolor{gray!6}{4.5} & \cellcolor{gray!6}{299.5} & \cellcolor{gray!6}{51.4} & \cellcolor{gray!6}{56.9} & \cellcolor{gray!6}{2103.1} & \cellcolor{gray!6}{27.7} & \cellcolor{gray!6}{1564.8} & \cellcolor{gray!6}{1955.5} & \cellcolor{gray!6}{2838.4} & \cellcolor{gray!6}{720.1}\\
M8 & 3.5 & 114.2 & 26.4 & 69.8 & 455.3 & 26.3 & 432.7 & 1634.9 & 2011.8 & 394.6\\
\cellcolor{gray!6}{M9} & \cellcolor{gray!6}{32.9} & \cellcolor{gray!6}{348.8} & \cellcolor{gray!6}{43.3} & \cellcolor{gray!6}{97.4} & \cellcolor{gray!6}{1169.2} & \cellcolor{gray!6}{83.4} & \cellcolor{gray!6}{340.4} & \cellcolor{gray!6}{4442.3} & \cellcolor{gray!6}{3306.7} & \cellcolor{gray!6}{1229.3}\\
\bottomrule
\end{tabular}}
\end{table}

\hypertarget{deep-dive-finetuning-emulators}{%
\subsection{Deep dive: finetuning emulators}\label{deep-dive-finetuning-emulators}}

All the RMC schemes invariably come with a host of regression and simulation (hyper-)parameters that can be optimized to maximize performance. Such fine-tuning is what makes the methods flexible and powerful, but also more of an art than a science (and therefore benchmarking invaluable). In this section I present two case studies that explore the role of tuning parameters for the M6 instance which features the 3D symmetric GBM max-Call.

\begin{figure}

{\centering \includegraphics[trim={0 0.5cm 0 0},clip]{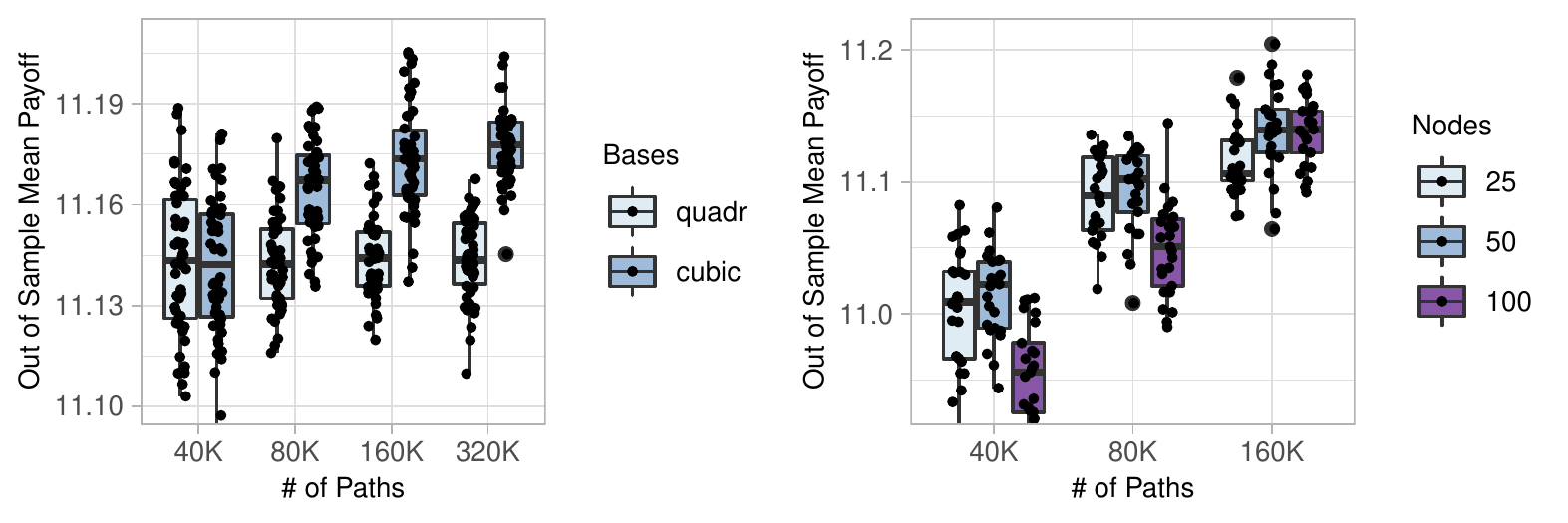}

}

\caption{\label{fig:l-boxplot}Left: Boxplots showing the joint impact of the number of basis functions $R$ and the number of paths $N$ for a LM-Poly emulator. Right: boxplots showing the joint impact of the number of hidden units and the number of paths $N$ for a NNet emulator. Both panels consider the M6 3D Max-Call OSP instance and visualize $\check{V}(0,X(0))$ across 50 algorithm runs.}\label{fig:lm-3dcall-table}
\end{figure}

Starting with the LM-Poly emulator, one has the choice of what degree polynomials to pick. Specifically, I explore the joint selection of either quadratic (\(R=10\) total bases) or cubic (\(R=20\)) basis functions, and the number of simulations \(N\). The left panel of Figure \ref{fig:l-boxplot} displays the resulting boxplots of out-of-sample estimates \(\check{V}(0,X(0))\) using different \texttt{lm} settings for the above LM-Poly emulator; I run 50 macro-replications for each case. According to the results, for low \(N\), there is no gain from having more basis functions, highlighting likely overfitting by the degree-3 polynomials.

The right panel of Figure \ref{fig:l-boxplot} shows a similar fine-tuning for a NNet emulator where I vary the number of hidden units (\texttt{nodes}) between 25, 50 and 100 and simultaneously change \(N\) again. I find that the NNet emulator tends to significantly overfit for a low number of paths (confirmed by a very large gap between in-sample and out-of-sample estimators, not displayed in the plot) and that there is no gain from increasing the number of hidden units beyond 50.

These two examples illustrate that there is infinite scope for further tuning \texttt{\{mlOSP\}} solvers and the template can be readily used as a basis for various in-depth studies of specific schemes. Implementing a detailed search for \emph{optimal} tuning parameters in the spirit of the \texttt{\{caret\}} package is left for future research.

\hypertarget{high-dimensional-optimal-stopping}{%
\subsection{High-dimensional optimal stopping}\label{high-dimensional-optimal-stopping}}

Recently, several articles \citet{CheriditoJentzen18}, \citet{becker2019solving}, \citet{chen2020deep}, \citet{GoudenegeZanette19}, \citet{herreraTeichmann}, \citet{reppen2022neural} have explored ultra high dimensional (UHD) OSP instances, with as many as \(d=500\) dimensions. While \texttt{mlOSP} can handle such case studies in principle, it is not currently geared for such use and the benchmarks above are limited to instances in dimension \(d \le 5\). First, handling high-dim.~OSP requires different RMC architectures, for example advanced deep learning approaches; the solvers listed in Appendix A are generally inappropriate for \(d \gg 5\). Second, UHD schemes tend to rely on Python-based frameworks like \texttt{TensorFlow} which are less convenient to access via \texttt{R} and are not implemented in the present version of \texttt{\{mlOSP\}}. Third, UHD problem instances require a lot of fine-tuning (e.g.~through exploiting specific problem structures and through experimenting with solver architectures) and therefore do not fit well into the envisioned automated nature of \textbf{mlOSP}. It remains an important open problem to extend the benchmarks to UHD case studies.

To demonstrate the issues that arise, consider a 10-dim example of an Asian Put with payoff
\[ h_{AsnPut}(x_{(k-10):k},{\cal K}) = \left({\cal K} - \frac{1}{10} \sum_{\ell =0}^{9} x_{k-\ell} \right)_+.\]

In this setting, most of the generic solvers from Section 6.1 are inapplicable. The LM-Poly solver either requires defining many dozens of basis functions (there are 55 polynomials of degree \(\le 2\) with \(d=10\)) or building a highly customized set of regressors. The BW solver cannot generate equi-partitions in \(d=10\) unless there are millions of training paths or just 2-3 partitions per coordinate. GP solvers will take hours to run unless one restricts to a couple thousand of unique training sites which underwhelm in emulator accuracy. This timing concern will be even more acute with the sequential design methods, unless one, again, fine-tunes very carefully.

Among the regression modules that can handle this problem off-the-shelf are the RF and NNet emulators. I run both with \(N=10^5\) paths and essentially default settings.

\begin{Shaded}
\begin{Highlighting}[]
\NormalTok{model10D }\OtherTok{\textless{}{-}} \FunctionTok{list}\NormalTok{(}\AttributeTok{dim=}\DecValTok{10}\NormalTok{, }
                \AttributeTok{sim.func=}\NormalTok{sim.gbm.moving.ave, }\CommentTok{\# Asian moving{-}average Call}
                \AttributeTok{r=}\FloatTok{0.05}\NormalTok{, }\AttributeTok{div=}\DecValTok{0}\NormalTok{, }\AttributeTok{sigma=}\FloatTok{0.2}\NormalTok{, }
                \AttributeTok{x0=}\FunctionTok{c}\NormalTok{(}\DecValTok{100}\NormalTok{,}\DecValTok{0}\NormalTok{,}\DecValTok{0}\NormalTok{,}\DecValTok{0}\NormalTok{,}\DecValTok{0}\NormalTok{,}\DecValTok{0}\NormalTok{,}\DecValTok{0}\NormalTok{,}\DecValTok{0}\NormalTok{,}\DecValTok{0}\NormalTok{,}\DecValTok{0}\NormalTok{),  }\CommentTok{\# start At{-}the{-}money}
                \AttributeTok{payoff.func=}\NormalTok{call.payoff, }
                \AttributeTok{K=}\DecValTok{100}\NormalTok{, }\AttributeTok{T=}\DecValTok{1}\NormalTok{, }\AttributeTok{dt=}\FloatTok{0.02}\NormalTok{) }\CommentTok{\# 50 steps}

\FunctionTok{set.seed}\NormalTok{(}\DecValTok{10}\NormalTok{) }
\NormalTok{test}\FloatTok{.10}\NormalTok{D }\OtherTok{\textless{}{-}} \FunctionTok{list}\NormalTok{()  }\CommentTok{\# Build a test set of 40,000 paths}
\NormalTok{init.cond }\OtherTok{\textless{}{-}} \FunctionTok{matrix}\NormalTok{(}\FunctionTok{rep}\NormalTok{(model10D}\SpecialCharTok{$}\NormalTok{x0, }\DecValTok{40000}\NormalTok{), }\AttributeTok{nrow=}\DecValTok{40000}\NormalTok{, }\AttributeTok{byrow=}\NormalTok{T)}
\NormalTok{test}\FloatTok{.10}\NormalTok{D[[}\DecValTok{1}\NormalTok{]] }\OtherTok{\textless{}{-}}\NormalTok{ model10D}\SpecialCharTok{$}\FunctionTok{sim.func}\NormalTok{( init.cond, model10D)}
\ControlFlowTok{for}\NormalTok{ (i }\ControlFlowTok{in} \DecValTok{2}\SpecialCharTok{:}\DecValTok{50}\NormalTok{)}
\NormalTok{    test}\FloatTok{.10}\NormalTok{D[[i]] }\OtherTok{\textless{}{-}}\NormalTok{ model10D}\SpecialCharTok{$}\FunctionTok{sim.func}\NormalTok{( test}\FloatTok{.10}\NormalTok{D[[i}\DecValTok{{-}1}\NormalTok{]], model10D)}
\FunctionTok{cat}\NormalTok{(}\FunctionTok{c}\NormalTok{(}\StringTok{"European Call price: "}\NormalTok{, }\FunctionTok{round}\NormalTok{(}\FunctionTok{mean}\NormalTok{( }\FunctionTok{exp}\NormalTok{(}\SpecialCharTok{{-}}\NormalTok{model10D}\SpecialCharTok{$}\NormalTok{r}\SpecialCharTok{*}\NormalTok{model10D}\SpecialCharTok{$}\NormalTok{T)}\SpecialCharTok{*}
\NormalTok{                                       model10D}\SpecialCharTok{$}\FunctionTok{payoff.func}\NormalTok{(test}\FloatTok{.10}\NormalTok{D[[}\DecValTok{50}\NormalTok{]],model10D)),}\DecValTok{3}\NormalTok{) ))}
\end{Highlighting}
\end{Shaded}

\begin{verbatim}
## European Call price:  9.755
\end{verbatim}

\begin{Shaded}
\begin{Highlighting}[]
\NormalTok{model10D}\SpecialCharTok{$}\NormalTok{nn.nodes }\OtherTok{=} \DecValTok{50}  \CommentTok{\# Neural Net solver parameters}
\NormalTok{nnSolve }\OtherTok{\textless{}{-}} \FunctionTok{osp.prob.design}\NormalTok{(}\AttributeTok{N=}\DecValTok{100000}\NormalTok{,model10D, }\AttributeTok{method=}\StringTok{"nnet"}\NormalTok{)}
\NormalTok{oos.nn }\OtherTok{\textless{}{-}} \FunctionTok{forward.sim.policy}\NormalTok{( test}\FloatTok{.10}\NormalTok{D, }\DecValTok{50}\NormalTok{, nnSolve}\SpecialCharTok{$}\NormalTok{fit, model10D)}
\FunctionTok{units}\NormalTok{(nnSolve}\SpecialCharTok{$}\NormalTok{timeElapsed) }\OtherTok{\textless{}{-}} \StringTok{"secs"}
\FunctionTok{cat}\NormalTok{(}\FunctionTok{c}\NormalTok{(}\StringTok{"NNet solver:"}\NormalTok{, }\FunctionTok{round}\NormalTok{(}\FunctionTok{mean}\NormalTok{(oos.nn}\SpecialCharTok{$}\NormalTok{payoff),}\DecValTok{3}\NormalTok{), }\FunctionTok{round}\NormalTok{(nnSolve}\SpecialCharTok{$}\NormalTok{time,}\DecValTok{1}\NormalTok{), }\StringTok{"secs"}\NormalTok{))}
\end{Highlighting}
\end{Shaded}

\begin{verbatim}
## NNet solver: 11.33 1159.7 secs
\end{verbatim}

\begin{Shaded}
\begin{Highlighting}[]
\NormalTok{model10D}\SpecialCharTok{$}\NormalTok{rf.ntree }\OtherTok{=} \DecValTok{100}\NormalTok{; model10D}\SpecialCharTok{$}\NormalTok{rf.maxnode}\OtherTok{=} \DecValTok{200} \CommentTok{\# RF solver parameters}
\NormalTok{rfSolve }\OtherTok{\textless{}{-}} \FunctionTok{osp.prob.design}\NormalTok{(}\AttributeTok{N=}\DecValTok{100000}\NormalTok{,model10D, }\AttributeTok{method=}\StringTok{"randomforest"}\NormalTok{)}
\NormalTok{oos.rf }\OtherTok{\textless{}{-}} \FunctionTok{forward.sim.policy}\NormalTok{( test}\FloatTok{.10}\NormalTok{D, }\DecValTok{50}\NormalTok{, rfSolve}\SpecialCharTok{$}\NormalTok{fit, model10D)}
\FunctionTok{units}\NormalTok{(rfSolve}\SpecialCharTok{$}\NormalTok{timeElapsed) }\OtherTok{\textless{}{-}} \StringTok{"secs"}
\FunctionTok{cat}\NormalTok{(}\FunctionTok{c}\NormalTok{(}\StringTok{"Random Forest solver:"}\NormalTok{, }\FunctionTok{round}\NormalTok{(}\FunctionTok{mean}\NormalTok{(oos.rf}\SpecialCharTok{$}\NormalTok{payoff),}\DecValTok{3}\NormalTok{), }\FunctionTok{round}\NormalTok{(rfSolve}\SpecialCharTok{$}\NormalTok{time,}\DecValTok{1}\NormalTok{), }\StringTok{"secs"}\NormalTok{))}
\end{Highlighting}
\end{Shaded}

\begin{verbatim}
## Random Forest solver: 11.24 419.6 secs
\end{verbatim}

\hypertarget{extensions}{%
\section{Extensions}\label{extensions}}

\hypertarget{valuing-swing-options}{%
\subsection{Valuing Swing Options}\label{valuing-swing-options}}

Swing options are common in commodity markets, such as natural gas or electricity. They are equivalent to a multiple-stopping problem or a bundle of regular Bermudan options. For example, in a Swing Put contract, the buyer has the right to exercise up to \(I\) Puts (e.g.~\(I=3\)) prior to the swing maturity \(T\). This means she has 3 distinct rights to sell the asset for the strike price, and can dynamically choose to use them one-by-one. The holder can exercise none or only some of her swings. Simultaneous exercising are not allowed: between consecutive exercises there is a postulated minimum interval of time that the holder must wait, called the \emph{refraction period} \(\Delta\).

To figure out the exercise strategy and ultimately the price of the aforementioned Swing Put, we need to keep track of the sequence of exercise times. Denote by \(\tau^1\le \tau^2\le \ldots\) the ordered exercise times and by \(V^{(i)}(t,x)\) the value of a swing Put with \(i\) \emph{remaining} exercise rights on \([t,T]\). \(V^{(1)}\) is simply the value of a standard Bermudan Put option with the given strike \({\cal K}\) and maturity \(T\), i.e.~payoff function \(\tilde{h}^{(1)}(t,x) = h(t,x) = e^{-r t}({\cal K}-x)_+\). Next, considering the problem embedded in \(V^{(2)}(t,X(t))\), the holder has two swing rights remaining and at an instant \(t\) has two choices:

-- Continue without exercising, keeping both rights;

-- Exercise one right, i.e.~collect \(({\cal K}-X(t))_+\); she will then have \(2-1=1\) rights remaining and cannot do anything until \(t+\Delta\). Since \(V^{(1)}\) captures the value of the last swing right, her overall payoff would be
\[
\tilde{h}^{(2)}(t,X(t)) := e^{-r t}({\cal K} -X(t))_+ + \E \left[ V^{(1)}(t+\Delta,X(t+\Delta)) \Big|\; X(t)\right],
\]
where the first term is the payoff from exercising the Put, and the second term is the value of the remaining exercise right once the refraction period passes. We obtain a coupling where the payoff for \(V^{(2)}\) depends on \(V^{(1)}\), or more generally the payoff for \(V^{(i)}\) depends on \(V^{(i-1)}\).

More formally to price a Swing option with payoffs \(h(k,x)\) we must solve the multiple-stopping problem (reverting back to discrete-time and setting \(k_\Delta := \Delta/ \Delta t\)):
\[
V^{(I)}(k,x) := \sup_{k \le \tau^1 < \tau^2 < \ldots < \tau^I \le K} \E \left[ \sum_{i=1}^I h(\tau^i, X({\tau^i})) \cdot 1_{\{\tau^i < K\}} + h(K,X(K)) \cdot 1_{\{\tau^I = K\}} \Big|\; X(k) = x \right],
\]
where each \(\tau^i\) is a stopping time. It is possible that multiple exercise rights go unused (if the options remain out-of-the-money) in which case we set \(\tau^i = \tau^{i+1} = \ldots = K\) for several \(\tau\)'s. The corresponding dynamic programming equation for \(i=1,\ldots\) is
\begin{align}
q^{(i)}(k,x) & = \E \left[ V^{(i)}(k+1,X({k+1})) \, \Big|\; X(k)=x \right]\\
V^{(i)}(k,x) & = \max \left( h(k,x) + \E \left[V^{(i-1)}(k+k_\Delta,X({k+k_\Delta})) \, \Big|\; X(k)=x \right], q^{(i)}(k,x)  \right), \label{eq:swing-dp}
\end{align}
with \(V^{(0)} \equiv 0\), and \(V^{(i)}(K,x)=h(K,x) \ \forall i \ge 1\). The first term in \eqref{eq:swing-dp} is the expected payoff from exercising followed by a refraction period of \(k_\Delta\) time-steps, and the second term is the \(q\)-value: expected payoff from waiting for one more time-step. Optimal swing times with a total of \(I\) rights are recursively defined by \(\tau^1 = \min \{ k : h(k,X(k)) > q^{(I)}(k,X({k})) \} \wedge K\) and for \(i=2,\ldots, I\),
\begin{align}
\tau^{i} = \min \left\{ k > \tau^{i-1}+ k_\Delta : \; h(k,X(k)) +  \E \left[ V^{(I-i)}(k+k_\Delta,X({k+k_\Delta}))\Big|\;X(k) \right] >  q^{(I-i)}(k,X({k})) \right\} \wedge K.
\end{align}
Similarly I define the timing value with \(i\) exercise rights
\(T^{(i)}(k,x)\) which is the difference between the two terms in \eqref{eq:swing-dp}, and \({\cal S}^{(i)}_k = \left\{ x : {T}^{(i)}(k,x) \le 0 \right\}\) the stopping region when \(i\) rights remain.

Based on the above discussion, applying RMC to price swing options reduces to replacing the Bermudan payoff \(h(k,x)\) with a functional that links \(V^{(i)}(k,x)\) to \(V^{(i-1)}(k+k_\Delta,\cdot)\). This implies that solving for \(V^{(i)}\) can be done \textbf{iteratively}, by starting with \(V^{(1)}\) (which is a regular Bermudan option) and then recursively solving for \(V^{(2)}, V^{(3)}, \ldots\).

The \textbf{mlOSP} Algorithm \ref{alg:1} can now be directly applied by
stacking all the regression objects \(\widehat{T}^{(i)}(k,\cdot)\) in terms of the number of rights \(i\) and re-defining the payoff structure.
Note that although we need to compute multiple \(\widehat{T}^{(i)}\)'s, we can do so \emph{in parallel} as part of the backward time-stepping over \(k\). Indeed, one can either proceed ``layer-by-layer'' (i.e.~loop over \(k\), then loop over \(i\)) or ``step-by-step'' (loop over \(i\), then loop over \(k\)), since fitting \(\widehat{T}^{(i)}(k,x)\) requires only knowing \(\widehat{T}^{(i-1)}({k+k_\Delta},\cdot)\). In \texttt{\{mlOSP\}}, I implement the first variant with explicitly specified space-filling training designs in the \texttt{swing.fixed.design} solver. The solver relies on the \texttt{swing.policy} function as the simulation device; the latter is called twice for each input \(x^n(k)\), in order to compute a pathwise reward in the case of immediate exercise of one right at \(k\), and to evaluate the expected payoff from no exercise and then acting optimally for steps \(k+1,\ldots, K\).

To illustrate the above, I present a short case study based on \citet{CarmonaTouzi} who considered a 1D Swing Put with GBM dynamics, \(T=1, X(0) = 100, {\cal K}=100, r=0.05, \sigma=0.3\) with \(\Delta t =0.02\) or 50 time steps and refraction period of \(\Delta = 0.1\) or \(k_\Delta=5\) time steps. They report results of \(V^{(i)}(0,100) = 9.85, 19.26, 28.802\) for \(i=1,2,3\) exercise rights. We see that \(V^{(2)} \le 2 V^{(1)}\) due to dis-economy of scale: the second swing is less valuable than the first one due to the inherent refraction between exercises. Moreover, the second swing (first chronologically) will be exercised sooner, the intuition being that as there are fewer and fewer swings left, the holder gets more and more protective about using them.
To compute the price of a \(I=3\)-right Swing Put with the above parameters, the \texttt{swing.fixed.design} solver is very similar to \texttt{osp.fixed.design}, again requiring to specify a \texttt{model} list with all the parameters, the regression emulation method, and the training design specification. Two new model parameters are \texttt{n.swing} (number of swing rights) and \texttt{refract} (refraction period \(\Delta\)). I select a smoothing spline emulator with automatically cross-validated degrees of freedom (\texttt{method=cvspline}).
For the simulation design I space-fill using option (iii) from Section 4.1.

\begin{Shaded}
\begin{Highlighting}[]
\FunctionTok{set.seed}\NormalTok{(}\DecValTok{10}\NormalTok{)}
\NormalTok{swingModel }\OtherTok{\textless{}{-}} \FunctionTok{list}\NormalTok{(}\AttributeTok{dim=}\DecValTok{1}\NormalTok{, }\AttributeTok{sim.func=}\NormalTok{sim.gbm, }\AttributeTok{x0=}\DecValTok{100}\NormalTok{,}
            \AttributeTok{swing.payoff=}\NormalTok{put.payoff, }\AttributeTok{n.swing=}\DecValTok{3}\NormalTok{,}\AttributeTok{K=}\DecValTok{100}\NormalTok{, }
            \AttributeTok{sigma=}\FloatTok{0.3}\NormalTok{, }\AttributeTok{r=}\FloatTok{0.05}\NormalTok{, }\AttributeTok{div=}\DecValTok{0}\NormalTok{,}
            \AttributeTok{T=}\DecValTok{1}\NormalTok{,}\AttributeTok{dt=}\FloatTok{0.02}\NormalTok{,}\AttributeTok{refract=}\FloatTok{0.1}\NormalTok{,}
            \AttributeTok{N=}\DecValTok{2500}\NormalTok{,}\AttributeTok{pilot.nsims=}\DecValTok{1000}\NormalTok{,}\AttributeTok{batch.nrep=}\DecValTok{10}\NormalTok{)}
\NormalTok{spl.swing }\OtherTok{\textless{}{-}} \FunctionTok{swing.fixed.design}\NormalTok{(swingModel,}\AttributeTok{input.domain=}\FloatTok{0.02}\NormalTok{, }\AttributeTok{method =}\StringTok{"cvspline"}\NormalTok{)}
\end{Highlighting}
\end{Shaded}

As usual, evaluating the expected payoff \(\check{V}^{(3)}(0,X(0))\) using the fitted \texttt{spl.swing} approximators \(\widehat{T}^{(i)}(\cdot,\cdot)\) is done by constructing a test set of out-of-sample forward paths.

\begin{Shaded}
\begin{Highlighting}[]
\FunctionTok{set.seed}\NormalTok{(}\DecValTok{10}\NormalTok{); test.swing }\OtherTok{\textless{}{-}} \FunctionTok{list}\NormalTok{()  }\CommentTok{\# 20000 forward scenarios}
\NormalTok{test.swing[[}\DecValTok{1}\NormalTok{]] }\OtherTok{\textless{}{-}} \FunctionTok{sim.gbm}\NormalTok{( }\FunctionTok{matrix}\NormalTok{(}\FunctionTok{rep}\NormalTok{(swingModel}\SpecialCharTok{$}\NormalTok{x0, }\DecValTok{20000}\NormalTok{),}\AttributeTok{nrow=}\DecValTok{20000}\NormalTok{), swingModel)}
\ControlFlowTok{for}\NormalTok{ (i }\ControlFlowTok{in} \DecValTok{2}\SpecialCharTok{:}\DecValTok{50}\NormalTok{)}
\NormalTok{  test.swing[[i]] }\OtherTok{\textless{}{-}} \FunctionTok{sim.gbm}\NormalTok{( test.swing[[i}\DecValTok{{-}1}\NormalTok{]], swingModel)}

\NormalTok{oos.spl3 }\OtherTok{\textless{}{-}} \FunctionTok{swing.policy}\NormalTok{(test.swing,}\DecValTok{50}\NormalTok{,spl.swing}\SpecialCharTok{$}\NormalTok{fit,swingModel,}\AttributeTok{n.swing=}\DecValTok{3}\NormalTok{)}
\FunctionTok{cat}\NormalTok{( }\FunctionTok{round}\NormalTok{(}\FunctionTok{mean}\NormalTok{(oos.spl3}\SpecialCharTok{$}\NormalTok{totPayoff),}\DecValTok{3}\NormalTok{))}
\end{Highlighting}
\end{Shaded}

\begin{verbatim}
## 27.813
\end{verbatim}

Note that the forward payoff evaluator \texttt{swing.policy} takes as input the number \(I\) of swing rights to start, and one
can similarly obtain, using the same functional approximators, that the value of having \texttt{n.swing=2} rights is 19.234 and of having a single right \texttt{n.swing=1} (which is equivalent to the standard Bermudan Put option with same parameters) is 9.912. Fine-tuning \textbf{swing.fixed.design} across various simulation design and regression modules is left for future work.

\emph{Remark:} In the very recent preprint \citet{ludkovski2022impulse}, I additionally extend \texttt{mlOSP} to stochastic impulse control problems.

\hypertarget{building-a-new-model}{%
\subsection{Building a New Model}\label{building-a-new-model}}

The \textbf{mlOSP} template is highly extensible and \texttt{\{mlOSP\}} easily accommodates the construction of new OSP instances, allowing straightforward addition of new benchmarks.
As an example of how the user can do so, I show the step-by-step process of implementing a new example based on the article by \citet{CheriditoJentzen18}, henceforth BCJ.

Specifically, BCJ consider a multivariate GBM model with asymmetric volatilities and constant correlation. The payoff functional \(h_{\text{maxCall}}\) is of the max-Call type already described. The dynamics of asset \(S_i\) are
\[ \label{eq:multi-gbm}
S_i(t) = s_i(0) \exp \left( [r-\delta_i-\frac{\sigma_i^2}{2}]t + \sigma_i W^i(t) \right), \qquad i=1,\ldots, d,
\]
where the instantaneous correlation between the Brownian motions \(W^i\) and \(W^j\) is \(\rho_{ij}\). In the example of \citep[Table 2, p16]{CheriditoJentzen18}, \(d=5, s_i(0) = X_i(0), \delta_i = \delta, \rho_{ij} = \rho\).

To import the above into \texttt{\{mlOSP\}} I first define a new simulation function for the multivariate non-i.i.d.~correlated Geometric Brownian motion using the \texttt{rmvnorm} function in the \texttt{\{mvtnorm\}} package. To avoid passing too many parameters, I introduce a new \texttt{model} field \texttt{rho} (taken to be a constant) and re-define the volatility \texttt{sigma} to be a vector of length \texttt{model\$dim}.

\begin{Shaded}
\begin{Highlighting}[]
\NormalTok{sim.corGBM }\OtherTok{\textless{}{-}} \ControlFlowTok{function}\NormalTok{( x0, model, dt)}
\NormalTok{\{   }\CommentTok{\# build a matrix of rho*sigma\_i*sigma\_j, plus correct the diagonal to be sigma\_i\^{}2}
\NormalTok{    sigm }\OtherTok{\textless{}{-}}\NormalTok{ model}\SpecialCharTok{$}\NormalTok{rho}\SpecialCharTok{*}\FunctionTok{kronecker}\NormalTok{(model}\SpecialCharTok{$}\NormalTok{sigma, }\FunctionTok{t}\NormalTok{(model}\SpecialCharTok{$}\NormalTok{sigma)) }\SpecialCharTok{+}
\NormalTok{            (}\DecValTok{1}\SpecialCharTok{{-}}\NormalTok{model}\SpecialCharTok{$}\NormalTok{rho)}\SpecialCharTok{*}\FunctionTok{diag}\NormalTok{(model}\SpecialCharTok{$}\NormalTok{sigma}\SpecialCharTok{\^{}}\DecValTok{2}\NormalTok{)  }

\NormalTok{    newX }\OtherTok{\textless{}{-}}\NormalTok{ x0}\SpecialCharTok{*}\FunctionTok{exp}\NormalTok{( }\FunctionTok{rmvnorm}\NormalTok{(}\FunctionTok{nrow}\NormalTok{(x0), }\AttributeTok{sig=}\NormalTok{sigm}\SpecialCharTok{*}\NormalTok{dt, }
                            \AttributeTok{mean=}\NormalTok{ (model}\SpecialCharTok{$}\NormalTok{r}\SpecialCharTok{{-}}\NormalTok{ model}\SpecialCharTok{$}\NormalTok{div}\SpecialCharTok{{-}}\NormalTok{ model}\SpecialCharTok{$}\NormalTok{sigma}\SpecialCharTok{\^{}}\DecValTok{2}\SpecialCharTok{/}\DecValTok{2}\NormalTok{)}\SpecialCharTok{*}\NormalTok{dt) )}
    \FunctionTok{return}\NormalTok{ (newX)}
\NormalTok{\}}
\end{Highlighting}
\end{Shaded}

In the referenced setting, the parameter values are \(\sigma_i = 0.08 i\), as well as \(\delta = 10\%, r = 5\%, \rho = 0\) and contract specification \(S_i(0) = 90 \forall i, T=3, {\cal K}=100, \Delta t=1/3\) (\(K=9\) time steps). I set up all the above and pass in the newly defined \texttt{sim.corGBM} as the simulation function:

\begin{Shaded}
\begin{Highlighting}[]
\NormalTok{modelBCJ }\OtherTok{\textless{}{-}} \FunctionTok{list}\NormalTok{(}\AttributeTok{dim=}\DecValTok{5}\NormalTok{,}\AttributeTok{sigma=}\FloatTok{0.08}\SpecialCharTok{*}\NormalTok{(}\DecValTok{1}\SpecialCharTok{:}\DecValTok{5}\NormalTok{), }\AttributeTok{r=} \FloatTok{0.05}\NormalTok{, }\AttributeTok{div=}\FloatTok{0.1}\NormalTok{, }\AttributeTok{rho=}\DecValTok{0}\NormalTok{, }
                    \AttributeTok{x0 =} \FunctionTok{rep}\NormalTok{(}\DecValTok{90}\NormalTok{,}\DecValTok{5}\NormalTok{), }\AttributeTok{T=}\DecValTok{3}\NormalTok{, }\AttributeTok{K=}\DecValTok{100}\NormalTok{, }\AttributeTok{dt=}\DecValTok{1}\SpecialCharTok{/}\DecValTok{3}\NormalTok{, }
                    \AttributeTok{sim.func=}\NormalTok{sim.corGBM, }\AttributeTok{payoff.func=}\NormalTok{maxi.call.payoff)}
\end{Highlighting}
\end{Shaded}

The package solver is now ready to be utilized. I select a \texttt{hetGP} emulator with Gaussian kernel and a design of size \(N_{unique}=640\), with \(N_{rep}=25\) replicates per site or \(N=16,000\), which requires a few more parameter specifications:

\begin{Shaded}
\begin{Highlighting}[]
\NormalTok{hetGP.params }\OtherTok{\textless{}{-}} \FunctionTok{list}\NormalTok{(}\AttributeTok{max.lengthscale=}\FunctionTok{rep}\NormalTok{(}\DecValTok{20000}\NormalTok{,}\DecValTok{5}\NormalTok{),}\AttributeTok{batch.nrep=}\DecValTok{25}\NormalTok{,}
                     \AttributeTok{kernel.family=}\StringTok{"Gaussian"}\NormalTok{,}\AttributeTok{pilot.nsims=}\DecValTok{10000}\NormalTok{,}\AttributeTok{N=}\DecValTok{640}\NormalTok{)}
\NormalTok{modelBCJ }\OtherTok{\textless{}{-}} \FunctionTok{c}\NormalTok{(modelBCJ,hetGP.params)}
\NormalTok{bcjFit }\OtherTok{\textless{}{-}} \FunctionTok{osp.fixed.design}\NormalTok{(modelBCJ,}\AttributeTok{input.dom=}\ConstantTok{NULL}\NormalTok{, }\AttributeTok{method=}\StringTok{"hetgp"}\NormalTok{)}
\end{Highlighting}
\end{Shaded}

All in all, just a few lines of code are necessary to construct such a case study; once done it is trivial to tackle it using the wide range of presented solvers and schemes.

Running on an out-of-sample test set of \(10^5\) scenarios I obtain a 95\% confidence interval for \(\check{V}(0,X(0))\) of {[}27.06, 27.459{]}. This can be compared against the reported interval of {[}27.63, 27.69{]} in \citep{CheriditoJentzen18}. The very high standard deviation of realized payoffs leads to a wide credible interval of \(\pm 0.199\) of the reported \(\check{V}(0,X(0))=\) 27.259. Thus, obtaining tighter bounds on the estimated option price would require a very large out-of-sample test set.

\hypertarget{conclusion}{%
\section{Conclusion}\label{conclusion}}

With dozens of RMC implementations having been proposed in the literature, there remains a gap in bringing everything ``under one roof'' to allow proper and reproducible comparisons. The described \texttt{\{mlOSP\}} package does exactly that via a publicly available, GitHub-hosted, free software. It offers an overview and a guide for exploring RMC versions. Indeed, writing the package opened numerous new possibilities that I have covered above:

\begin{itemize}
\item
  New regression surrogates leveraging the unparalleled range of statistical methods implemented in \texttt{R};
\item
  New mix-and-match options thanks to the modular nature of \textbf{mlOSP} around its core template;
\item
  Transparent and reproducible benchmarks that offer apples-to-apples comparison of different schemes;
\item
  Extension of the framework to multiple optimal stopping.
\end{itemize}

By design, the accompanying software library is a work in progress (and might have additional contributors in the future, via the GitHub pull request mechanism). Consequently, the scope of the implementations will continue to evolve, as will the precise behavior of the benchmarks which is relative to the respective package versions, and will change as the utilized packages (such as \texttt{\{DiceKriging\}} or \texttt{\{hetGP\}}) get updated. I look forward to feedback and requests for functionality to be added or bugs to be fixed. For example, a work in progress is to implement the new scheme of \citet{reppen2022neural} that directly searches for the optimal stopping boundary \(\mathfrak{S}_k\) instead of first learning \(\widehat{T}(k,\cdot)\).

As demonstrated throughout the article, \texttt{\{mlOSP\}} offers almost limitless opportunities for designing and fine-tuning new RMC variants. Many of the mix-and-match options available are just that---experiments to be tried and discarded, being clearly inferior to better choices. Identifying state-of-the-art requires specifying the objective (e.g.~speed as measured in seconds of runtime), the problem(s) of interest (e.g.~OSP dimension) and additional forethought on how to combine all the moving parts. This is beyond the scope of the present already-long article. Two more projects for future work are to investigate in-depth neural network solvers that have exploded in popularity recently, and to design self-tuning solvers that require as few tuning parameters as possible.

\hypertarget{appendix-a-benchmark-solvers}{%
\section*{Appendix A: Benchmark Solvers}\label{appendix-a-benchmark-solvers}}
\addcontentsline{toc}{section}{Appendix A: Benchmark Solvers}

\begin{longtable}[]{@{}
  >{\raggedright\arraybackslash}p{(\columnwidth - 6\tabcolsep) * \real{0.0667}}
  >{\raggedright\arraybackslash}p{(\columnwidth - 6\tabcolsep) * \real{0.2667}}
  >{\raggedright\arraybackslash}p{(\columnwidth - 6\tabcolsep) * \real{0.5400}}
  >{\raggedright\arraybackslash}p{(\columnwidth - 6\tabcolsep) * \real{0.1267}}@{}}
\toprule()
\begin{minipage}[b]{\linewidth}\raggedright
Solver
\end{minipage} & \begin{minipage}[b]{\linewidth}\raggedright
Regression Emulator
\end{minipage} & \begin{minipage}[b]{\linewidth}\raggedright
Solver \& Simulation Design
\end{minipage} & \begin{minipage}[b]{\linewidth}\raggedright
Package
\end{minipage} \\
\midrule()
\endhead
S1 & Linear model emulator with polynomial basis functions & \texttt{osp.prob.design} w/size \(N\) of \(40K/100K\). Up to degree 3 polynomials in 1-2 dimensions, else up to degree 2 & \texttt{lm} \\
S2 & Multivariate adaptive regression splines & \texttt{osp.prob.design} w/size \(N\) of \(40K/100K\). Degree=2: bases consist of linear/quadratic hinge functions; 100 knots & {\small\texttt{\{earth\}}} \\
S3 & Random Forest & \texttt{osp.prob.design} w/size \(N\) of \(40K/100K/200K\). 200 trees and $100/200$ nodes per tree & {\small\texttt{\{randomForest\}}} \\
S4 & Neural Network & \texttt{osp.prob.design} w/size \(N\) of \(40K/100K\). Single-layer neural net with 20/40/50 nodes & \texttt{\{nnet\}} \\
S5 & MARS  using the TvR scheme & \texttt{osp.tvr} w/size \(N\) of \(40K/100K/200K\). Degree=2 hinge functions; 100 knots & \texttt{\{earth\}} \\
S6 & Treed hierarchical equi-probable partitioning w/constant tree leaf fits & \texttt{osp.probDesign.piecewisebw} with \texttt{nBins} of 8/5/4 and \(N= 4000/100000/204800\) (5000 inputs/leaf in 1D, 625 per leaf in 2D, 800 in 3D, 200 in 5D) & Based on \citep{BouchardWarin10} \\
S7 & GP-trainkm with Matérn-5/2 kernel & \texttt{osp.fixed.design} with replicated design with \texttt{batch.nRep=100} design size of \(N_{unique}=400/800/1000\) and LHS space-filling using pilot paths and \texttt{input.dom=0.02} & {\small \texttt{\{DiceKriging\}}} \\
S8 & GP-trainkm with Matérn-5/2 kernel & \texttt{osp.seq.batch.design} with adaptive sequential batching via ADSA. The initial designs are space-filling; additional sites added using AMCU criterion up to a total of 200/250/500. & {\small\texttt{\{DiceKriging\}}} \\
S9 & GP-trainkm with Matérn-5/2 kernel & \texttt{osp.seq.design} with adaptive sequential design via SUR criterion. Initial designs are Sobol sequences. \texttt{batch.nrep}, total design size and \texttt{cand.len} vary across models. & \citep{Lu18} \\
S10 & GP-hetGP with Gaussian kernel & \texttt{osp.fixed.design} with replicated fixed design (pseudo-regression) based on Sobol sequences for ITM regions. Both \texttt{batch.nRep} and design size (100-1000) vary across models. & \texttt{\{hetGP\}} \\
\bottomrule()
\end{longtable}

\hypertarget{appendix-b-details-for-the-toy-rmc-example-of-a-1d-bermudan-put}{%
\section*{Appendix B: Details for the Toy RMC Example of a 1D Bermudan Put}\label{appendix-b-details-for-the-toy-rmc-example-of-a-1d-bermudan-put}}
\addcontentsline{toc}{section}{Appendix B: Details for the Toy RMC Example of a 1D Bermudan Put}

Figure \ref{fig:tvr-panels} illustrates the full backward progression of an RMC solver for the toy 1-D Put example with 5 exercise periods from Section \ref{sec:toy} . We see that in this case the earlier fits are problematic as we obtain that \(\hat{q}(k,x) > h(k,x)\) and so the resulting stopping rule will \emph{never} stop at \(k=2\). This is the direct result of picking a poor approximation space \(\cH_k\): the quadratic functions are unable to properly fit the true \(q(k,\cdot)\).

\begin{figure}[htb]
\centering
\includegraphics{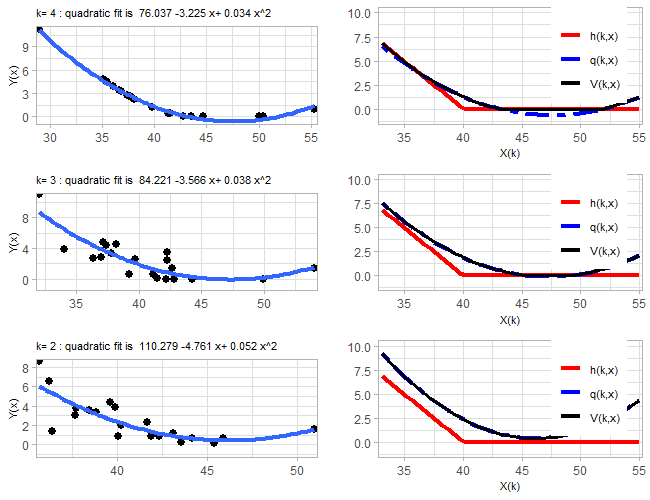}
\caption{\label{fig:ls-full-loop}\label{fig:tvr-panels}Illustration of the TvR RMC scheme with quadratic fits and 20 training inputs.}
\end{figure}

\begin{verbatim}
## [1] 2.504367
\end{verbatim}

\hypertarget{appendix-c.1-simulation-and-payoff-functions-provided-with-mlosp}{%
\section*{\texorpdfstring{Appendix C.1: Simulation and payoff functions provided with \texttt{\{mlOSP\}}}{Appendix C.1: Simulation and payoff functions provided with \{mlOSP\}}}\label{appendix-c.1-simulation-and-payoff-functions-provided-with-mlosp}}
\addcontentsline{toc}{section}{Appendix C.1: Simulation and payoff functions provided with \texttt{\{mlOSP\}}}

\begin{longtable}[]{@{}
  >{\raggedright\arraybackslash}p{(\columnwidth - 4\tabcolsep) * \real{0.2500}}
  >{\raggedright\arraybackslash}p{(\columnwidth - 4\tabcolsep) * \real{0.4583}}
  >{\raggedright\arraybackslash}p{(\columnwidth - 4\tabcolsep) * \real{0.2917}}@{}}
\toprule()
\begin{minipage}[b]{\linewidth}\raggedright
Simulator
\end{minipage} & \begin{minipage}[b]{\linewidth}\raggedright
Description
\end{minipage} & \begin{minipage}[b]{\linewidth}\raggedright
Parameters
\end{minipage} \\
\midrule()
\endhead
\texttt{sim.gbm} & Geometric Brownian Motion (GBM) & \texttt{r} (drift), \texttt{sigma} (volatility), \texttt{div} (dividend yield) \\
\texttt{sim.gbm.cor} & Correlated multi-d GBM with a single constant correlation parameter & \texttt{rho} \\
\texttt{sim.gbm.matrix} & Correlated multi-d GBM specified via a covariance matrix & \texttt{sigma} (matrix) \\
\texttt{sim.ouExp} & Geometric Ornstein Uhlenbeck process & \texttt{alpha,\ meanRev,\ sigma} \\
\texttt{sim.expOU.sv} & 1- or 2-factor Exponential O-U process with stochastic volatility (Vasicek) & \\
\texttt{sim.gbm.asian} & GBM model which also keeps track of the arithmetic and geometric averages of \(S_k\) & \texttt{r,\ sigma} \\
\texttt{sim.gbm.moving.ave} & GBM model which also keeps track of the lagged \(S_k\) to implement moving-average options & \\
\bottomrule()
\end{longtable}

\begin{longtable}[]{@{}
  >{\raggedright\arraybackslash}p{(\columnwidth - 4\tabcolsep) * \real{0.2059}}
  >{\raggedright\arraybackslash}p{(\columnwidth - 4\tabcolsep) * \real{0.5000}}
  >{\raggedright\arraybackslash}p{(\columnwidth - 4\tabcolsep) * \real{0.2941}}@{}}
\toprule()
\begin{minipage}[b]{\linewidth}\raggedright
Payoff
\end{minipage} & \begin{minipage}[b]{\linewidth}\raggedright
Description
\end{minipage} & \begin{minipage}[b]{\linewidth}\raggedright
Formula (\({\cal K}\) needed for all)
\end{minipage} \\
\midrule()
\endhead
\texttt{put.payoff} & (basket) Put on the arithmetic average & \(({\cal K}- \mean_i X_i)_+\) \\
\texttt{call.payoff} & (basket) Call on the arithmetic average & \((\mean_i X_i - {\cal K})_+\) \\
\texttt{digital.put.payoff} & Digital Put on the geometric average & \(1_{ \prod X_i^{1/d} < {\cal K}}\) \\
\texttt{geom.put.payoff} & (basket) Put on the geometric average & \(({\cal K} - \prod X_i^{1/d})_+\) \\
\texttt{mini.put.payoff} & Put on the minimum asset & \(({\cal K} - \min_i X_i)_+\) \\
\texttt{maxi.call.payoff} & Call on the maximum asset & \((\max_i X_i - {\cal K})_+\) \\
\texttt{sv.put.payoff} & Put for a stochastic volatility model (asset assumed to be the first coordinate) & \(({\cal K}-X_1)_+\) \\
\bottomrule()
\end{longtable}

\hypertarget{appendix-c.2-mlosp-solvers}{%
\section*{\texorpdfstring{Appendix C.2: \texttt{\{mlOSP\}} Solvers}{Appendix C.2: \{mlOSP\} Solvers}}\label{appendix-c.2-mlosp-solvers}}
\addcontentsline{toc}{section}{Appendix C.2: \texttt{\{mlOSP\}} Solvers}

\begin{itemize}
\item
  \texttt{osp.prob.design} -- the original Longstaff-Schwartz scheme \citep{LS}. Thus, the simulation design is randomized with the training density being the density of \(X(k)\) and is constructed from forward paths \(x^{1:N}\). The latter are re-used as the forward paths \(X^{(k),n}(k:k+w)\). No replication is used. The solver supports a variety of regression methods.
\item
  \texttt{osp.tvr} -- the Tsitsiklis-van Roy scheme \citep{TsitsiklisVanRoy}. This is equivalent to \texttt{osp.prob.design} except that the regression is applied to the step-ahead value function, rather than pathwise payoffs.
\item
  \texttt{osp.fixed.design} -- mlOSP with a user-specified simulation design. The design can be replicated using the \texttt{batch.nrep} parameter. The typical usage is to create a space-filling design on a given hyper-rectangular domain of approximation \(\tilde{\cal X}\).
  \texttt{osp.fixed.design} generates \emph{fresh} forward paths \(X^{(k),n}(k:k+w)\) at each time-step.
\item
  \texttt{osp.probDesign.piecewisebw} -- the Bouchard-Warin \citep{BouchardWarin10} implementation of RMC that utilizes a hierarchical (piecewise) linear model based on an equi-probable partition of the forward trajectory sites. The simulation design is the same as in \texttt{osp.prob.design}. This solver does not interface with \texttt{forward.sim.policy}; instead it directly accepts a collection of test trajectories that are evaluated in parallel with the backward DP iteration.
\item
  \texttt{osp.seq.design} -- sequential RMC with Gaussian Process-based emulators that are used to construct sequential design acquisition functions via the posterior emulator variance. Due to the overhead of adaptive simulation design, use of replication is strongly encouraged and number of sequential rounds should not exceed a couple of hundred.
\item
  \texttt{osp.seq.batch.design} -- sequential RMC with adaptive batching based on the heuristics in \citep{Lyu20}.
\end{itemize}

\hypertarget{appendix-c.3-supported-regression-modules}{%
\section*{Appendix C.3: Supported Regression Modules}\label{appendix-c.3-supported-regression-modules}}

The \texttt{method} field of the \textbf{osp.prob.design} solver supports the following regression modules:

\begin{itemize}
\item
  \texttt{lm}: linear model, specified through the set of user-provided basis functions, defined via the \texttt{bases} parameter;
\item
  \texttt{loess}: local linear regression (only works in 1D and 2D), specified via \texttt{lo.span};
\item
  \texttt{rf}: random forest model, specified through the number of trees \texttt{rf.ntree} and \texttt{rf.maxnode};
\item
  \texttt{earth}: multivariate adaptive regression spline (MARS) model from the \texttt{\{earth\}} package, specified through degree \texttt{earth.deg}, number of knots \texttt{earth.nk} and backward fit threshold \texttt{earth.thresh};
\item
  \texttt{spline}: smoothing splines (only in 1D), specified via number of knots \texttt{nk}; The \texttt{cvspline} variant
  automatically picks number of knots via cross-validation.
\item
  \texttt{km}: Gaussian process model from the \texttt{\{DiceKriging\}} \citep{kmPackage-R} package with fixed GP hyperparameters given through \texttt{km.var}, \texttt{km.cov} and \texttt{kernel.family};
\item
  \texttt{trainkm}: GP model trained via the MLE optimizer as in \texttt{\{DiceKriging\}};
\item
  \texttt{hetgp}: heteroskedastic Gaussian process regression based on the eponymous \texttt{\{hetGP\}} package from \citep{binois2016practical};
\item
  \texttt{homtp}: homoskedastic \(t\)-Process regression also implemented in \texttt{\{hetGP\}};
\item
  \texttt{rvm}: relevance vector machine kernel regression from \texttt{\{kernlab\}}. The kernel family is specified via \texttt{rvm.kernel}; if not specified uses \texttt{rbfdot} kernel by default;
\item
  \texttt{nnet}: single-layer neural network specified via the number of neurons \texttt{nn.nodes}. Uses a linear activation function and the \texttt{\{nnet\}} package;
\item
  \texttt{npreg}: kernel regression using the \texttt{\{np\}} package. The bandwidth is estimated using least squares cross-validation (default \texttt{npreg} option). The respective parameters are \texttt{np.kertype} (default ``gaussian''); \texttt{np.regtype} (default ``lc'') and \texttt{np.kerorder} (default ``2'').
\item
  \texttt{dynatree}: dynamic trees using the \texttt{\{dynaTree\}} package. The respective regression parameters are \texttt{dt.type} (``constant''
  or ``linear'' fits at the leafs), \texttt{dt.Npart} (number of trees), \texttt{dt.minp} (minimum size of each partition) and \texttt{dt.ab} (the tree prior parameter).
\end{itemize}

\hypertarget{appendix-d-benchmarked-osp-instances-specifications-m1-m9}{%
\section*{Appendix D: Benchmarked OSP Instances Specifications: M1-M9}\label{appendix-d-benchmarked-osp-instances-specifications-m1-m9}}
\addcontentsline{toc}{section}{Appendix D: Benchmarked OSP Instances Specifications: M1-M9}

\hypertarget{d}{%
\subsubsection{1D}\label{d}}

\textbf{M1}: 1D at-the-money Put \(X(0) = 40 = {\cal K}\) with Black-Scholes GBM dynamics, and 25 time periods, cf.~\citep{LS}.

\begin{Shaded}
\begin{Highlighting}[]
\NormalTok{BModel }\OtherTok{\textless{}{-}} \FunctionTok{list}\NormalTok{()}
\NormalTok{BModel[[}\DecValTok{1}\NormalTok{]] }\OtherTok{\textless{}{-}} \FunctionTok{list}\NormalTok{(}\AttributeTok{dim=}\DecValTok{1}\NormalTok{,}
            \AttributeTok{sim.func=}\NormalTok{sim.gbm,}
            \AttributeTok{K=}\DecValTok{40}\NormalTok{, }
            \AttributeTok{payoff.func=}\NormalTok{put.payoff,}
            \AttributeTok{x0=}\DecValTok{40}\NormalTok{,}
            \AttributeTok{sigma=}\FloatTok{0.2}\NormalTok{,}
            \AttributeTok{r=}\FloatTok{0.06}\NormalTok{,}
            \AttributeTok{div=}\DecValTok{0}\NormalTok{,}
            \AttributeTok{T=}\DecValTok{1}\NormalTok{,}\AttributeTok{dt=}\FloatTok{0.04}\NormalTok{)}
\end{Highlighting}
\end{Shaded}

\textbf{M2}: 1D out-of-the-money Put \(X(0)=44 > {\cal K}=40\) with same Black-Scholes GBM dynamics as M1.

\begin{Shaded}
\begin{Highlighting}[]
\NormalTok{BModel[[}\DecValTok{2}\NormalTok{]] }\OtherTok{\textless{}{-}} \FunctionTok{list}\NormalTok{(}\AttributeTok{dim=}\DecValTok{1}\NormalTok{,}
            \AttributeTok{sim.func=}\NormalTok{sim.gbm,}
            \AttributeTok{K=}\DecValTok{40}\NormalTok{, }
            \AttributeTok{payoff.func=}\NormalTok{put.payoff,}
            \AttributeTok{x0=}\DecValTok{44}\NormalTok{,}
            \AttributeTok{sigma=}\FloatTok{0.2}\NormalTok{,}
            \AttributeTok{r=}\FloatTok{0.06}\NormalTok{,}
            \AttributeTok{div=}\DecValTok{0}\NormalTok{,}
            \AttributeTok{T=}\DecValTok{1}\NormalTok{,}\AttributeTok{dt=}\FloatTok{0.04}\NormalTok{)}
\end{Highlighting}
\end{Shaded}

\hypertarget{d-1}{%
\subsubsection{2D}\label{d-1}}

\textbf{M3}: 2D Basket Put with independent and identical GBM dynamics for each asset. At-the-money initial condition \(X_1(0)=X_2(0)=40={\cal K}\) and 25 time steps.

\begin{Shaded}
\begin{Highlighting}[]
\NormalTok{BModel[[}\DecValTok{3}\NormalTok{]] }\OtherTok{\textless{}{-}} \FunctionTok{list}\NormalTok{(}\AttributeTok{dim=}\DecValTok{2}\NormalTok{,}
                    \AttributeTok{K=}\DecValTok{40}\NormalTok{,}
                    \AttributeTok{x0=}\FunctionTok{rep}\NormalTok{(}\DecValTok{40}\NormalTok{,}\DecValTok{2}\NormalTok{),}
                    \AttributeTok{sigma=}\FunctionTok{rep}\NormalTok{(}\FloatTok{0.2}\NormalTok{,}\DecValTok{2}\NormalTok{),}
                    \AttributeTok{r=}\FloatTok{0.06}\NormalTok{,}\AttributeTok{div=}\DecValTok{0}\NormalTok{,}
                    \AttributeTok{T=}\DecValTok{1}\NormalTok{,}\AttributeTok{dt=}\FloatTok{0.04}\NormalTok{,}
                    \AttributeTok{sim.func=}\NormalTok{sim.gbm, }
                    \AttributeTok{payoff.func=}\NormalTok{put.payoff)}
\end{Highlighting}
\end{Shaded}

\textbf{M4}: 2D Max Call with independent and identical GBM dynamics for each asset with high dividend yield \(\delta = 0.1 > r = 0.06\). In-the-money initial condition, \(X_1(0)=X_2(0)=110 > {\cal K} = 100\) and 9 time steps.

\begin{Shaded}
\begin{Highlighting}[]
\NormalTok{BModel[[}\DecValTok{4}\NormalTok{]]}\OtherTok{\textless{}{-}} \FunctionTok{list}\NormalTok{(}\AttributeTok{dim=}\DecValTok{2}\NormalTok{,}
                   \AttributeTok{K=}\DecValTok{100}\NormalTok{, }
                   \AttributeTok{r=}\FloatTok{0.05}\NormalTok{, }
                   \AttributeTok{div=}\FloatTok{0.1}\NormalTok{, }
                   \AttributeTok{sigma=}\FunctionTok{rep}\NormalTok{(}\FloatTok{0.2}\NormalTok{,}\DecValTok{2}\NormalTok{),}
                   \AttributeTok{T=}\DecValTok{3}\NormalTok{, }\AttributeTok{dt=}\DecValTok{1}\SpecialCharTok{/}\DecValTok{3}\NormalTok{,}
                   \AttributeTok{x0=}\FunctionTok{rep}\NormalTok{(}\DecValTok{110}\NormalTok{,}\DecValTok{2}\NormalTok{),}
                   \AttributeTok{sim.func=}\NormalTok{sim.gbm,}
                   \AttributeTok{payoff.func=}\NormalTok{ maxi.call.payoff)}
\end{Highlighting}
\end{Shaded}

\textbf{M5}: Put within a 1-factor Stochastic Volatility model from \citep{Rambharat11}. In-the-money initial condition \(S(0)=90 < {\cal K}=100\) and 50 time-steps.

\begin{Shaded}
\begin{Highlighting}[]
\NormalTok{BModel[[}\DecValTok{5}\NormalTok{]] }\OtherTok{\textless{}{-}} \FunctionTok{list}\NormalTok{(}\AttributeTok{K=}\DecValTok{100}\NormalTok{,}
                    \AttributeTok{x0=}\FunctionTok{c}\NormalTok{(}\DecValTok{90}\NormalTok{, }\FunctionTok{log}\NormalTok{(}\FloatTok{0.35}\NormalTok{)),}
                    \AttributeTok{r=}\FloatTok{0.0225}\NormalTok{,}\AttributeTok{div=}\DecValTok{0}\NormalTok{,}\AttributeTok{sigma=}\DecValTok{1}\NormalTok{,}
    \AttributeTok{T=}\DecValTok{50}\SpecialCharTok{/}\DecValTok{252}\NormalTok{,}\AttributeTok{dt=}\DecValTok{1}\SpecialCharTok{/}\DecValTok{252}\NormalTok{,}
    \AttributeTok{svAlpha=}\FloatTok{0.015}\NormalTok{,}\AttributeTok{svEpsY=}\DecValTok{1}\NormalTok{,}\AttributeTok{svVol=}\DecValTok{3}\NormalTok{,}\AttributeTok{svRho=}\SpecialCharTok{{-}}\FloatTok{0.03}\NormalTok{,}\AttributeTok{svMean=}\FloatTok{2.95}\NormalTok{,}
    \AttributeTok{eulerDt=}\DecValTok{1}\SpecialCharTok{/}\DecValTok{2520}\NormalTok{, }\AttributeTok{dim=}\DecValTok{2}\NormalTok{,}
    \AttributeTok{sim.func=}\NormalTok{sim.expOU.sv,}
    \AttributeTok{payoff.func =}\NormalTok{sv.put.payoff)}

\CommentTok{\#putPr \textless{}{-} osp.probDesign.piecewisebw(40000,modelSV5)}
 \CommentTok{\# get putPr$price= 16.81677}
\end{Highlighting}
\end{Shaded}

\hypertarget{d-and-above}{%
\subsubsection{3D and above}\label{d-and-above}}

\textbf{M6}: 3D Max Call from \citep[\citet{BroadieCao08}]{Broadie}. Independent and identical GBM dynamics for each asset with high dividend yield \(\delta = 0.1 > r = 0.05\). Out-of-the-money initial condition \(X_1(0)=X_2(0)=X_3(0) = 90 < {\cal K} = 100\) and nine time-steps.

\begin{Shaded}
\begin{Highlighting}[]
\NormalTok{BModel[[}\DecValTok{6}\NormalTok{]]}\OtherTok{\textless{}{-}} \FunctionTok{list}\NormalTok{(}\AttributeTok{dim=}\DecValTok{3}\NormalTok{,}
                   \AttributeTok{K=}\DecValTok{100}\NormalTok{, }
                   \AttributeTok{r=}\FloatTok{0.05}\NormalTok{, }
                   \AttributeTok{div=}\FloatTok{0.1}\NormalTok{, }
                   \AttributeTok{sigma=}\FunctionTok{rep}\NormalTok{(}\FloatTok{0.2}\NormalTok{,}\DecValTok{3}\NormalTok{),}
                   \AttributeTok{T=}\DecValTok{3}\NormalTok{, }\AttributeTok{dt=}\DecValTok{1}\SpecialCharTok{/}\DecValTok{3}\NormalTok{,}
                   \AttributeTok{x0=}\FunctionTok{rep}\NormalTok{(}\DecValTok{90}\NormalTok{,}\DecValTok{3}\NormalTok{),}
                   \AttributeTok{sim.func=}\NormalTok{sim.gbm,}
                   \AttributeTok{payoff.func=}\NormalTok{ maxi.call.payoff)}
\end{Highlighting}
\end{Shaded}

\textbf{M7}: 5D Max Call from \citep{BroadieCao08}; matches the setting of M6 but with 5 assets rather than three and different initial condition. Independent and identical GBM dynamics for each asset with high dividend yield \(\delta = 0.1 > r = 0.05\). At-the-money initial condition \(X_1(0)=X_2(0)=X_3(0) = X_4(0) = X_5(0) = 100 = {\cal K}\) and nine time-steps.

\begin{Shaded}
\begin{Highlighting}[]
\NormalTok{BModel[[}\DecValTok{7}\NormalTok{]] }\OtherTok{\textless{}{-}} \FunctionTok{list}\NormalTok{(}\AttributeTok{dim=}\DecValTok{5}\NormalTok{, }
                \AttributeTok{sim.func=}\NormalTok{sim.gbm, }
                \AttributeTok{r=}\FloatTok{0.05}\NormalTok{, }
                \AttributeTok{div=}\FloatTok{0.1}\NormalTok{, }
                \AttributeTok{sigma=}\FunctionTok{rep}\NormalTok{(}\FloatTok{0.2}\NormalTok{,}\DecValTok{5}\NormalTok{), }
                \AttributeTok{x0=}\FunctionTok{rep}\NormalTok{(}\DecValTok{100}\NormalTok{, }\DecValTok{5}\NormalTok{),  }\CommentTok{\# also 70, 130}
                \AttributeTok{payoff.func=}\NormalTok{maxi.call.payoff, }
                \AttributeTok{K=}\DecValTok{100}\NormalTok{, }
                \AttributeTok{T=}\DecValTok{3}\NormalTok{, }
                \AttributeTok{dt=}\DecValTok{1}\SpecialCharTok{/}\DecValTok{3}\NormalTok{)}

\CommentTok{\# paper\_result=c(3.892,26.12,59.235) for S\_0 = 70, 100, 130}
\end{Highlighting}
\end{Shaded}

\textbf{M8}: 5D asymmetric Max Call from \citep{CheriditoJentzen18}. Independent GBM dynamics for each asset with progressively higher volatilities and identical dividend yields \(\delta_i = 0.1 > r = 0.05\). Out-of-the-money initial condition \(X_i(0) = 70 < {\cal K} = 100\) for \(i=1,\ldots, 5\) and 9 time-steps.

\begin{Shaded}
\begin{Highlighting}[]
\NormalTok{BModel[[}\DecValTok{8}\NormalTok{]] }\OtherTok{\textless{}{-}} \FunctionTok{list}\NormalTok{(}\AttributeTok{dim=}\DecValTok{5}\NormalTok{, }
                \AttributeTok{sim.func=}\NormalTok{sim.gbm, }
                \AttributeTok{r=}\FloatTok{0.05}\NormalTok{, }
                \AttributeTok{div=}\FloatTok{0.1}\NormalTok{, }
                \AttributeTok{sigma=}\FunctionTok{c}\NormalTok{(}\FloatTok{0.08}\NormalTok{,}\FloatTok{0.16}\NormalTok{,}\FloatTok{0.24}\NormalTok{,}\FloatTok{0.32}\NormalTok{,}\FloatTok{0.4}\NormalTok{), }
                \AttributeTok{x0=}\FunctionTok{rep}\NormalTok{(}\DecValTok{70}\NormalTok{, }\DecValTok{5}\NormalTok{),  }
                \AttributeTok{payoff.func=}\NormalTok{maxi.call.payoff, }
                \AttributeTok{K=}\DecValTok{100}\NormalTok{, }
                \AttributeTok{T=}\DecValTok{3}\NormalTok{, }
                \AttributeTok{dt=}\DecValTok{1}\SpecialCharTok{/}\DecValTok{3}\NormalTok{)}

\CommentTok{\# paper\_result=c(11.756,37.730,73.709) for S\_0 = 70, 100, 130}
\end{Highlighting}
\end{Shaded}

\textbf{M9}: 5D Basket Average Put from \citep{Lelong19}. Correlated GBM dynamics with identical volatilities \(\sigma_i = 0.2\) and constant cross-asset correlation \(\rho_{ij}=0.2\) for all \(i,j=1,\ldots, 5\). Zero dividend yield. At-the-money initial condition \(X_i(0)=100 = {\cal K}\) and 20 time-steps.

\begin{Shaded}
\begin{Highlighting}[]
\NormalTok{BModel[[}\DecValTok{9}\NormalTok{]] }\OtherTok{\textless{}{-}} \FunctionTok{list}\NormalTok{(}\AttributeTok{dim=}\DecValTok{5}\NormalTok{, }
                \AttributeTok{sim.func=}\NormalTok{sim.gbm.cor, }
                \AttributeTok{r=}\FloatTok{0.05}\NormalTok{, }
                \AttributeTok{div=}\DecValTok{0}\NormalTok{, }
                \AttributeTok{sigma=}\FloatTok{0.2}\NormalTok{, }
                \AttributeTok{x0=}\FunctionTok{rep}\NormalTok{(}\DecValTok{100}\NormalTok{, }\DecValTok{5}\NormalTok{), }
                \AttributeTok{rho=}\FloatTok{0.2}\NormalTok{,}
                \AttributeTok{K=}\DecValTok{100}\NormalTok{,}
                \AttributeTok{payoff.func=}\NormalTok{put.payoff, }
                \AttributeTok{T=}\DecValTok{3}\NormalTok{, }
                \AttributeTok{dt=}\DecValTok{3}\SpecialCharTok{/}\DecValTok{20}\NormalTok{) }\CommentTok{\# 20 steps}
\CommentTok{\#paper\_result=4.254 }
\end{Highlighting}
\end{Shaded}

\renewcommand\refname{References}
  \bibliography{rmcBook}

\begin{thebibliography}{36}
\providecommand{\natexlab}[1]{#1}
\providecommand{\url}[1]{\texttt{#1}}
\expandafter\ifx\csname urlstyle\endcsname\relax
  \providecommand{\doi}[1]{doi: #1}\else
  \providecommand{\doi}{doi: \begingroup \urlstyle{rm}\Url}\fi

\bibitem[Andersen and Broadie(2004)]{Broadie}
L.~Andersen and M.~Broadie.
\newblock A primal-dual simulation algorithm for pricing multi-dimensional
  {A}merican options.
\newblock \emph{Management Science}, 50\penalty0 (9):\penalty0 1222--1234,
  2004.

\bibitem[Ankenman et~al.(2010)Ankenman, Nelson, and
  Staum]{ankenman2010stochastic}
B.~Ankenman, B.~L. Nelson, and J.~Staum.
\newblock Stochastic kriging for simulation metamodeling.
\newblock \emph{Operations Research}, 58\penalty0 (2):\penalty0 371--382, 2010.

\bibitem[Becker et~al.(2019{\natexlab{a}})Becker, Cheridito, and
  Jentzen]{CheriditoJentzen18}
S.~Becker, P.~Cheridito, and A.~Jentzen.
\newblock Deep optimal stopping.
\newblock \emph{Journal of Machine Learning Research}, 20:\penalty0 74,
  2019{\natexlab{a}}.

\bibitem[Becker et~al.(2019{\natexlab{b}})Becker, Cheridito, Jentzen, and
  Welti]{becker2019solving}
S.~Becker, P.~Cheridito, A.~Jentzen, and T.~Welti.
\newblock Solving high-dimensional optimal stopping problems using deep
  learning.
\newblock \emph{arXiv preprint arXiv:1908.01602}, 2019{\natexlab{b}}.

\bibitem[Beketov(2013)]{LSMonteCarlo}
M.~A. Beketov.
\newblock \emph{LSMonteCarlo: {A}merican options pricing with Least Squares
  {M}onte {C}arlo method}, 2013.
\newblock URL \url{https://CRAN.R-project.org/package=LSMonteCarlo}.
\newblock R package version 1.0.

\bibitem[Belomestny(2011)]{Belomestny11}
D.~Belomestny.
\newblock Pricing {B}ermudan options by nonparametric regression: optimal rates
  of convergence for lower estimates.
\newblock \emph{Finance and Stochastics}, 15\penalty0 (4):\penalty0 655--683,
  2011.

\bibitem[Belomestny and Schoenmakers(2018)]{BelomestnyBook}
D.~Belomestny and J.~Schoenmakers.
\newblock \emph{Advanced Simulation-Based Methods for Optimal Stopping and
  Control: With Applications in Finance}.
\newblock Palgrave Macmillan, London, 2018.

\bibitem[Binois et~al.(2018)Binois, Gramacy, and
  Ludkovski]{binois2016practical}
M.~Binois, R.~B. Gramacy, and M.~Ludkovski.
\newblock Practical heteroscedastic {G}aussian process modeling for large
  simulation experiments.
\newblock \emph{Journal of Computational and Graphical Statistics}, 27\penalty0
  (4):\penalty0 808--821, 2018.

\bibitem[Bouchard and Warin(2011)]{BouchardWarin10}
B.~Bouchard and X.~Warin.
\newblock Monte-{C}arlo valorisation of {A}merican options: facts and new
  algorithms to improve existing methods.
\newblock In R.~Carmona, P.~D. Moral, P.~Hu, and N.~Oudjane, editors,
  \emph{Numerical Methods in Finance}, volume~12 of \emph{Springer Proceedings
  in Mathematics}. Springer, 2011.

\bibitem[Broadie and Cao(2008)]{BroadieCao08}
M.~Broadie and M.~Cao.
\newblock Improved lower and upper bound algorithms for pricing {A}merican
  options by simulation.
\newblock \emph{Quantitative Finance}, 8\penalty0 (8):\penalty0 845--861, 2008.

\bibitem[Carmona and Touzi(2008)]{CarmonaTouzi}
R.~Carmona and N.~Touzi.
\newblock Optimal multiple stopping and valuation of swing options.
\newblock \emph{Mathematical Finance}, 18\penalty0 (2):\penalty0 239--268,
  2008.

\bibitem[Chen and Wan(2020)]{chen2020deep}
Y.~Chen and J.~W. Wan.
\newblock Deep neural network framework based on backward stochastic
  differential equations for pricing and hedging {A}merican options in high
  dimensions.
\newblock \emph{Quantitative Finance}, pages 1--23, 2020.

\bibitem[Egloff et~al.(2007)Egloff, Kohler, and
  Todorovic]{EgloffKohlerTodorovic07}
D.~Egloff, M.~Kohler, and N.~Todorovic.
\newblock A dynamic look-ahead {M}onte {C}arlo algorithm for pricing {B}ermudan
  options.
\newblock \emph{The Annals of Applied Probability}, 17\penalty0 (4):\penalty0
  1138--1171, 2007.

\bibitem[Gevret et~al.(2018)Gevret, Langren{\'e}, Lelong, Warin, and
  Maheshwari]{StOpt}
H.~Gevret, N.~Langren{\'e}, J.~Lelong, X.~Warin, and A.~Maheshwari.
\newblock \emph{{STochastic OPTimization library in C++}}, 2018.

\bibitem[Goudenege et~al.(2019)Goudenege, Molent, Zanette,
  et~al.]{GoudenegeZanette19}
L.~Goudenege, A.~Molent, A.~Zanette, et~al.
\newblock Machine learning for pricing {A}merican options in high dimension.
\newblock \emph{arXiv preprint arXiv:1903.11275}, 2019.

\bibitem[Gouden{\`e}ge et~al.(2020)Gouden{\`e}ge, Molent, and
  Zanette]{GoudenegeZanette20}
L.~Gouden{\`e}ge, A.~Molent, and A.~Zanette.
\newblock Machine learning for pricing {A}merican options in high-dimensional
  {M}arkovian and non-{M}arkovian models.
\newblock \emph{Quantitative Finance}, 20\penalty0 (4):\penalty0 573--591,
  2020.

\bibitem[{Gramacy} and {Apley}(2015)]{GramacyApley13}
R.~{Gramacy} and D.~{Apley}.
\newblock Local {G}aussian process approximation for large computer
  experiments.
\newblock \emph{Journal of Computational and Graphical Statistics}, 24\penalty0
  (2):\penalty0 561--578, 2015.

\bibitem[Gramacy and Ludkovski(2015)]{GL13}
R.~Gramacy and M.~Ludkovski.
\newblock Sequential design for optimal stopping problems.
\newblock \emph{SIAM Journal on Financial Mathematics}, 6\penalty0
  (1):\penalty0 748--775, 2015.
\newblock URL \url{http://arXiv.org/abs/1309.3832}.

\bibitem[Herrera et~al.(2021)Herrera, Krach, Ruyssen, and
  Teichmann]{herreraTeichmann}
C.~Herrera, F.~Krach, P.~Ruyssen, and J.~Teichmann.
\newblock Optimal stopping via randomized neural networks.
\newblock \emph{arXiv preprint arXiv:2104.13669}, 2021.

\bibitem[Kohler(2008)]{Kohler08spline}
M.~Kohler.
\newblock A regression-based smoothing spline {M}onte {C}arlo algorithm for
  pricing {A}merican options in discrete time.
\newblock \emph{Advances in Statistical Analysis}, 92\penalty0 (2):\penalty0
  153--178, 2008.

\bibitem[Kohler(2010)]{Kohler10review}
M.~Kohler.
\newblock A review on regression-based {M}onte {C}arlo methods for pricing
  {A}merican options.
\newblock In \emph{Recent Developments in Applied Probability and Statistics},
  pages 37--58. Springer, 2010.

\bibitem[Kohler and Krzy{\.z}ak(2012)]{Kohler12lasso}
M.~Kohler and A.~Krzy{\.z}ak.
\newblock Pricing of {A}merican options in discrete time using least squares
  estimates with complexity penalties.
\newblock \emph{Journal of Statistical Planning and Inference}, 142\penalty0
  (8):\penalty0 2289--2307, 2012.

\bibitem[Kohler et~al.(2010)Kohler, Krzy{\.z}ak, and Todorovic]{Kohler10nn}
M.~Kohler, A.~Krzy{\.z}ak, and N.~Todorovic.
\newblock Pricing of high-dimensional {A}merican options by neural networks.
\newblock \emph{Mathematical Finance}, 20\penalty0 (3):\penalty0 383--410,
  2010.

\bibitem[Lelong(2019)]{Lelong19}
J.~Lelong.
\newblock Pricing path-dependent {B}ermudan options using {W}iener chaos
  expansion: an embarrassingly parallel approach.
\newblock \emph{arXiv preprint arXiv:1901.05672}, 2019.

\bibitem[Longstaff and Schwartz(2001)]{LS}
F.~Longstaff and E.~Schwartz.
\newblock Valuing {A}merican options by simulations: a simple least squares
  approach.
\newblock \emph{The Review of Financial Studies}, 14:\penalty0 113--148, 2001.

\bibitem[Ludkovski(2018)]{Lu18}
M.~Ludkovski.
\newblock Kriging metamodels and experimental design for {B}ermudan option
  pricing.
\newblock \emph{Journal of Computational Finance}, 22\penalty0 (1):\penalty0
  37--77, 2018.

\bibitem[Ludkovski(2020)]{mlOSP}
M.~Ludkovski.
\newblock \emph{mlOSP: Regression {M}onte {C}arlo Algorithms for Optimal
  Stopping}, 2020.
\newblock R package version 1.0.

\bibitem[Ludkovski(2022)]{ludkovski2022impulse}
M.~Ludkovski.
\newblock Regression {M}onte {C}arlo for impulse control.
\newblock \emph{arXiv preprint arXiv:2203.06539}, 2022.

\bibitem[Lyu and Ludkovski(2020)]{Lyu20}
X.~Lyu and M.~Ludkovski.
\newblock Adaptive batching for {G}aussian process surrogates with application
  in noisy level set estimation.
\newblock \emph{arXiv preprint arXiv:2003.08579}, 2020.

\bibitem[Nadarajah et~al.(2017)Nadarajah, Margot, and
  Secomandi]{NadarajahSecomandi17}
S.~Nadarajah, F.~Margot, and N.~Secomandi.
\newblock Comparison of least squares {M}onte {C}arlo methods with applications
  to energy real options.
\newblock \emph{European Journal of Operational Research}, 256\penalty0
  (1):\penalty0 196--204, 2017.

\bibitem[Rambharat and Brockwell(2010)]{Rambharat11}
B.~R. Rambharat and A.~E. Brockwell.
\newblock Sequential {M}onte {C}arlo pricing of {A}merican-style options under
  stochastic volatility models.
\newblock \emph{The Annals of Applied Statistics}, 4\penalty0 (1):\penalty0
  222--265, 2010.

\bibitem[Reppen et~al.(2022)Reppen, Soner, and
  Tissot-Daguette]{reppen2022neural}
A.~M. Reppen, H.~M. Soner, and V.~Tissot-Daguette.
\newblock Neural optimal stopping boundary.
\newblock \emph{arXiv preprint arXiv:2205.04595}, 2022.

\bibitem[Roustant et~al.(2012)Roustant, Ginsbourger, Deville,
  et~al.]{kmPackage-R}
O.~Roustant, D.~Ginsbourger, Y.~Deville, et~al.
\newblock Dicekriging, diceoptim: Two r packages for the analysis of computer
  experiments by kriging-based metamodeling and optimization.
\newblock \emph{Journal of Statistical Software}, 51\penalty0 (1), 2012.

\bibitem[Tompaidis and Yang(2013)]{TompaidisYang13}
S.~Tompaidis and C.~Yang.
\newblock Pricing {A}merican-style options by {M}onte {C}arlo simulation:
  Alternatives to ordinary least squares.
\newblock \emph{Journal of Computational Finance}, 18\penalty0 (1):\penalty0
  121--143, 2013.

\bibitem[Tsitsiklis and {Van Roy}(2001)]{TsitsiklisVanRoy}
J.~Tsitsiklis and B.~{Van Roy}.
\newblock Regression methods for pricing complex {A}merican-style options.
\newblock \emph{IEEE Transactions on Neural Networks}, 12\penalty0
  (4):\penalty0 694--703, July 2001.

\bibitem[Yee(2018)]{rlsm}
J.~Yee.
\newblock rlsm: {R} package for least squares {M}onte {C}arlo.
\newblock \emph{arXiv preprint arXiv:1801.05554}, 2018.

\end{thebibliography}

\end{document}